\documentclass{mn2e}
\input{psfig.sty}
\usepackage{mnras_cite}
\newcommand{\apj}{Astrophys. J.}
\newcommand{\aj}{Astron. J.}
\newcommand{\apjs}{Astrophys. J. Supp.}

\newcommand{\mnras}{Mon. Not. R. Astron. Soc.}

\def\sun{\hbox{$\odot$}}

\newdimen\hssize
\newdimen\hdsize 
\begin{document}

\title[Redshift Evolution of the Galaxy LF]{A Brighter Past: Galaxy Luminosity Function At High Redshifts}
\author[Cooray]{Asantha Cooray\\ 
Center for Cosmology, Department of Physics and Astronomy, University of California, Irvine, CA 92697\\
E-mail:acooray@uci.edu}

\maketitle

\begin{abstract}
Using the conditional luminosity function --- the luminosity distribution of
 galaxies in a dark matter halo as a function of the halo mass  ---
we present an empirical model to describe the redshift evolution of the rest B-band galaxy luminosity function (LF).
The model is compared to various estimates of the LF, in rest UV- and B-bands, out to a redshift of 6, including estimates of 
LFs of galaxy types separated to red and blue galaxies. 
Using the observed LFs out to $z \sim 5$, we present a general constraint on the redshift
evolution of the central galaxy-halo mass relation.
The increase in the number density of luminous galaxies, at the bright-end of the LF,
can be explained as due to a brightening of the luminosity of galaxies present in 
dark matter halo centers, relative to the luminosity of central galaxies in similar mass halos today.
The lack of strong evolution in the faint-end of the LF, however,  argues against a model involving pure
luminosity evolution at all halo mass scales. The increase in luminosity at the bright-end
compensates the rapid decline in the number density of massive halos as the redshift is increased.
The decline in group to cluster-mass dark matter halos out to a redshift of $\sim$ 2 is not important
as the central galaxy luminosity flattens at halo masses around 10$^{13}$ M$_{\sun}$.
At redshifts $\sim$ 2 to 3, however, the density of bright galaxies begins to decrease due to the rapid decline in the number density
of dark matter halos at mass scales around and below 10$^{13}$ M$_{\sun}$.

Based on a comparison of our predictions to the measured UV LF of galaxies at redshifts $\sim$ 3,
we estimate the probability distribution of halo masses to host Lyman break galaxies (LBGs).
This probability for galaxies brighter than AB-absolute magnitudes of -21, with a number density of
$\sim 5  \times 10^{-3}$ h$^{3}$ Mpc$^{-3}$, peaks at a halo mass of $\sim 7 \times 10^{11}$ h$^{-1}$ M$_{\sun}$
with a 68\% confidence level of $(4$ -- $21) \times 10^{11}$ h$^{-1}$ M$_{\sun}$.
These estimates are consistent with the mass estimates for LBGs using two-point clustering statistics
and recent estimates of halo masses based on spectroscopic observations.
For galaxies brighter than AB-absolute magnitudes of -21 at $z \sim 6$, the halo mass scale is a factor of $\sim$ 2 smaller;
the LF predictions at $z \sim 6$ are consistent with measured estimates in the literature.
Based on the models, we also predict the LF of galaxies at redshifts greater than 6 and also the bias factor of
galaxies at redshifts greater than 3; these predictions will soon be tested with observational data.
In general, to explain high-redshift LFs, galaxies in dark matter halos around 10$^{12}$ M$_{\sun}$
must increase in luminosity by a factor of $\sim$ 4 to 6 between today and redshift of 6. 
\end{abstract}


\section{Introduction}

The luminosity function (LF) of field galaxies at the present day can be described using
a simple empirical model that involves the dark matter halo mass function and the relation between
central galaxy luminosity and the mass of the halo occupied (Cooray \& Milosavljevic 2005b; Cooray 2005).
The model utilizes an approach based on the  conditional luminosity function 
(CLF; Yang et al. 2003b, 2005), or the luminosity
distribution of galaxies as a function of the halo mass, $\Phi(L|M)$, to construct the  galaxy LF, $\Phi(L)$. The CLFs
are basically an extension of the halo approach to galaxy statistics 
(Seljak 2000; Scoccimarro et al. 2001; Cooray et al. 2000; see, Cooray \& Sheth 2002 for a review)
where instead of the halo occupation number as a function of the halo mass, $\langle N_g(M) \rangle$ ---
which describes the average number of galaxies in a dark matter halo  ---
we consider the number of galaxies in a given dark matter halo conditioned in terms of 
the galaxy luminosity such that $\Phi(L|M)=d\langle N_g(M) \rangle/dL$.
Other conditions of the halo occupation number has been suggested (such as in terms of the stellar mass in
Zheng et al. 2004), but for comparison with observational data, a halo occupation number conditioned in terms of
galaxy luminosity and type is most useful.

As the halo occupation statistics are best described based on a division to central and satellite
galaxies (Kravtsov et al. 2003), we also divide the CLF to central galaxies and
satellites; central galaxies are assigned a log-normal distribution in the luminosity, centered
around the mean central galaxy luminosity given the halo mass (Cooray \& Milosavljevic 2005a), 
while satellite galaxies are assigned a power-law distribution in luminosity.  
This empirical approach has the main advantage that it can elucidate important aspects associated with the
galaxy distribution as measured by surveys of the large scale structure.
Statistical measurements from observations mostly include galaxy LFs, both field
and cluster galaxies, and statistics related to galaxy
clustering such as the large-scale bias factor or the correlation length. 
Thus, we use CLFs to model same statistics here. 

Previous studies using the halo model to describe galaxy statistics concentrated primarily on average 
clustering properties, such as the galaxy power spectrum 
(e.g., Seljak 2000; Scoccimarro et al. 2001; Cooray 2002; Berlind et al. 2003); 
for these statistics the conditional occupation number, either
in luminosity or other galaxy property, is not needed as the statistic
depends simply on the total number of galaxies in a given halo.
With large data sets, on the other hand, measurements can be made with the galaxy sample divided to various physical
properties such as the luminosity, color, environment, etc (e.g., Zehavi et al. 2004 who considered clustering of
Sloan galaxies as a function of the luminosity). 
In this scenario, to compare
observations with analytical or numerical models, the average halo occupation number
must be conditioned in terms of the physical property. Similarly, wide-field surveys at high redshifts have
now allowed detailed measurements related to the redshift evolution of the LF. Thus, one must 
also account for redshift variations in the CLF. Here, we consider the latter application and
improve prior analytical models of the LF by discussing redshift dependence of the CLF.

In Cooray \& Milosavljevi\'c (2005b), we used the
CLF  to explain  why the LF can be  described with the Schechter (1976) form of 
$\Phi(L) \propto (L/L_\star)^\alpha \exp(-L/L_\star)$. In this approach
the main ingredient, in addition to the halo mass function,
is the relation between central galaxy luminosity and the halo mass, hereafter called the $L_c(M)$ relation.
This relation, as appropriate for galaxies at low-redshifts, was established in 
Cooray \& Milosavljevic (2005a)  from a combination of weak lensing (e.g., Yang et al. 2003a)
and direct measurements of galaxy luminosity and mass in groups and clusters (e.g., Lin et al. 2004).
The same relation has been established with a statistical analysis of the 2dFGRS $b_J$-band LF 
(e.g., Norberg et al. 2002) by Vale \& Ostriker (2004) and, 
independently, by Yang et al. (2005) based on the 2dF Galaxy Redshift Survey (2dFGRS; Colles et al. 2001) galaxy
group catalog.  The shape of the $L_c(M)$ relation, where luminosities grow rapidly with increasing mass but
flattens at a mass scale around $\sim$ 10$^{13}$ M$_{\sun}$ is best explained through dissipationless 
merging history of central galaxies (Cooray \& Milosavljevic 2005a).

In Cooray \& Milosavljevi\'c (2005b), we related the two parameters of the
Schechter (1976) LF involving the slope at low luminosities, $\alpha$, and the exponential cut-off
luminosity, $L_*$, to a combination of the mass function slope and the slope of the $L_c(M)$ relation.
In this analytical model, we expect $\alpha < -1.25$, consistent with observations that  indicate
 $\alpha \approx -1.3$ (Blanton et al. 2004; Huang et al. 2003). The characteristic scale $L_*$ comes
about when the luminosity scatter in the $L_c(M)$ relation dominates over the increase in the
luminosity with mass or when $d \ln L_c/d \ln M \approx \ln(10) \Sigma$ where
$\Sigma$ is the dispersion in the $L_c(M)$ relation. Given the observed dispersion, we find
$M_\star \approx 2 \times 10^{13}$ M$_{\sun}$ and $L_\star = L_c(M_\star)$ to be consistent
with observed $L_\star$ values from Schechter (1976) function fits to the LF (Cooray \& Milosavljevic
2005a in the case of k-band observations and Cooray 2005 in the case of 2dFGRS  $b_J$ band).

The empirical modeling approach can easily be extended to consider statistics of galaxy types or color as well.
For example, in Cooray (2005), we studied the environmental dependence of galaxy colors, broadly
divided to blue and red types given the bimodal nature of the color distribution 
(e.g., Baldry et al. 2004; Balogh et al. 2004). There, we described the
conditional type-dependent LFs from 2dFGRS (Croton et al. 2004) 
as a function of the galaxy overdensity based on an empirical description of blue-to-red galaxy
fraction in dark matter halos as a function of the halo mass. With an increasing fraction  of
early-type, or red, galaxies with increasing halo mass, the simple analytical model
considered in Cooray (2005) explain why the LF of galaxies in
dense environments are dominated by these galaxies.

In addition to galaxy statistics today from wide-field  
redshift surveys such as 2dFGRS or
 Sloan Digital Sky Survey (SDSS; York et al. 2000),  various techniques, such as the Lyman drop-out
method (e.g., Steidel et al. 1999),  have now allowed the study of galaxy LF and related statistics on the
galaxy distribution at high redshifts. Spectroscopic redshift surveys, such as 
DEEP2 (Davis et al. 2003), and photometric data based redshift selections,
such COMBO-17 survey (Wolf et al. 2001, 2003),
 have now allowed detailed studies of galaxy properties at redshifts
around unity (e.g., Willmer et al. 2005; Faber et a. 2005 with DEEP2 and Bell et al. 2004 with COMBO-17).
Similar photometric redshift based studies extend the LF statistics to higher redshifts using
ground-based data (e.g., Gabasch et al. 2004 using ESO VLT's FORS Deep Field),
space-based data (e.g., Bouwens et al. 2004 using HST NICMOS Ultra Deep Field), or
a combination of space-based imaging and ground-based spectroscopic followup data (e.g., Giallongo et al. 2005).
While the LF of galaxies today can be described through the observationally established
$L_c(M)$ relation,  it is also useful to understand the extent to which the same empirical approach can be
applied at high redshifts. In return, using the measured galaxy LFs out to a redshift of 6,
we can attempt to extract information on how galaxy luminosities evolve as a function of redshift.
The model may then allow one to address if the galaxy formation was efficient in the past and how
galaxy properties are different when compared to properties today. When comparing to LF measurements,
we ignore any potential systematics in these data such as due to selection effects and
biases that may have affected the LF measurements. We make the assumption that, if any biases
or selection effects exist, these effects have been considered and that the LFs are properly corrected to
account for them. Thus, our model comparisons may only be accurate to the same extent that
measurements of high redshift LFs can be considered reliable.

Here, we compare our predictions to the measured rest frame B-band
LF of galaxies, including red and blue galaxies, out to a redshift of 1.2
from DEEP2 (Willmer et al. 2005) and COMBO-17 (Bell et al. 2004), out to a redshift of 5 from
Gabasch et al. (2004) and out to a redshift of 3.5 from Giallongo et al. (2005).
We use the latter data set from Giallongo et al. (2005) to build out models but then perform
a detailed model fit to other data sets by varying some of the parameters related to redshift evolution of
the $L_c(M)$ relation.
We also make comparisons to rest-UV LFs of galaxies at redshifts 3 to 6 from Steidel et al. (1999) and Bouwens et al. (2004a),
since there surveys provide an additional data sets to compare with models at the highest redshift ranges surveyed so far.
While previous studies have measured the redshift dependence of the LF, say in the
K-band (e.g., Drory et al. 2003), given that the LF corresponds to
different rest wavelengths as a function of redshift, any evolutionary aspects associated with galaxy properties,
at a given wavelength, must be distinguished from evolutionary effects resulting from color differences.
The B-band LFs considered for modeling here have the advantage that one can directly address how galaxy properties
change with redshift at the same band, regardless of color differences.

The paper is organized as follows: In the next
 section, we will outline the basic ingredients in the empirical model  for CLFs
and how it is modified to model the LF at high redshifts. We refer the reader to Cooray (2005) and Cooray
\& Milosavljevi\'c (2005b) for initial discussions related to
this empirical modeling approach. In Section~3, we will describe the z-dependent LF and compare with measurements by
discussed above. We also compare our models to rest-UV LFs of galaxies between redshifts of 3 to 6,
and galaxy clustering bias around the same redshift ranges. We conclude with a summary of our main results
and implications related to the galaxy distribution at redshifts $\sim$ 3 to 6 in \S~4. 
Throughout the paper we assume cosmological parameters consistent with observational analyses of LF
measurements modeled here and take $\Omega_m=0.3$,
$\Omega_\Lambda=0.7$, and  a scaled Hubble constant of  $h=0.7$ in units of 100 km s$^{-1}$ Mpc$^{-1}$.
The matter power spectrum is normalized to a $\sigma_8$, rms fluctuations at 8 $h^{-1}$ Mpc scales, of
0.84 consistent with WMAP data (Spergel et al. 2003). 

\begin{figure*}
\centerline{\psfig{file=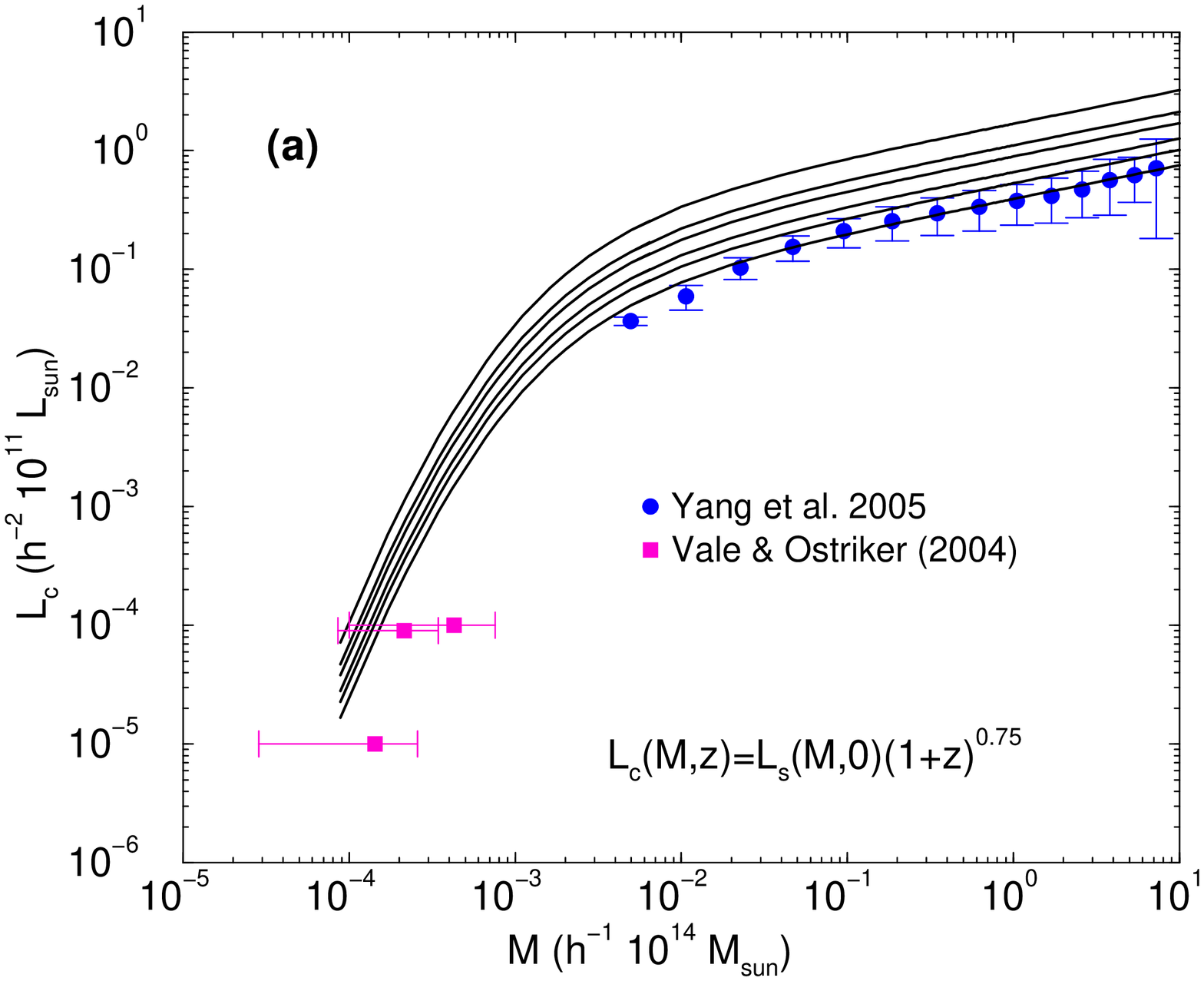,width=\hssize,angle=-90}
\psfig{file=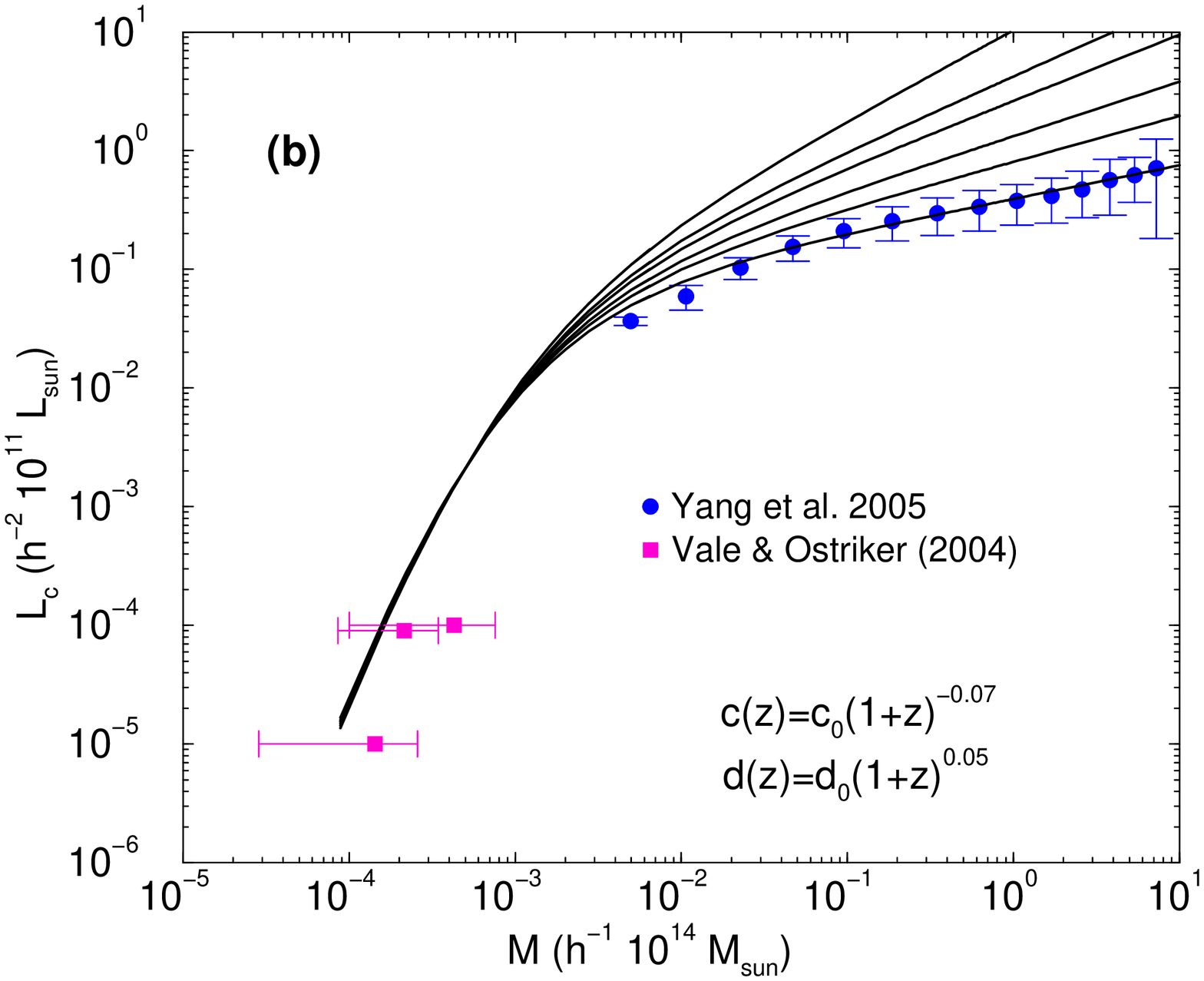,width=\hssize,angle=-90}}
\caption{Central galaxy luminosity as a function of the
halo mass as appropriate for 2dFGRS $b_J$-band. 
The data points at the low mass end are from Vale \& Ostriker (2004; {\it squares})
while at the high end are from Yang et al. (2005). The solid curves  are
the relation as a function of redshift; from bottom to top in each of the two panels,
the redshifts are 0, 0.5, 1, 2, 3, and 6, respectively. The relation at $z=0$ comes from
Vale \& Ostriker (2004) in the $b_J$ band of 2dFGRS by converting the luminosity function to
extract the plotted relation based on the sub-halo mass function. 
We use this relation to construct our
the CLF, and include a scatter in this relation in our description of central galaxy CLFs.
}
\end{figure*}

\section{Conditional Luminosity Function: Empirical Model}

In order to construct the redshift evolution of the luminosity function (LF), we follow Cooray \& Milosavljevi\'c (2005b) and Cooray (2005).
The redshift-dependent conditional luminosity function (CLF; Yang et al. 2003b, 2005), denoted by $\Phi(L|M,z)$, 
is the average number of galaxies with luminosities between $L$ and $L+dL$ that 
reside in halos of mass $M$ at a redshift of $z$. As in our application to 2dFGRS at $z=0$ (Cooray 2005), the CLF is separated 
into terms associated with central and satellite galaxies,  such that
\begin{eqnarray}
\Phi(L|M,z)&=&\Phi_{\rm c}(L|M,z)+\Phi_{\rm s}(L|M,z) \nonumber \\
\Phi_{\rm c}(L|M,z)  &=& \frac{f_{\rm c}(M,z)}{\sqrt{2 \pi} \ln(10)\Sigma(z) L} \times \nonumber \\
&& \quad \quad \exp \left\{-\frac{\log_{10} [L /L_{\rm c}(M,z)]^2}{2 \Sigma^2(z)}\right\}  \nonumber \\
\Phi_{\rm s}(L|M) &=& A(M,z) L^{\gamma(M,z)}\, .
\label{eqn:clf}
\end{eqnarray}
Here $L_c(M,z)$ is the relation between central galaxy luminosity of a given dark matter halo and it's halo mass, taken to be
a function of redshift, while $\ln(10) \Sigma(z)$ is the dispersion in this relation, again a function of redshift. 
In equation~(1), $f_c(M,z)$ is an additional selection function  introduce to
account for the efficiency for galaxy formation as a function of the halo mass, given the
fact that at low mass halos galaxy formation is inefficient and not all dark matter halos may host a galaxy.
Given limited statistics to extract information on $f_c(M,z)$, and given that our primary goal is to
study the evolution associated with $L_c(M,z)$ relation,  we set $f_c(M,z)=1$ throughout
here. This assumption also leads to reasonably well model fits to LFs either at low redshifts as 
studied in Cooray \& Milosavljevic (2005b) or at high redshifts as in the case here.

The central galaxy CLF takes a log-normal form, while the satellite galaxy CLF takes a power-law form in luminosity.
Such a separation describes the LF best, with an overall better fit to the data in the K-band as explored by
Cooray \& Milosavljevi\'c (2005b) and 2dFGRS $b_J$-band in Cooray (2005).  Our motivation for log-normal distribution also comes
from measured conditional LFs, such as galaxy cluster LFs that include bright central galaxies, where 
a log-normal component, in addition to the Schechter (1976) form, is required to fit the data (e.g., Trentham \& Tully 2002). 
Similarly, the stellar mass function, as a function of halos mass in semi-analytical models, 
is best described with a log-normal component for  central galaxies (Zheng et al. 2004).
To simplify the modeling approach, we assume $\Sigma(z) =\Sigma(0)=0.17$, and $\gamma(M,z)=-0.5$;
The value for $\Sigma$ comes from a comparison to low-redshift LF in 2dFGRS (Norberg et al. 2002),
while the latter ignores the mass dependence of the satellite luminosity distribution suggested in
Cooray (2005). 

To describe galaxies as a function of color in this analytical description, we must further divide central and satellite
galaxies as a function of their color given the luminosity. Here, motivated by the bimodality of color (e.g., Baldry et al. 2004)
that extends out to high redshifts (e.g., Giallongo et al. 2005), we consider models in terms of galaxy types.
The description in terms of galaxy types is also useful since measurements at high redshifts,
so far, involve the division of galaxy samples to two broad categories involving early-type, or red, and
late-type, or blue, galaxies. Thus, in the case of early type galaxies, we write CLFs as
\begin{eqnarray}
\Phi_{\rm early-cen}(L|M,z) &=& \Phi_{\rm c}(L|M,z) f_{\rm early-cen}(M,z) \nonumber \\
\Phi_{\rm early-sat}(L|M) &=& \Phi_{\rm s}(L|M,z) f_{\rm early-sat}(M,L,z) \, ,
\end{eqnarray}
where the two functions that divide between early- and late-types are taken to be
functions of mass, in the case of central galaxies,
and both mass and luminosity in the case of satellites. As fractions are defined
with respect to the total galaxy number of a halo,  late-type fractions are simply $[1-f_{\rm early-cen}(M,z)]$
and $[1-f_{\rm early-sat}(M,L,z)]$ for central and satellite galaxies, respectively. 
These functions, as appropriate for 2dFGRS $b_J$-band galaxy type LFs, are described in Cooray (2005).

Here, we will use these functions to describe the type dependence of the high redshift galaxy LF
and ignore all redshift dependences and assume that regardless of the redshift, the fraction of
of early-to-late type galaxies, in a halo of fixed mass, is same as the fraction of that halo mass today. 
While this assumption may seem contradictory with observations at the first instance,
given that observations indicate that ellipticals are older than spirals or the fact that mass
correlates with age, this is not necessarily the case given the strong evolution associated with the halo
mass distribution. While our assumption is that the ratio of red-to-blue galaxies, of a given halo mass,
at a redshift $z$ is same as the fraction today, the mass function evolves such that 
at high redshifts, the universe is dominated by small mass halos while the density of
high mass halos increases to low redshifts. Given that in the model of Cooray (2005)
the fraction of early-type galaxies increases with increasing mass, the relative fraction of
early-to-late type galaxies, when averaged over the mass distribution of halos at a given redshift,
is not the same, but increases to lower redshifts. Thus, while we have not assumed or specified
important astrophysical processes such as galaxy mergers that may be responsible for galaxy types,
given that the halo mass function evolves through merging, galaxies must undergo merging as well.
While the specific process may remain hidden, what we can extract with this empirical modeling approach
is the exact fraction of early-to-late type galaxies as a function of halo mass; It remains a task for
numerical simulators and other approaches to understand galaxy distribution, such as semi-analytical models of
the galaxy formation, to explain the results extracted from the empirical model here when compared to
observations.

After various simplifications, which can be ignored as observational statistics at high redshifts improve,
the only redshift dependence
in our empirical model comes from redshift variations associated with the $L_c(M,z)$ relation, and when describing satellites,
$L_{\rm tot}(M,z)$, the total luminosity of galaxies as a function of the halo mass,
relation; the latter, however, is not an important ingredient since the LF is primarily determined by statistics of
central galaxies (Cooray \& Milosavljevic 2005b; Cooray 2005).

\subsection{Central Galaxy Luminosity-Halo Mass Relation}

For $L_{\rm c}(M,z=0)$ relation, here we make use of the  relation derived in Vale \& Ostriker (2004).
These authors established this relation by inverting the 2dFGRS luminosity function given an analytical description for the sub-halo
mass function of the Universe (e.g., De Lucia et al. 2004; Oguri \& Lee 2004). We used this relation to describe the
2dFGRS $b_J$-band LF in Cooray (2005), and we will use the same relation, at $z=0$, as
an approximation to describe the B-band LF. The relation is described with a general fitting formula given by
\begin{equation}
\label{eqn:lcm}
L(M,z=0) = L_0 \frac{(M/M_1)^{a}}{[b+(M/M_1)^{cd}]^{1/d}}\, .
\end{equation}
For central galaxy luminosities, the parameters are $L_0=5.7\times10^{9} L_{\sun}$, $M_1=10^{11} M_{\sun}$,
$a=4.0$, $b=0.57$, $c=3.72$, and $d=0.23$ (Vale \& Ostriker 2004).
For the total galaxy luminosity, as a function of the halo mass,
we also use the fitting formula in equation (\ref{eqn:lcm}), but with
$c=3.57$. As discussed in Cooray (2005), 
the overall shape of the LF is {\it strongly} sensitive to the shape of the $L_{\rm c}$--$M$ relation,
and it's scatter, and less on details related to the $L_{\rm tot}$--$M$ relation.

To describe the redshift evolution, first, we consider two possibilities, but using
a large sample of datasets will combine them to consider a general model fit.
First, we describe the high-z LFs with
$L(M,z)=L(M,z=0)(1+z)^\alpha$. This is a scenario in where all luminosities either increase or decrease depending on the value 
and sign of $\alpha$, which we take to be
mass independent. Such an evolution provides an acceptable description of the LFs, say of
Giallongo et al. (2005), though the increase in luminosity of galaxies in less massive dark matter halos overestimates the
LF at the faint-end at high redshifts. At $z \sim 6$,  this overestimate becomes significant
and even the LF at the bright-end is overestimated relative to the UV LF measured by Bouwens et al. (2004a).
A preferred description may be a case where low luminosity end of the $L_c(M,z)$ relation remains
independent of the redshift, while the bright-end increases with increasing redshift. 
To describe this behavior, we take parameters $c$ and $d$ in equation~(\ref{eqn:lcm}) 
to be dependent on the redshift with $c(z)=c(z=0)(1+z)^\beta$ and $d(z)=d(z=0)(1+z)^\eta$,
where $\beta$ and $\eta$ are taken to be free parameters. When combined, as we discuss later,
we find that the redshift evolution related to $d(z)$ is not strongly constrained by observational data
while $\alpha$ and $\beta$ are. Using the LF measurements from DEEP2, COMBO-17, and those extending
to $z \sim  5$ by Gabasch et al. (2004), we will provide general constraints on $\alpha$ and $\beta$.

In Figure~1, we show the $L_c(M,z)$ relation 
 as a function of the halo mass and for redshifts from 0 to 6.  For comparison, we also show 
measurements from the 2dFGRS galaxy group catalog from Yang et al. (2005).  In the left panel, we show the pure-luminosity evolution case
with $\alpha=0.75$ and in the right-panel, we show the evolution of the bright-end with $\beta=-0.07$
and $\eta=0.05$; These numerical values were selected based on a comparison to the high-redshift
LFs of Giallongo et al. (2005).

\begin{figure*}
\centerline{\psfig{file=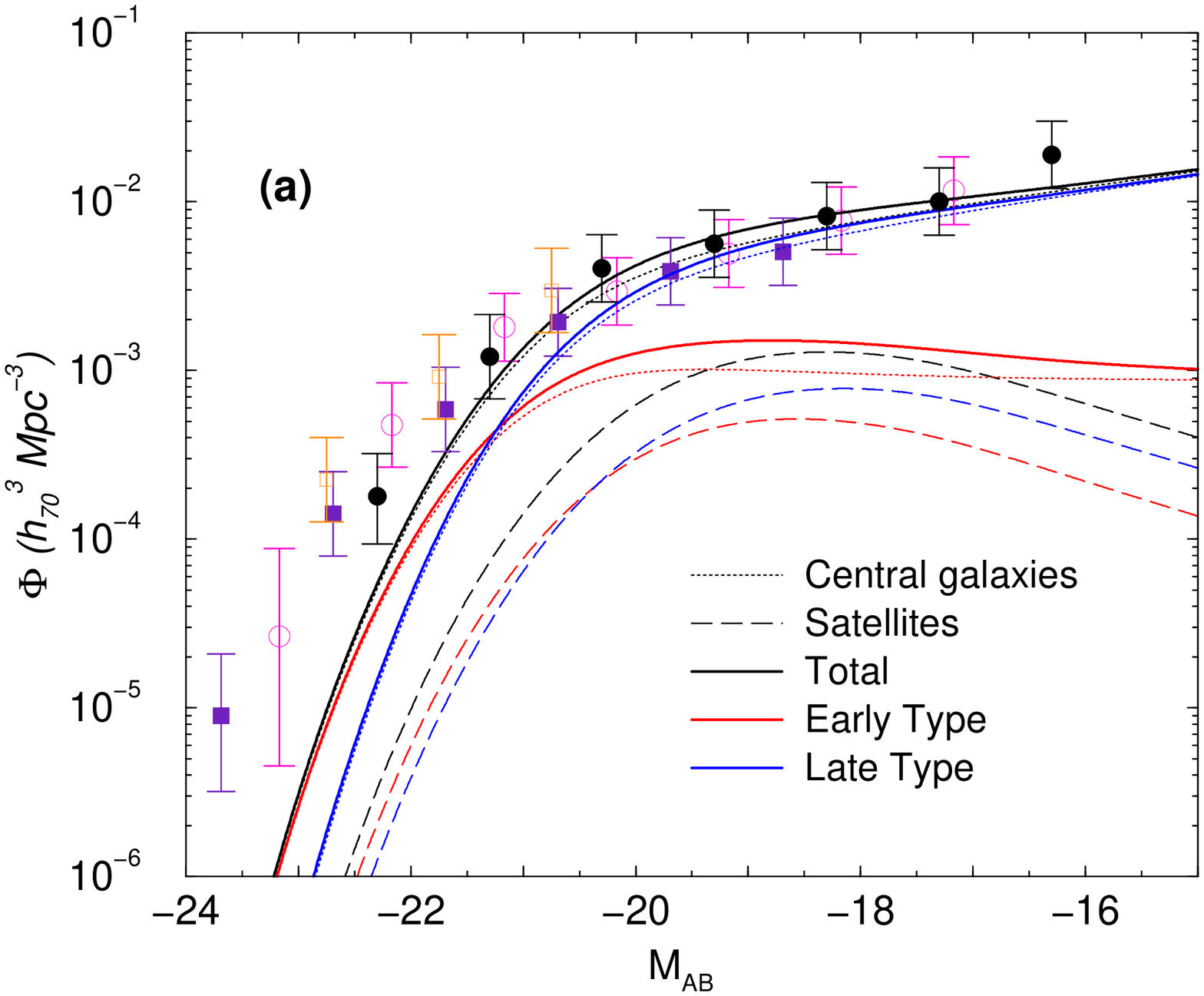,width=\hssize,angle=-90}
\psfig{file=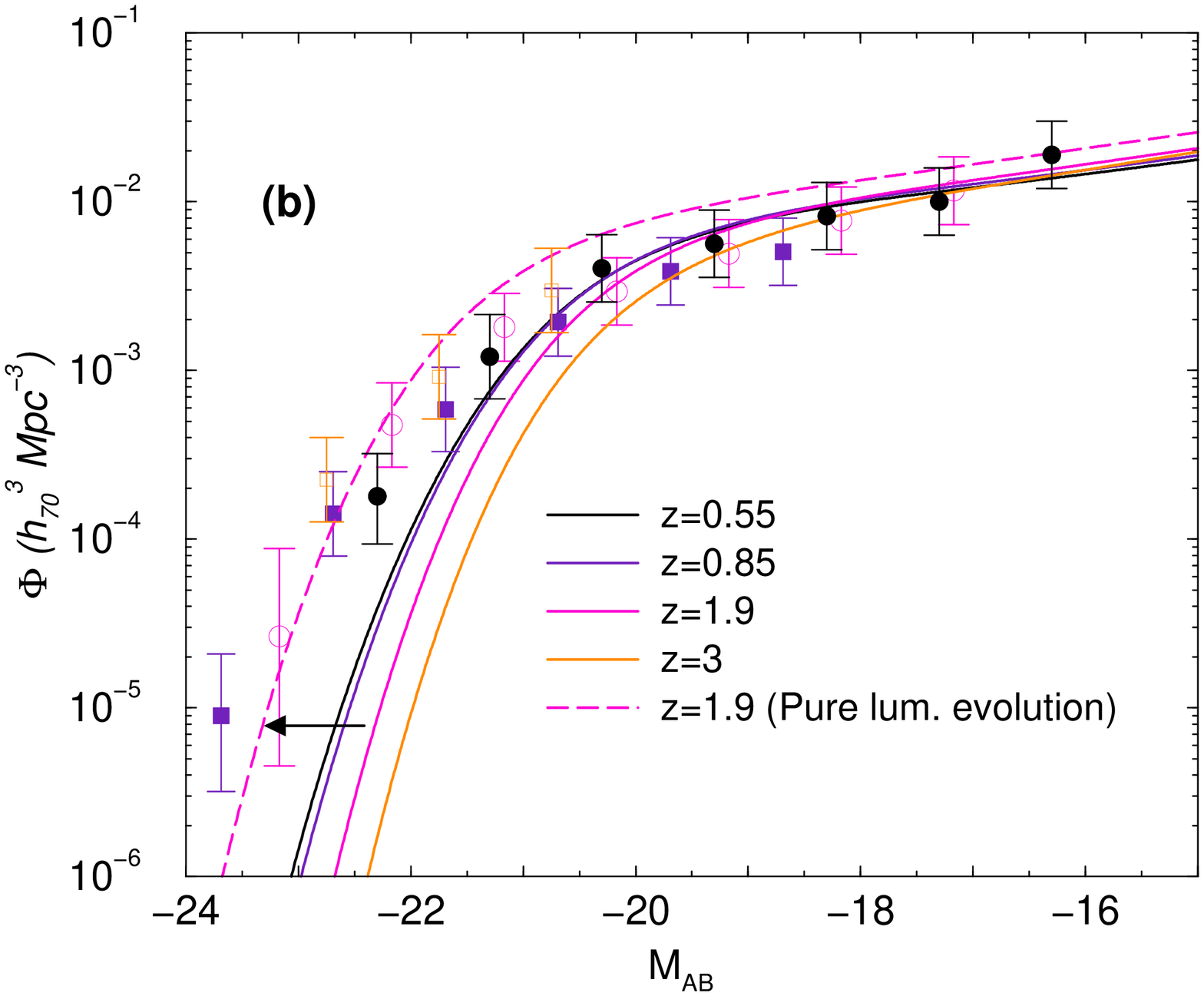,width=\hssize,angle=-90}}
\caption{The LF of galaxies. In both panels, the data shown are the total LF from Giallongo et al. (2005)
corresponding to redshifts between 0.4 and 0.7 (filled circles), 0.7 to 1.0 (open circles),
1.3 to 2.5 (filled squares) and 2.5 to 3.5 (open squares).
In (a), the plotted curves  are the LF at $z=0$ divided to central galaxies (dotted lines),
satellites (dashed lines), and the total (solid line). We also further subdivide the sample
to early (red lines) and late type (blue lines) galaxies, based on the model
description of Cooray (2005). Note that the high-z LF, when compared to $z \sim 0$,
shows an increase in the number density at the bright end, while the faint-end density remains
the same. In (b), we show the expected LF at the mid redshift of redshift ranges considered in Giallongo et al. (2005) and
 under the assumption that $L_c(M,z)=L_c(M,z=0)$;
the resulting redshift variations are associated with evolution of the dark matter halo mass function (see,
e.g., Reed et al. 2003).  A simple modification under the assumption of a pure luminosity evolution,
a constant shift in magnitude, as shown by a long-dashed line, cannot describe the high-redshift LF.
In the CLF-based approach, since the halo mass function at the high-mass
end decreases as the redshift is increased, the increase in the bright-end density could only be associated
with an increase in the luminosity of galaxies, for a given halo mass. To avoid increasing
the faint-end density, however, the increase in luminosity should only be associated with 
dark matter halos with masses corresponding to the bright-end of the LF.}
\end{figure*}

\begin{figure*}
\centerline{\psfig{file=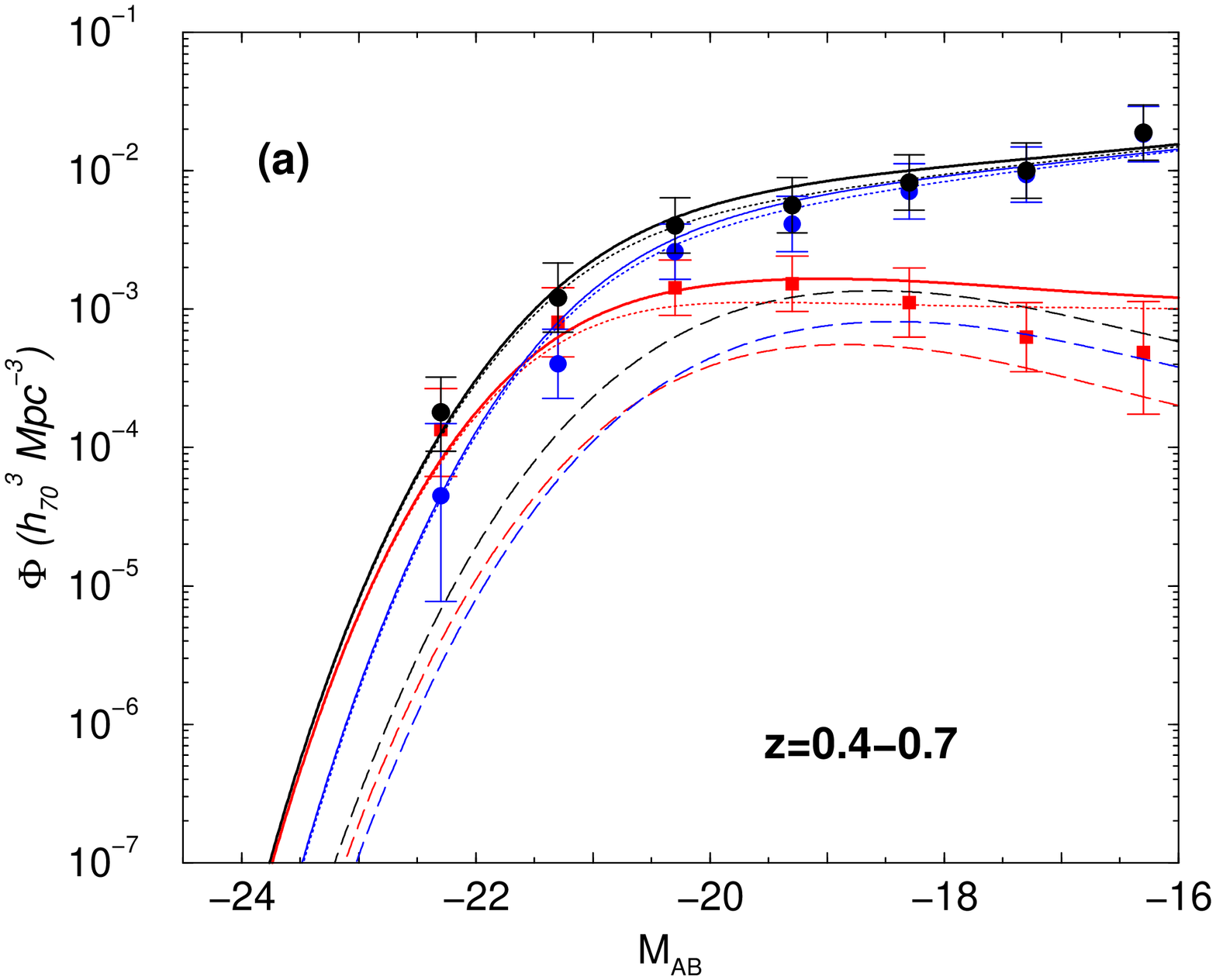,width=\hssize,angle=-90}
\psfig{file=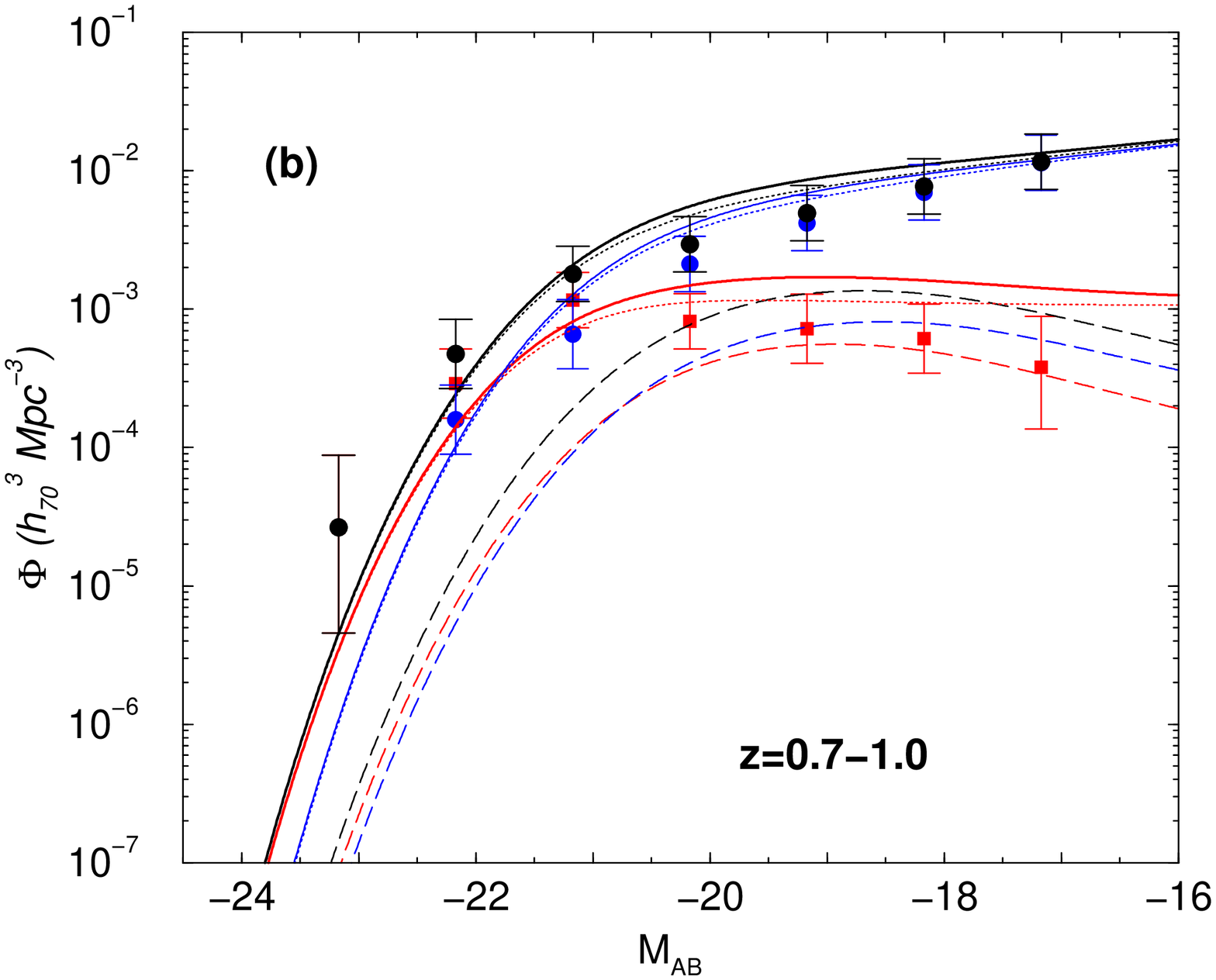,width=\hssize,angle=-90}}
\centerline{\psfig{file=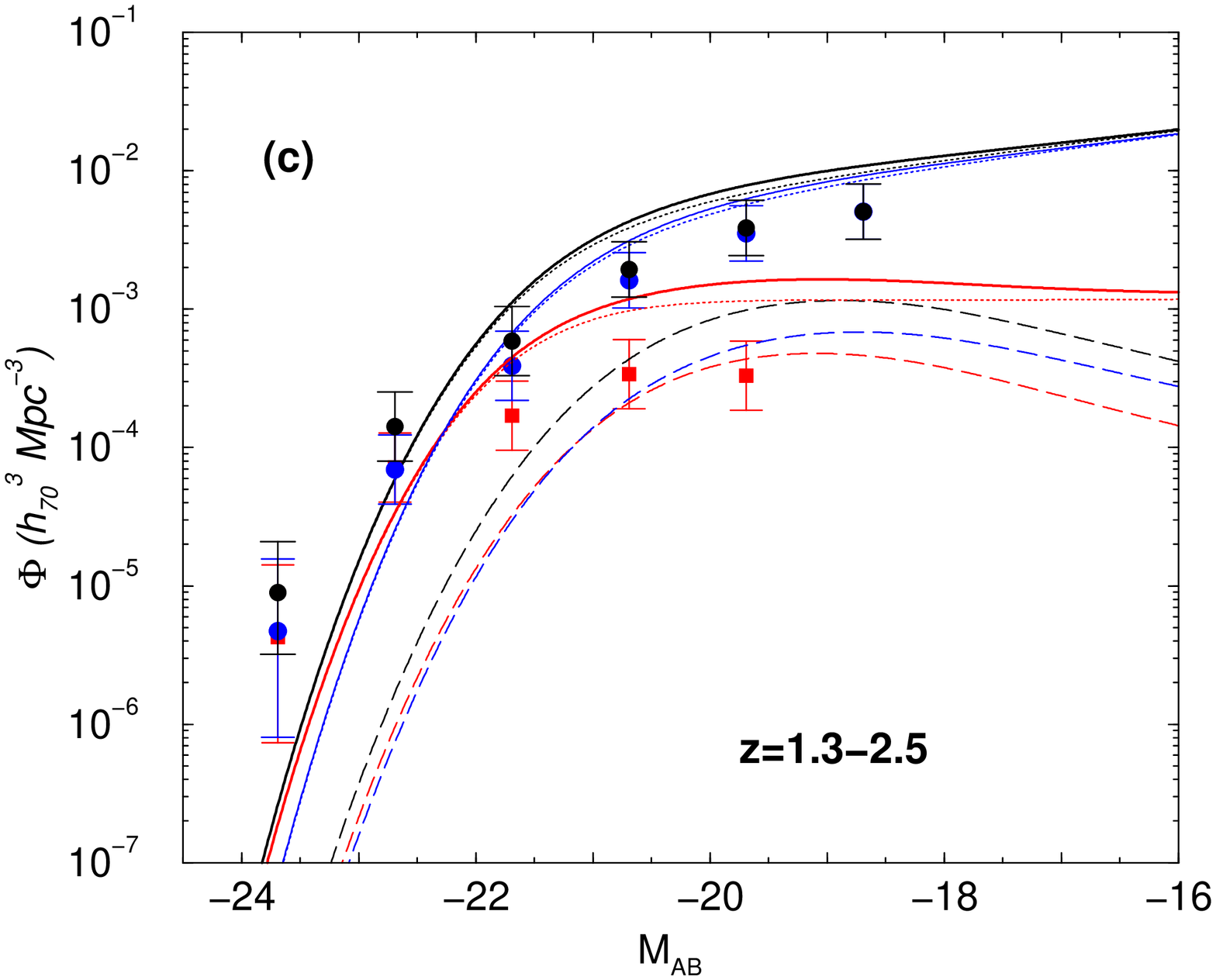,width=\hssize,angle=-90}
\psfig{file=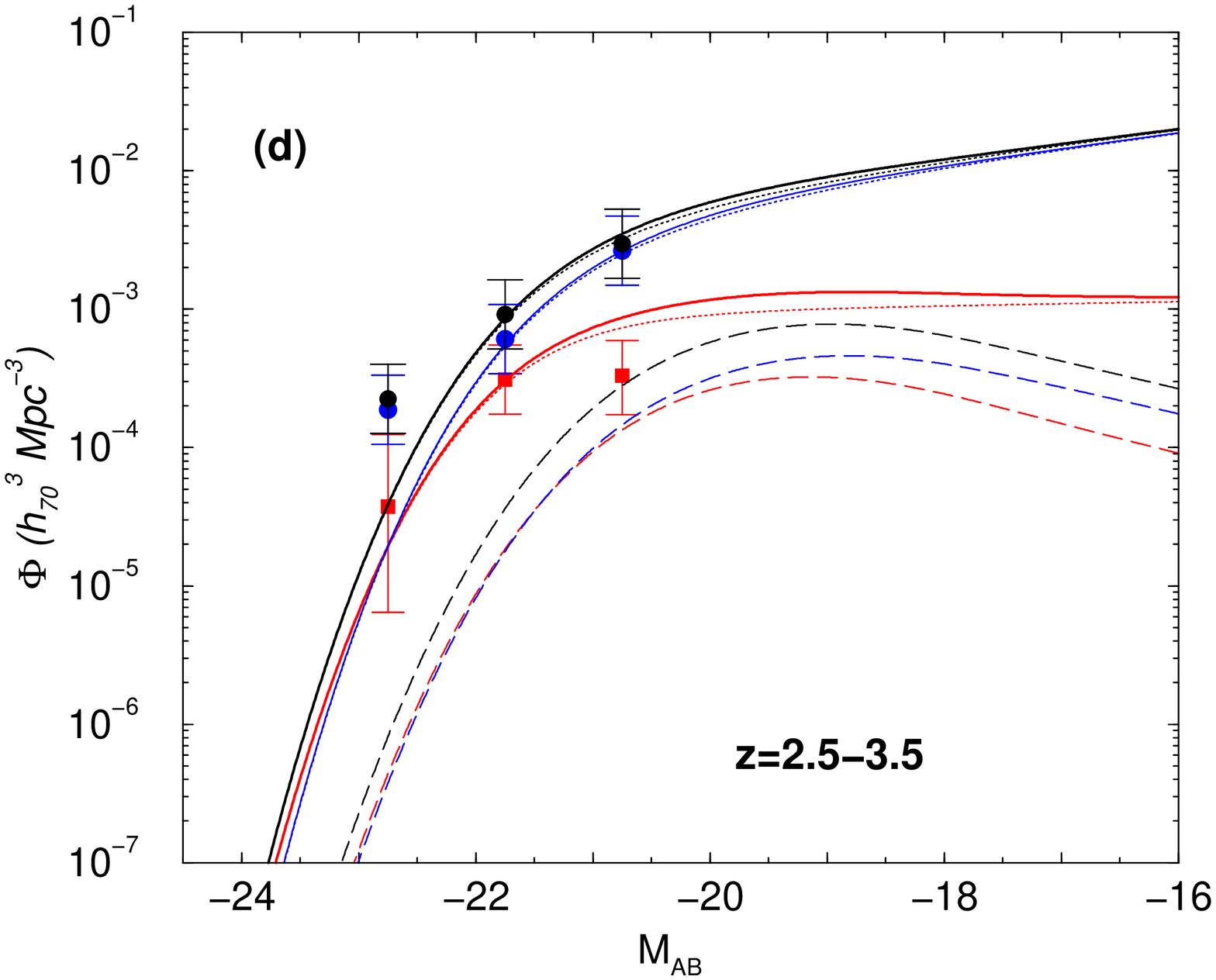,width=\hssize,angle=-90}}
\caption{
Luminosity function in the B-band as a function of redshift. From (a) to (d), we show the LF
in four redshift ranges considered by Giallongo et al. (2005) with (a) 0.4 to 0.7, (b) 0.7 to 1.0,
(c) 1.3 to 2.5, and (d) 2.5 to 3.5. In addition to the total LF, we also show the division to
red and blue galaxies. The plotted curves are predictions based on the luminosity evolution shown in
Figure~1(a), with $L_c(M,z)=L_c(M,z=0)(1+z)^{0.75}$. The lines follow Figure~2(a),
with dotted lines for central galaxies, dashed lines for satellites, and solid lines for the total sample.
The red lines are for the red galaxy LF and the blue lines show the blue galaxy LF. We will use this line conventions
through out the paper when we plot the LF.}
\end{figure*}

\begin{figure*}
\centerline{\psfig{file=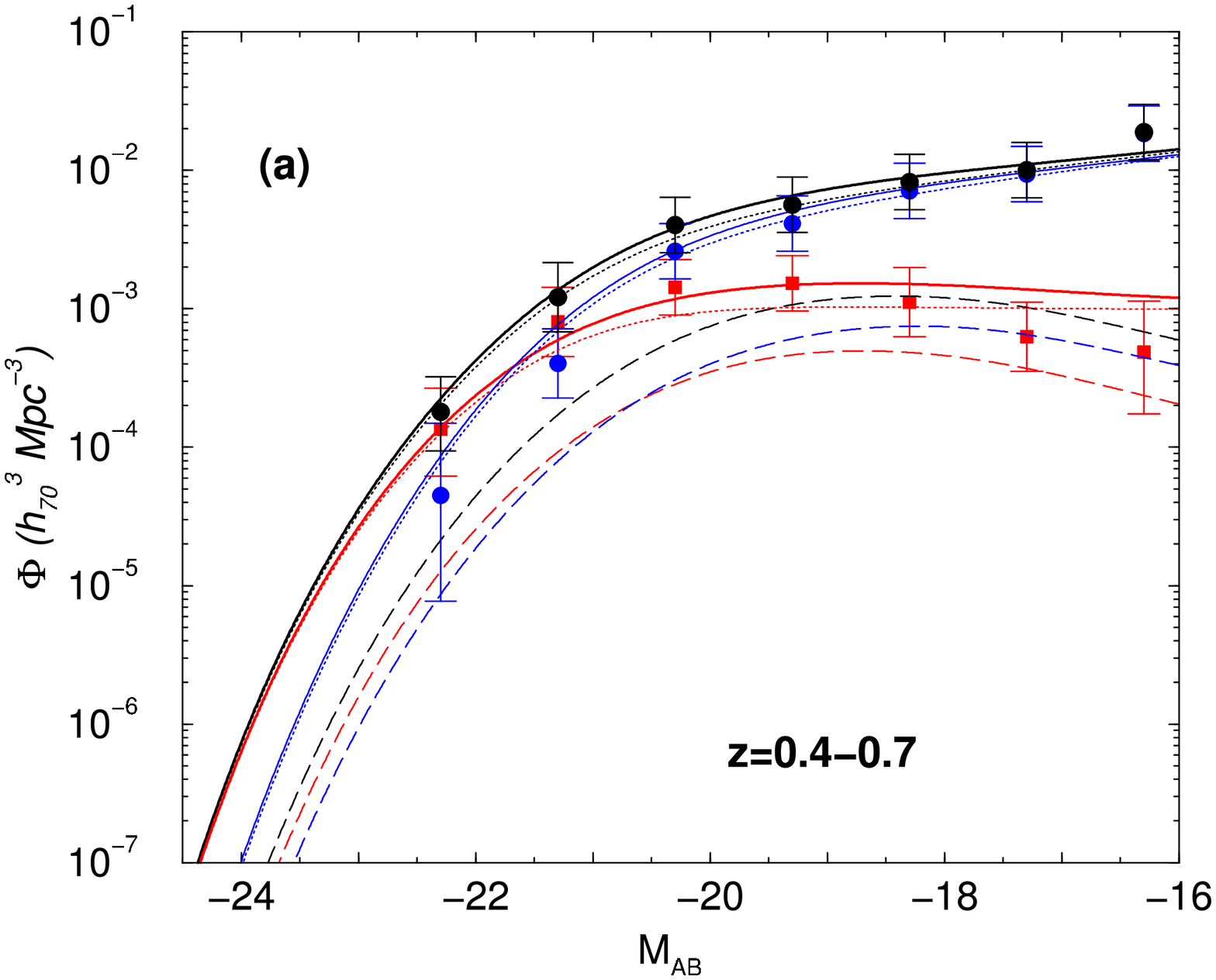,width=\hssize,angle=-90}
\psfig{file=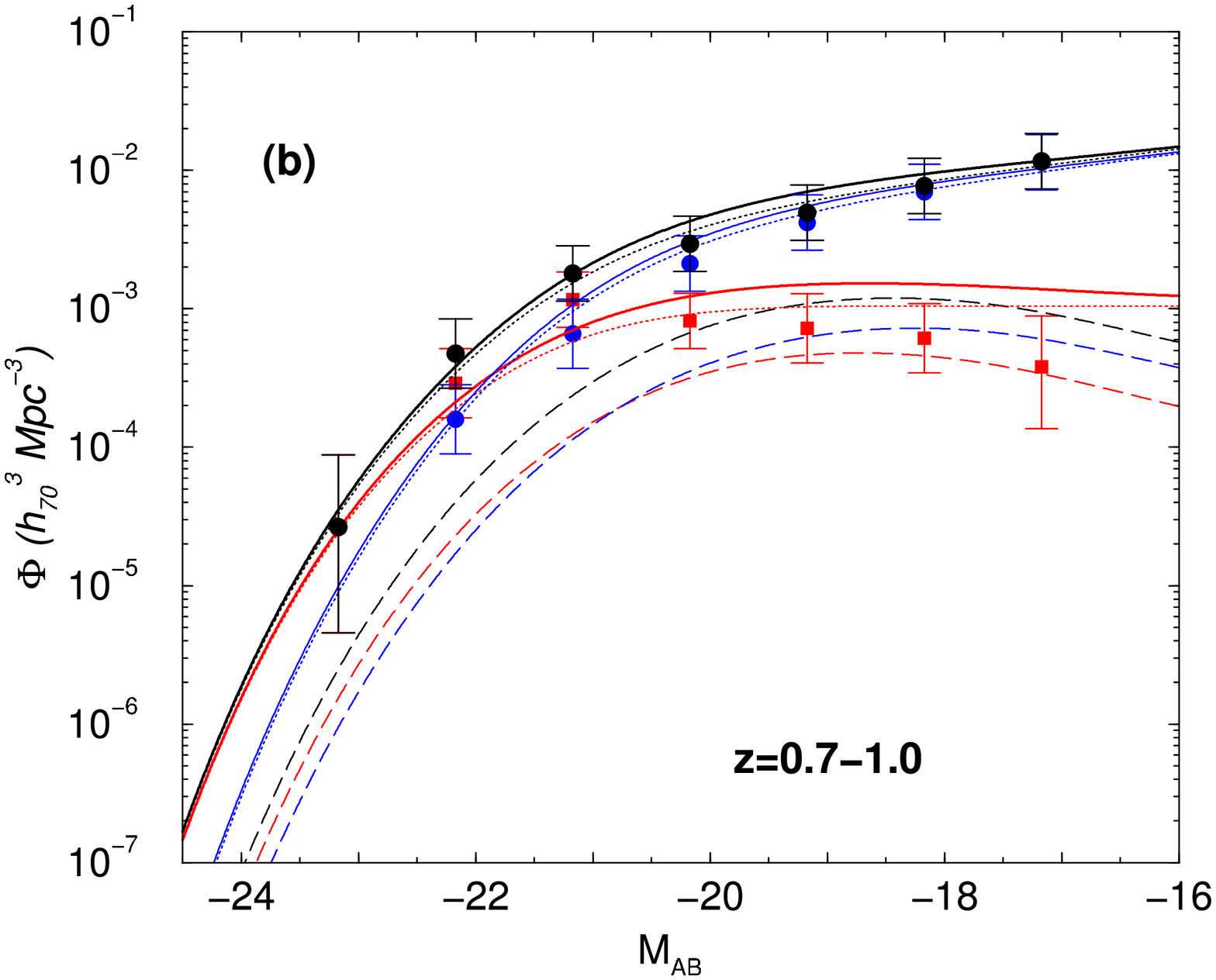,width=\hssize,angle=-90}}
\centerline{\psfig{file=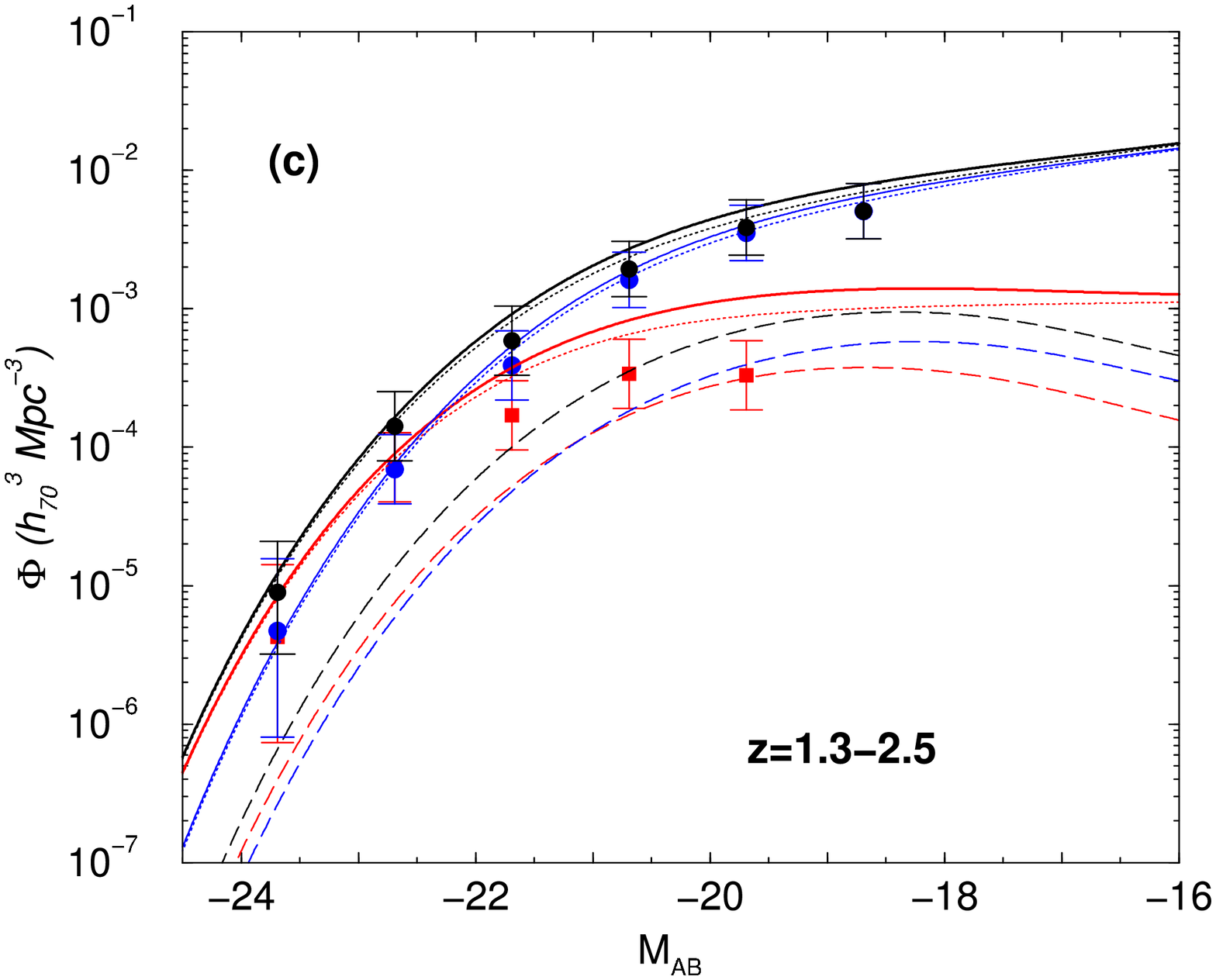,width=\hssize,angle=-90}
\psfig{file=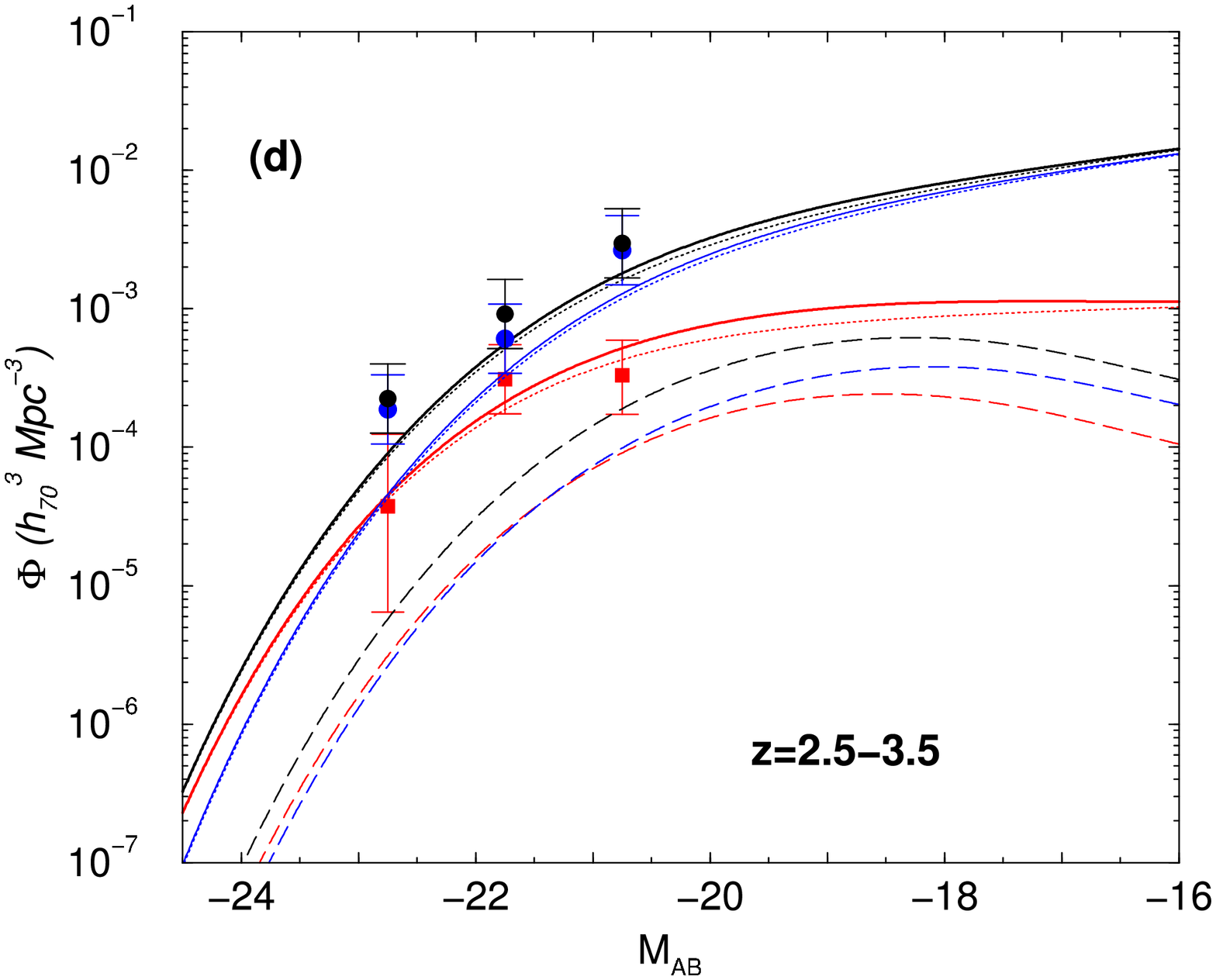,width=\hssize,angle=-90}}
\caption{
Luminosity function in the B-band as a function of redshift. The figure is same as Figure~3,
except that the plotted curves are the prediction based on the mass dependent luminosity evolution shown in
Figure~1(b).}
\end{figure*}

\begin{figure*}
\psfig{file=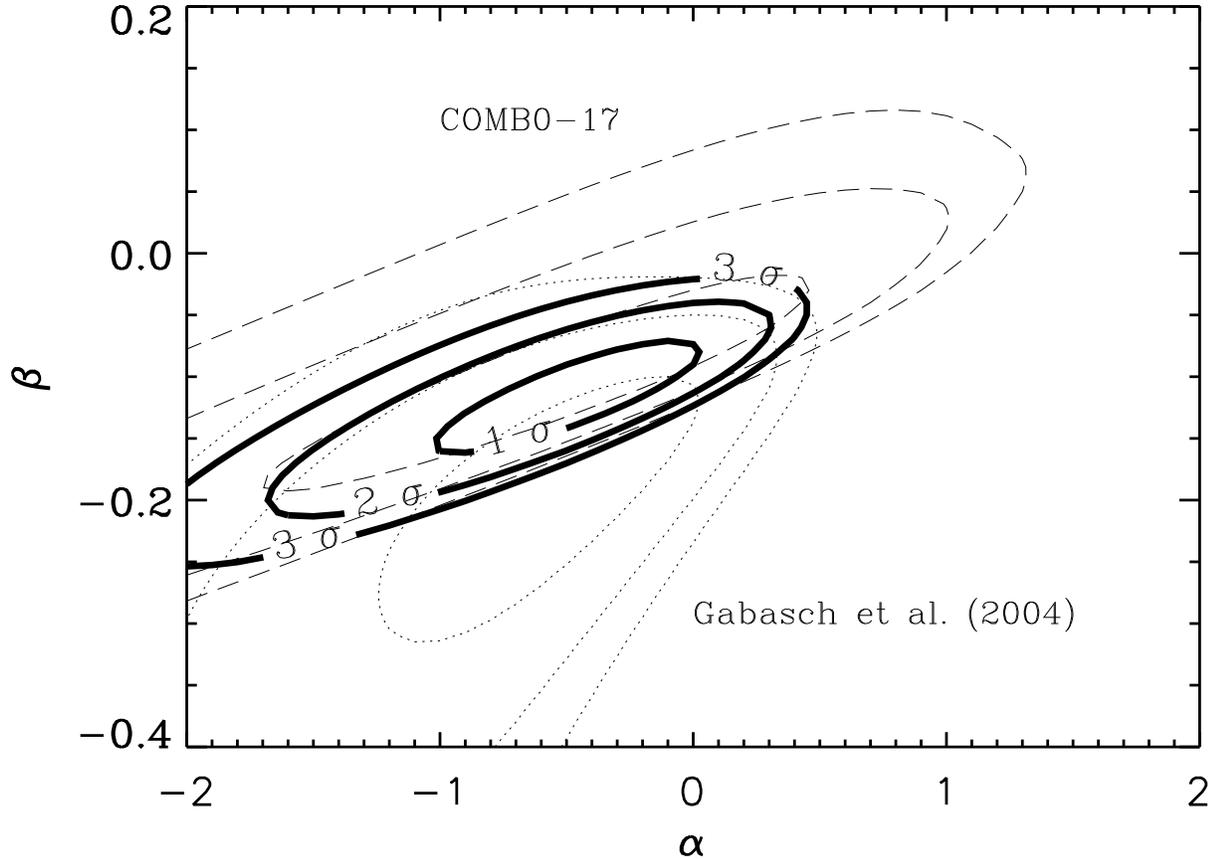,width=\hdsize,angle=90}
\caption{
Constraints on $L_c(M,z)$ evolution parameters, $\alpha$ and $\beta$ (see, equation~\ref{eqn:lcmz}),
where we plot $1\sigma$ and $2\sigma$ and 3$\sigma$ allowed regions based on a likelihood analysis.
In dashed lines, we show constraints based on a model comparison to data from DEEP2 (Willmer et al. 2005)
and COMBO-17 (Bell et al. 2004) LFs, while in dotted lines, we show constraints from
Gabasch et al. (2004) LFs out to a redshift of 5. The region allowed by models for the combined data
is shown in solid lines. Here, $\beta >0$ is rule out at $\sim$ 3$\sigma$ level and the allowed region strongly
suggests evidence for mass-dependent luminosity evolution in the $L_c(M,z)$ relation.
}
\end{figure*}

\begin{figure*}
\centerline{\psfig{file=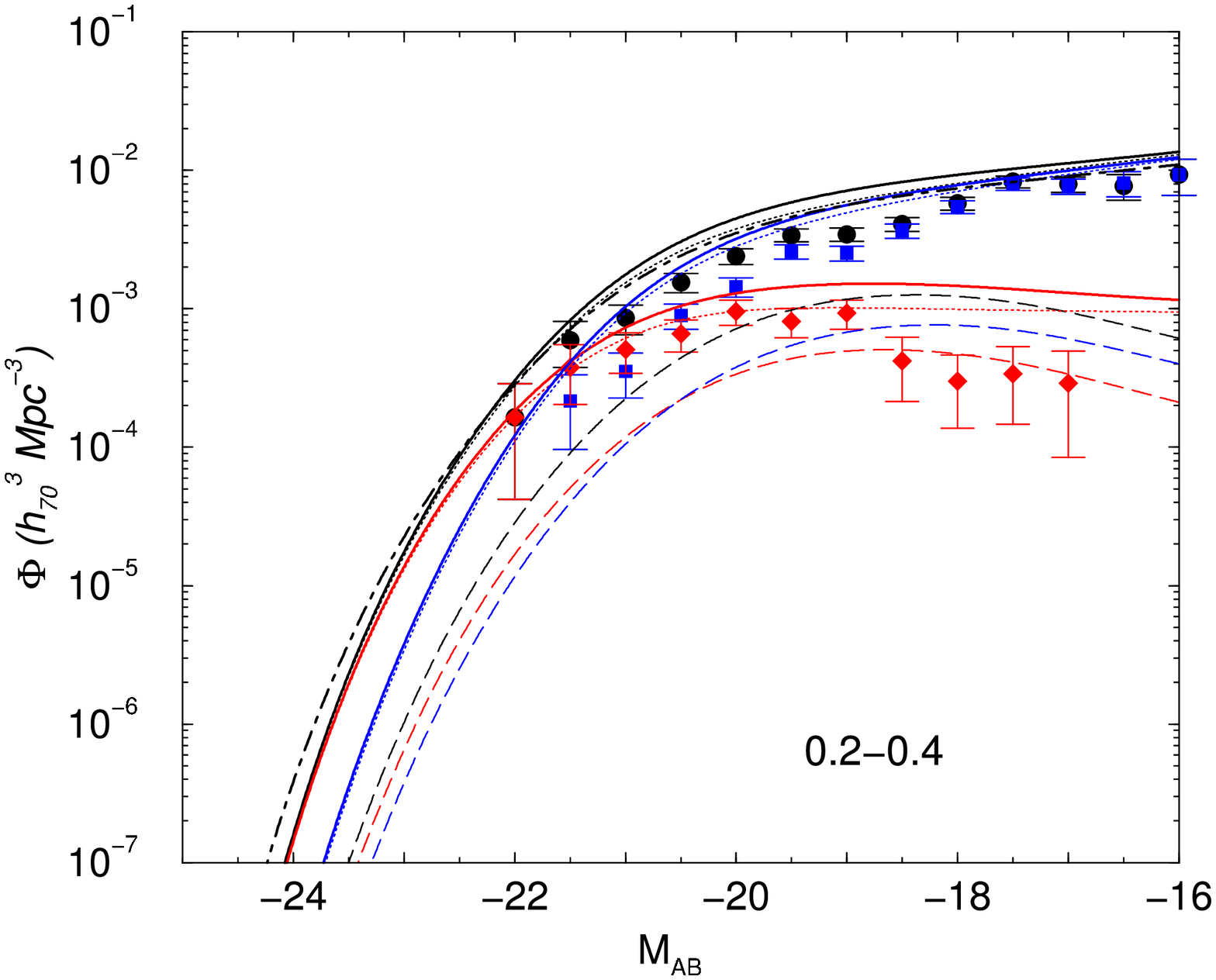,width=\hssize,angle=-90}
\psfig{file=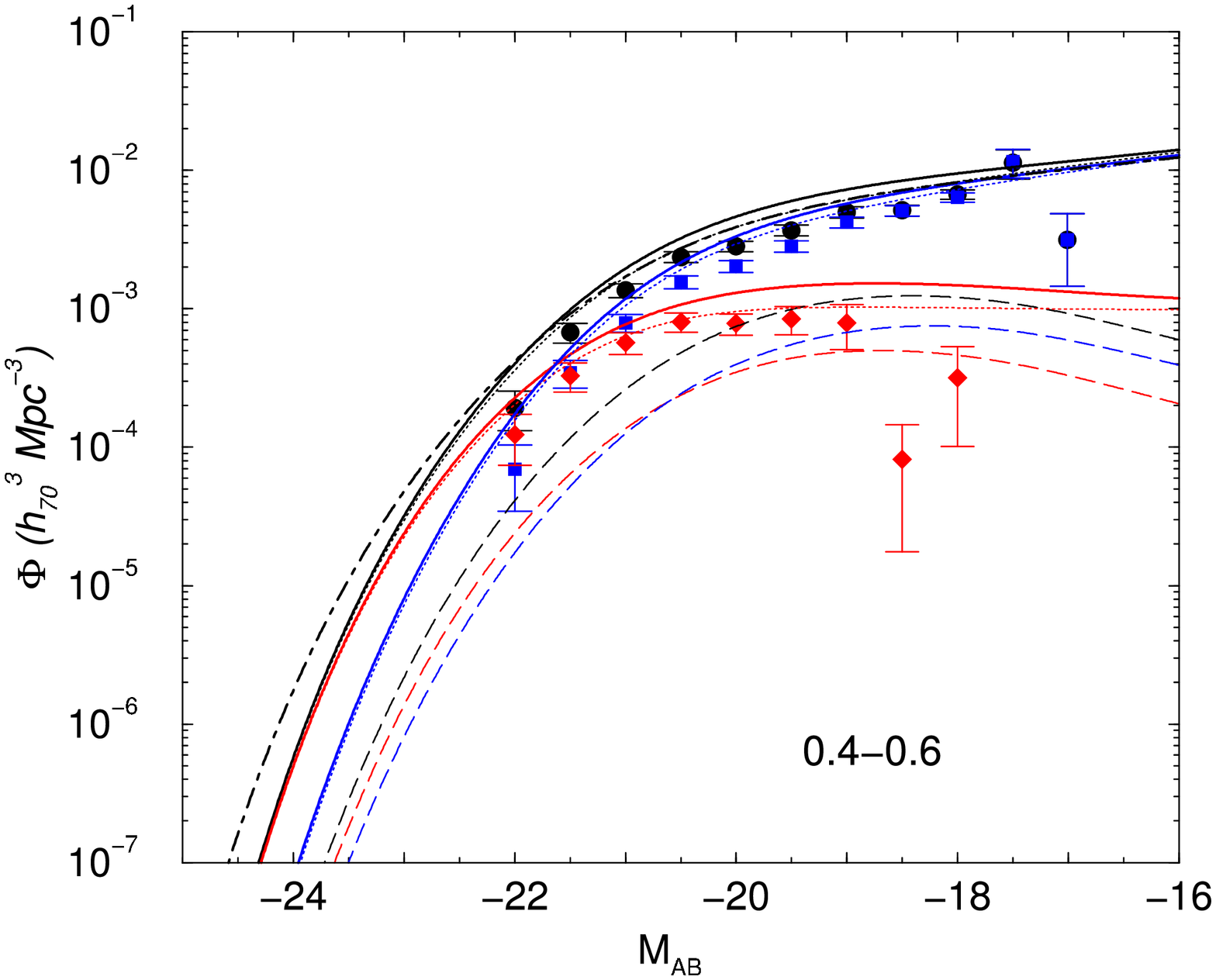,width=\hssize,angle=-90}}
\centerline{\psfig{file=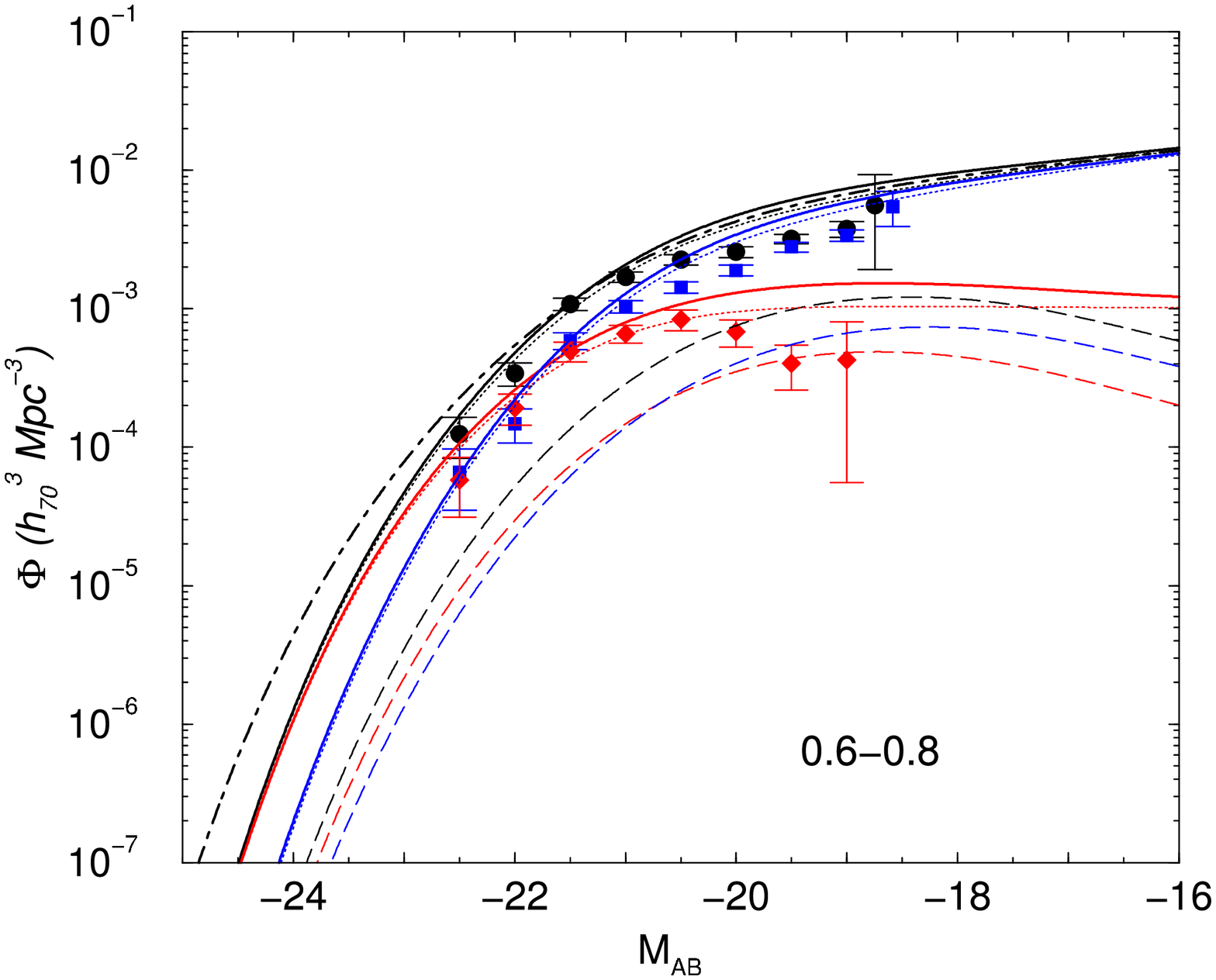,width=\hssize,angle=-90}
\psfig{file=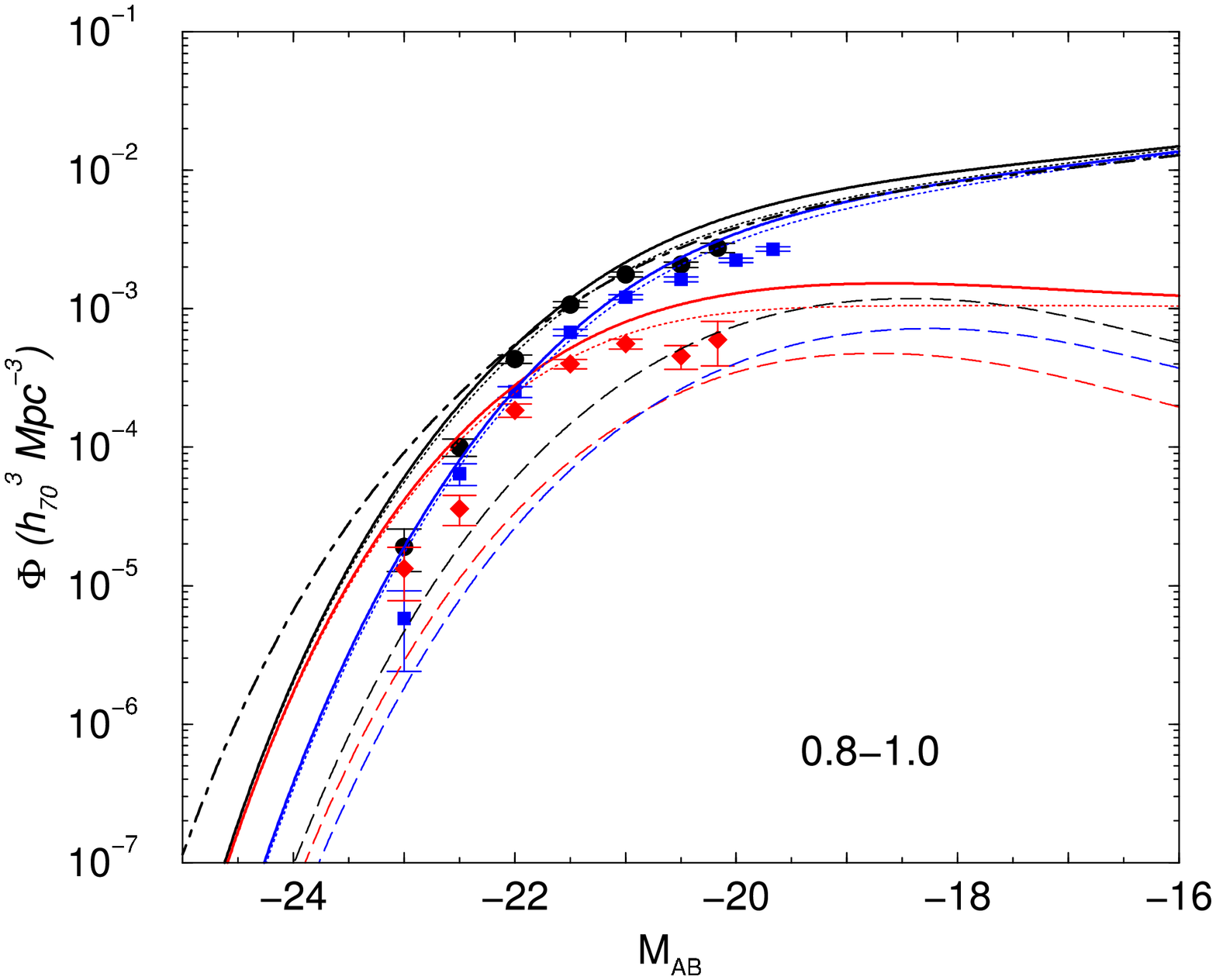,width=\hssize,angle=-90}}
\centerline{\psfig{file=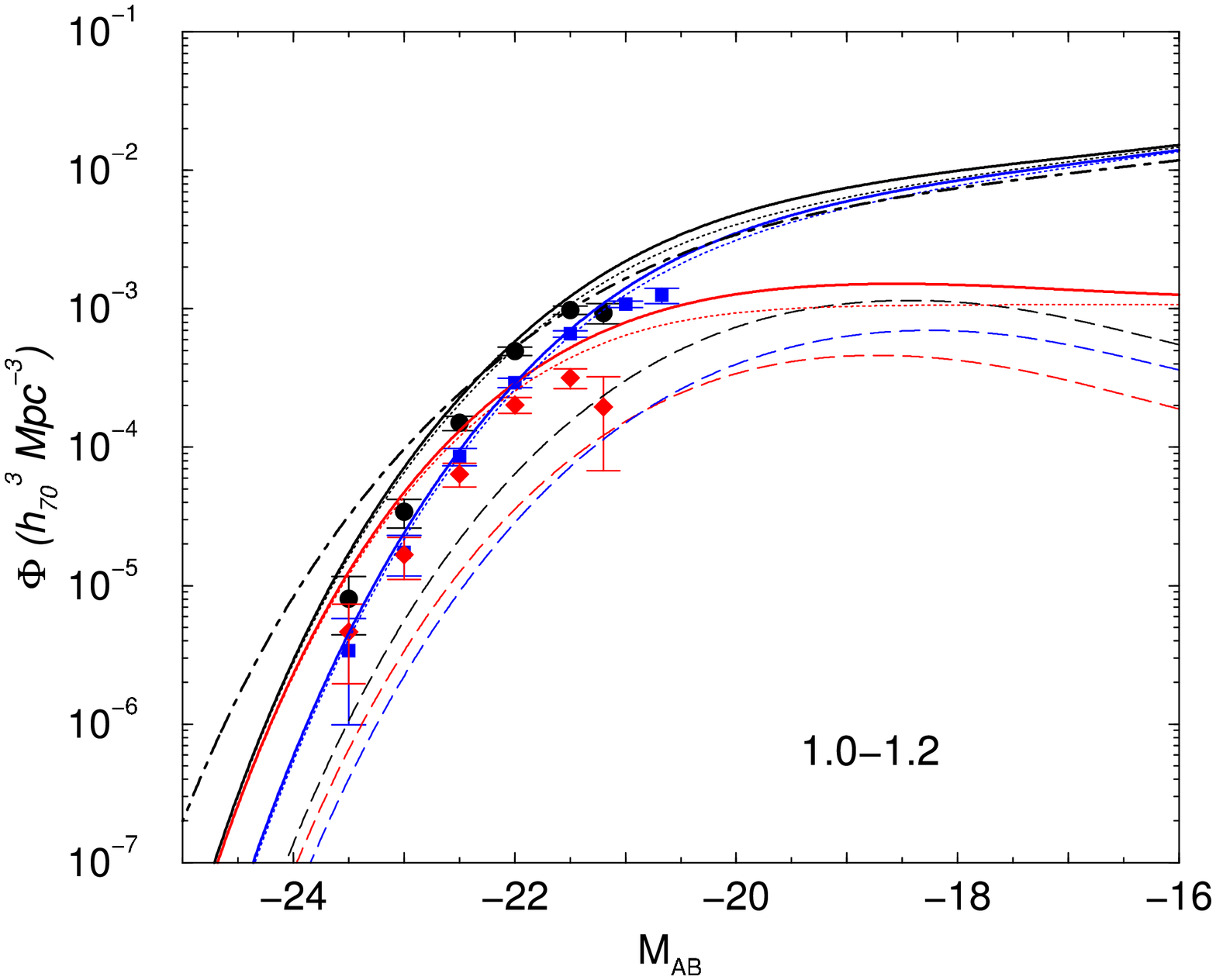,width=\hssize,angle=-90}}
\caption{
Luminosity function in the B-band as a function of redshift, from the DEEP2 survey (from Willmer et al. 2005).
The lines show the overall best fit model description while the dot-dashed line show the total LF
that best describes this data set alone. Other curves are same as the ones in Figure~3.
}
\end{figure*} 

\begin{figure*}
\centerline{\psfig{file=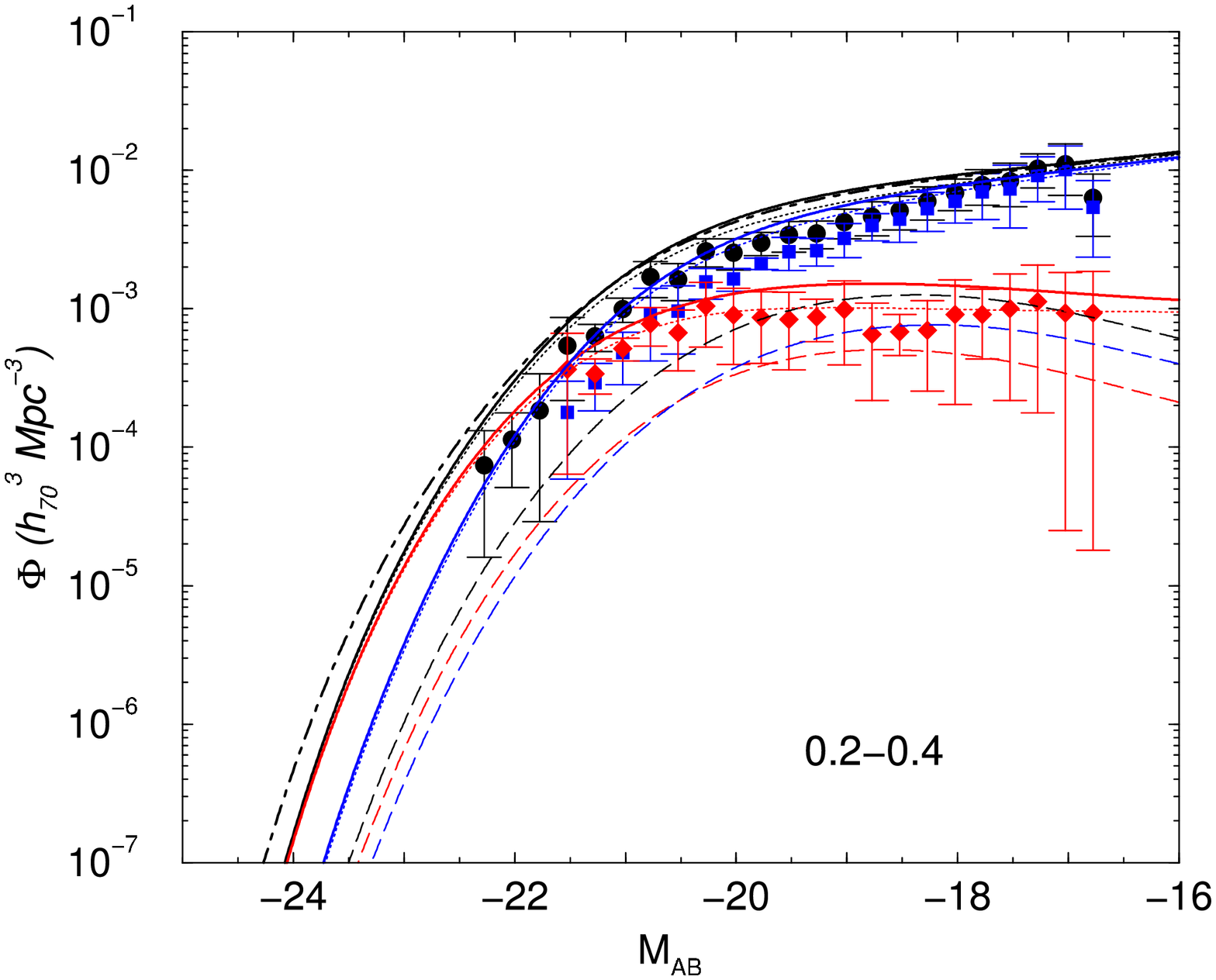,width=\hssize,angle=-90}
\psfig{file=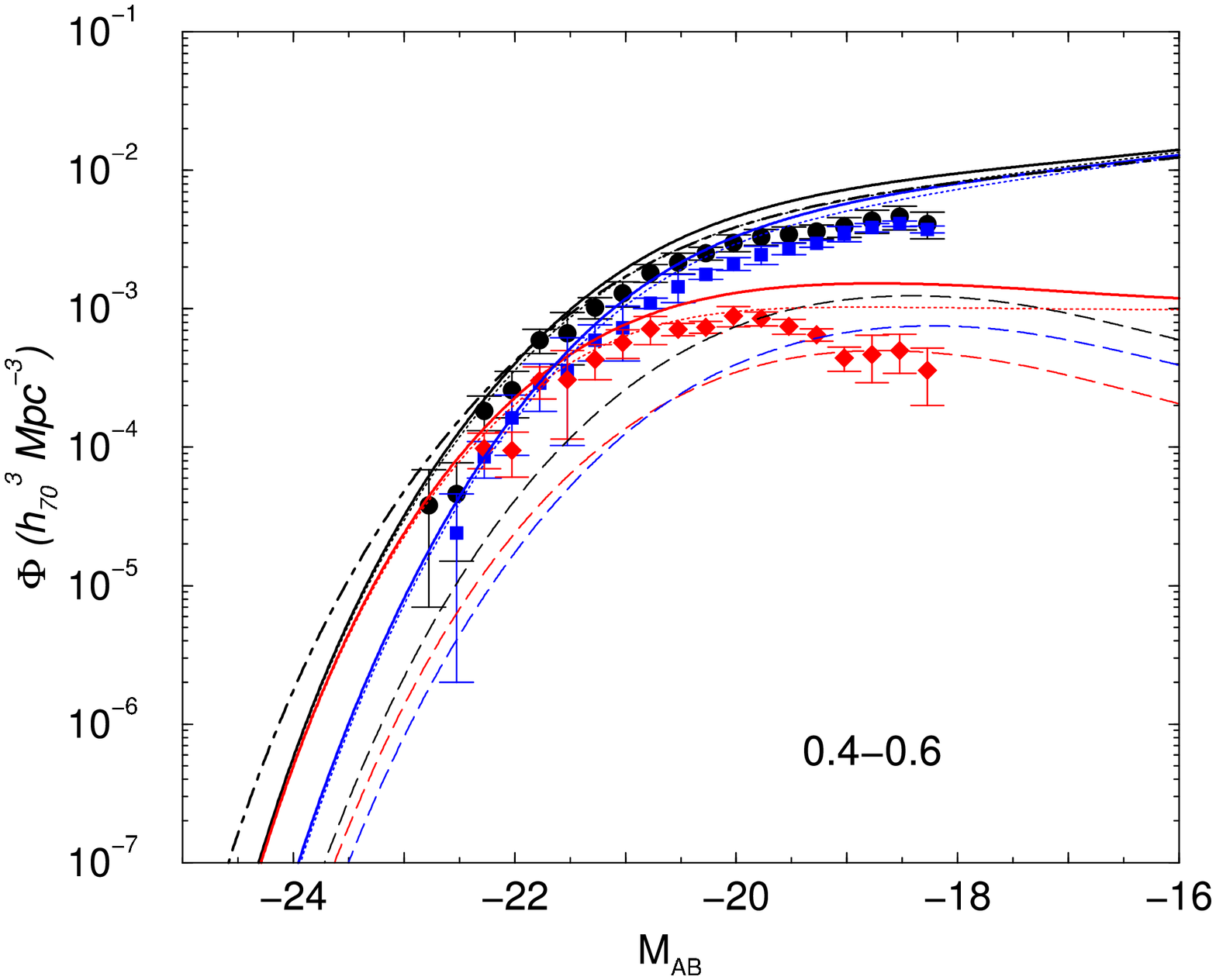,width=\hssize,angle=-90}}
\centerline{\psfig{file=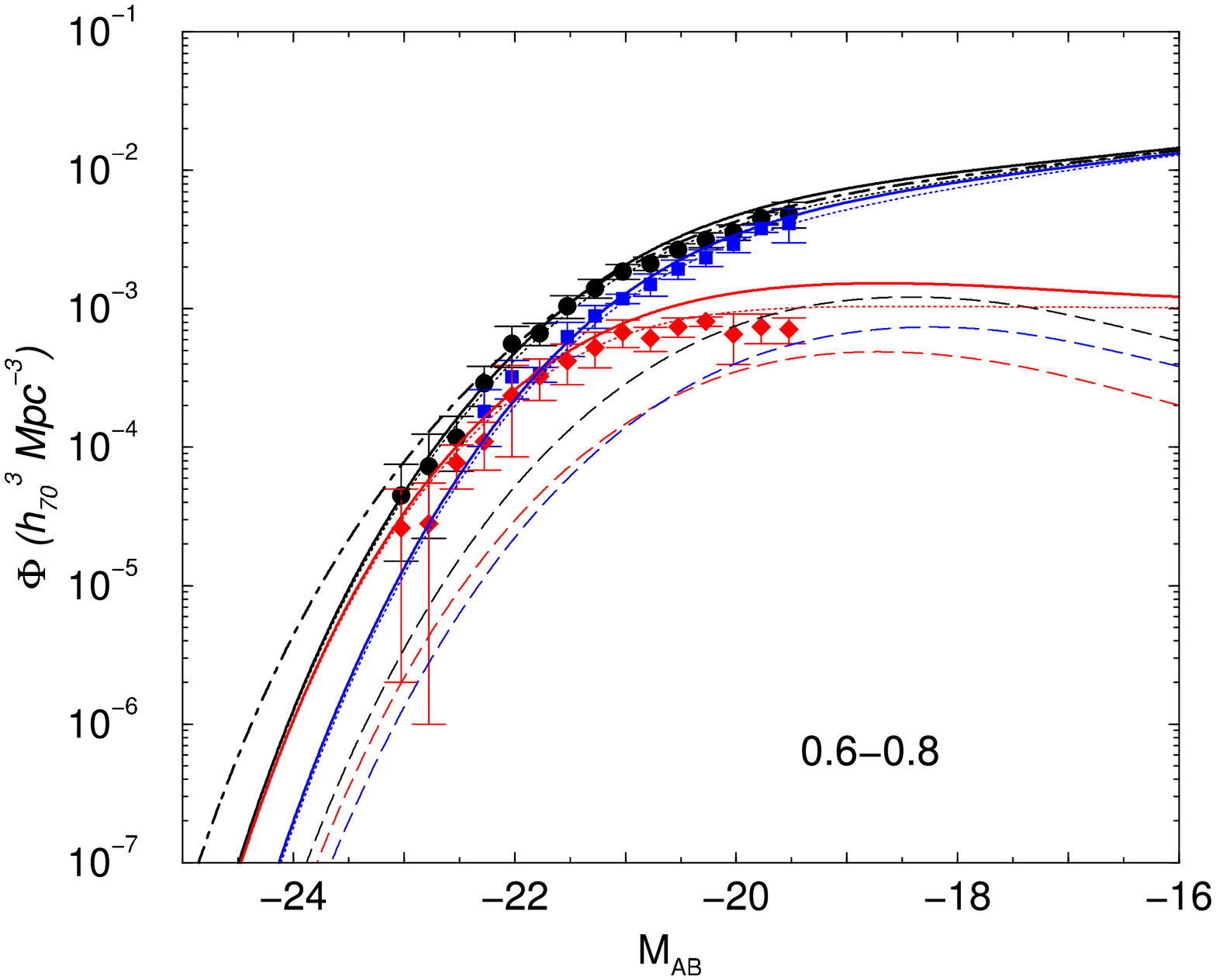,width=\hssize,angle=-90}
\psfig{file=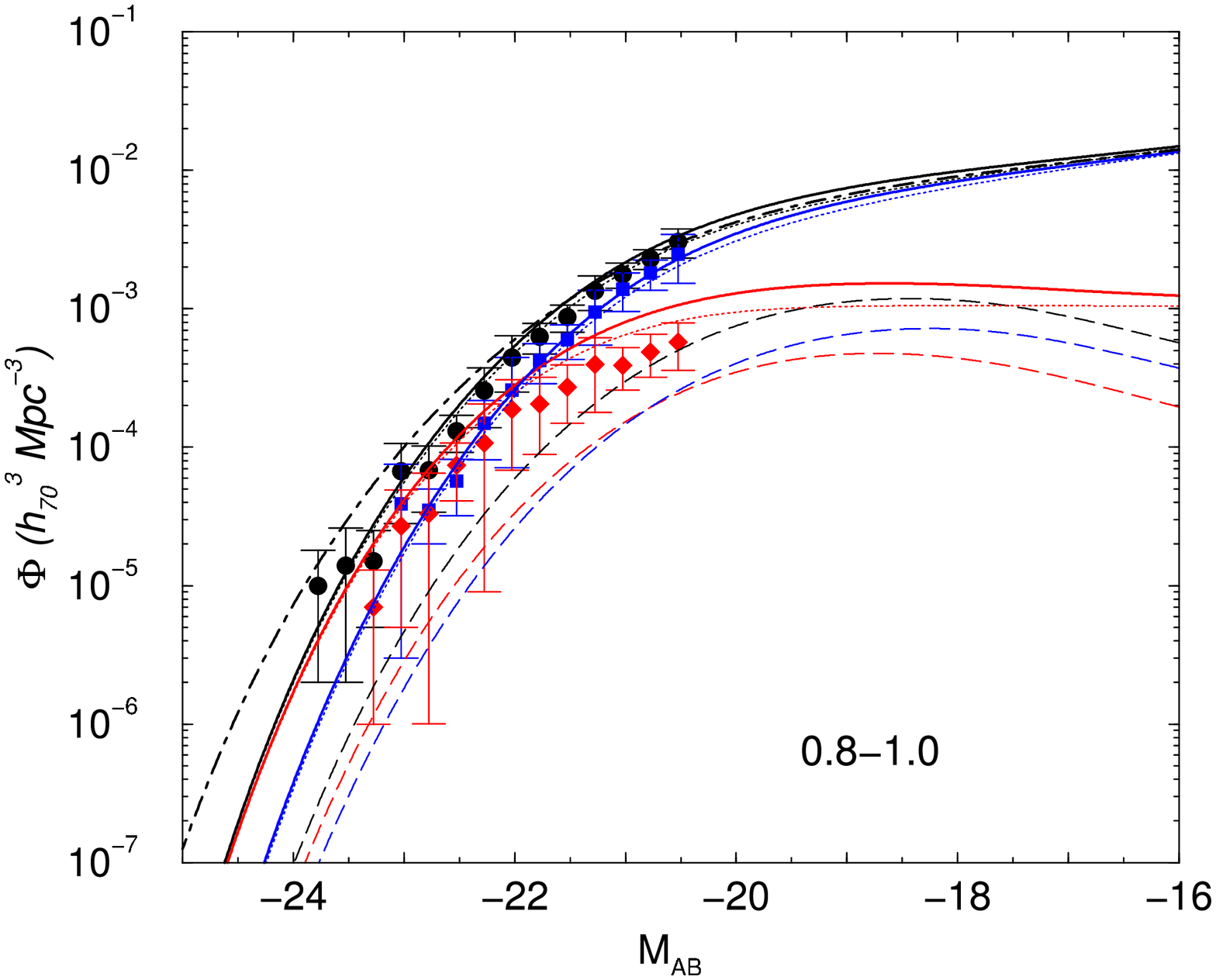,width=\hssize,angle=-90}}
\centerline{\psfig{file=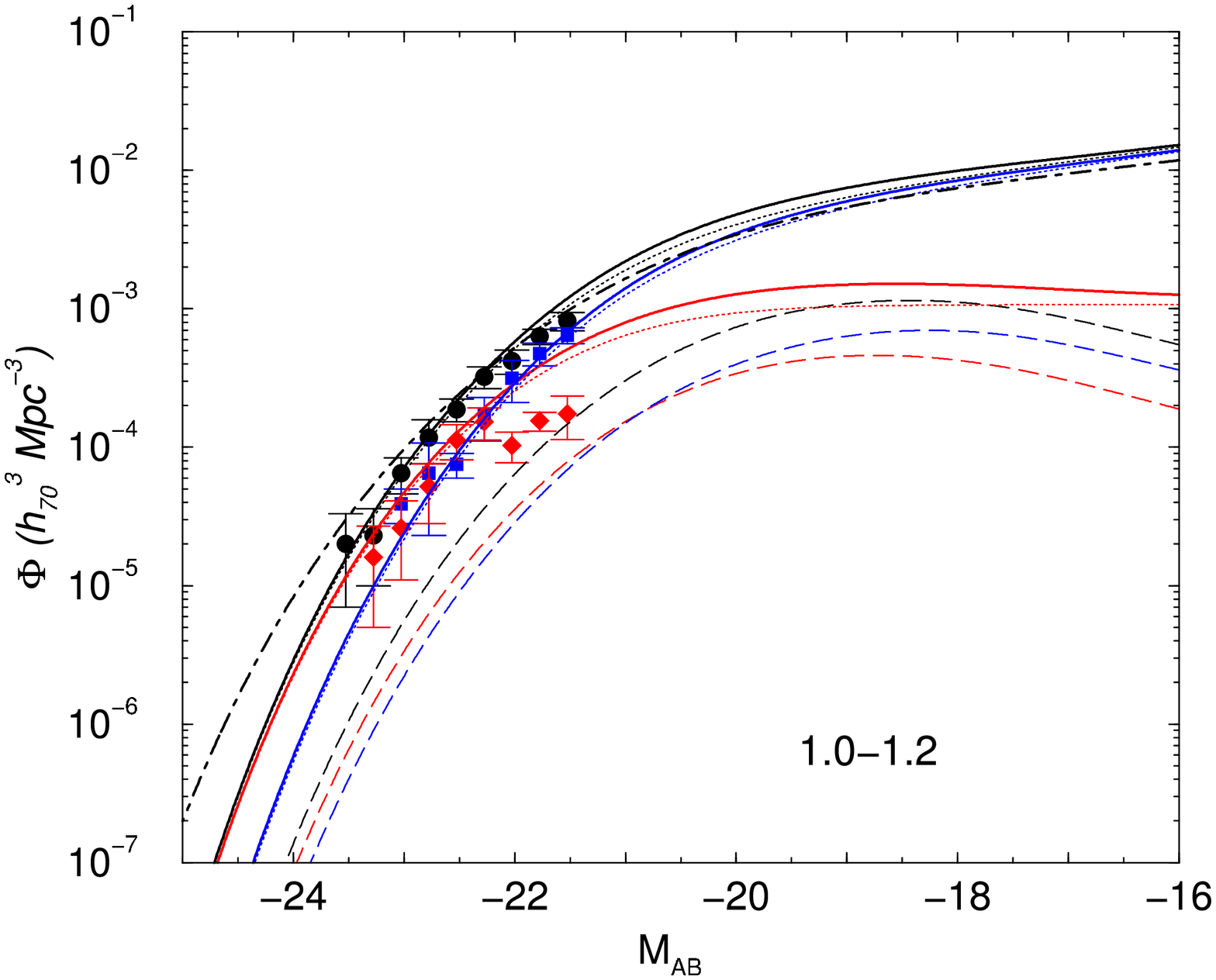,width=\hssize,angle=-90}}
\caption{
Luminosity function in the B-band as a function of redshift, from the COMBO-17 survey (from Bell et al. 2004).
The lines show the overall best fit model description while the dot-dashed line show the total LF
that best describes this data set alone. Other curves are same as the ones in Figure~3.
}
\end{figure*} 

\begin{figure*}
\centerline{\psfig{file=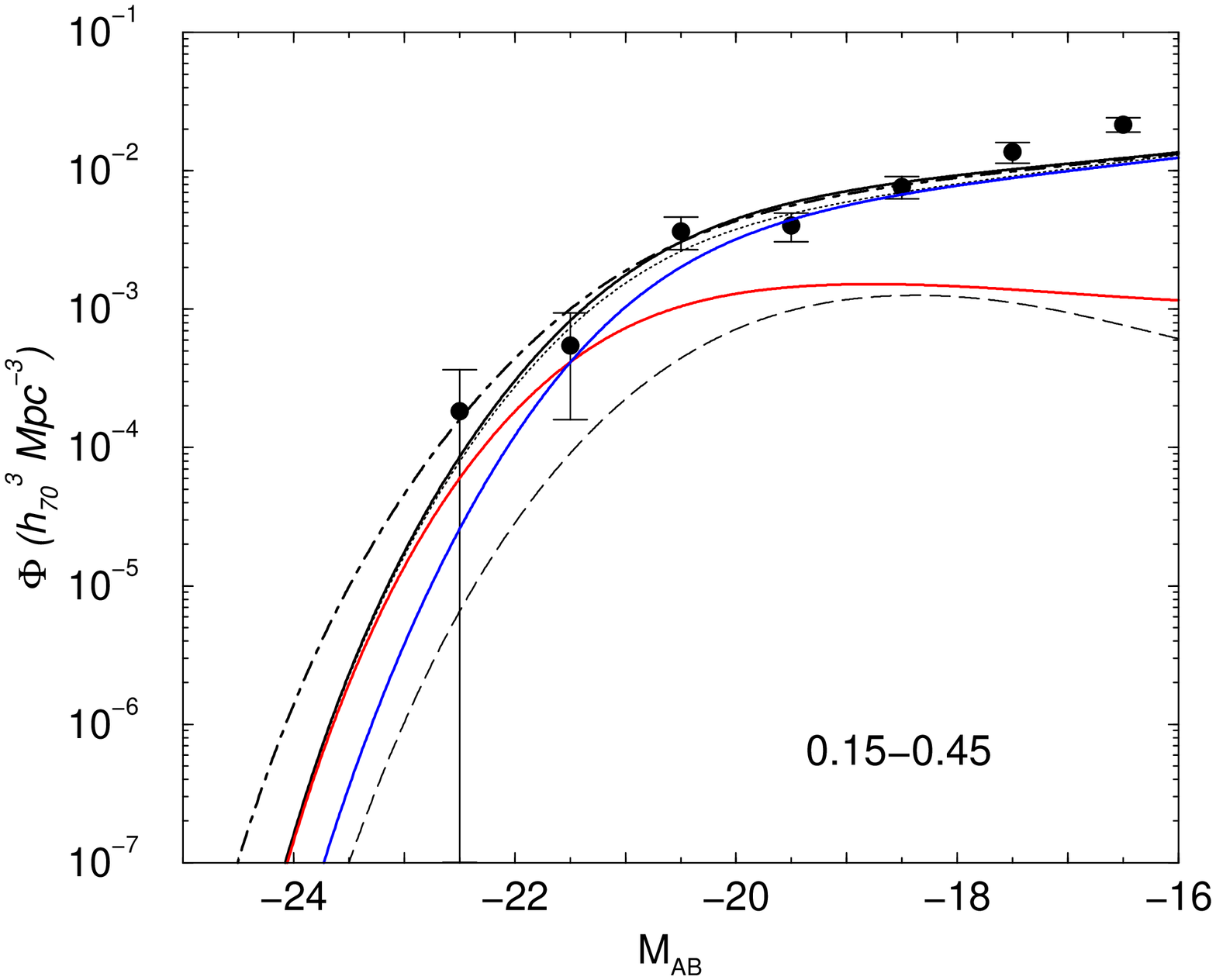,width=5cm,angle=-90}
\psfig{file=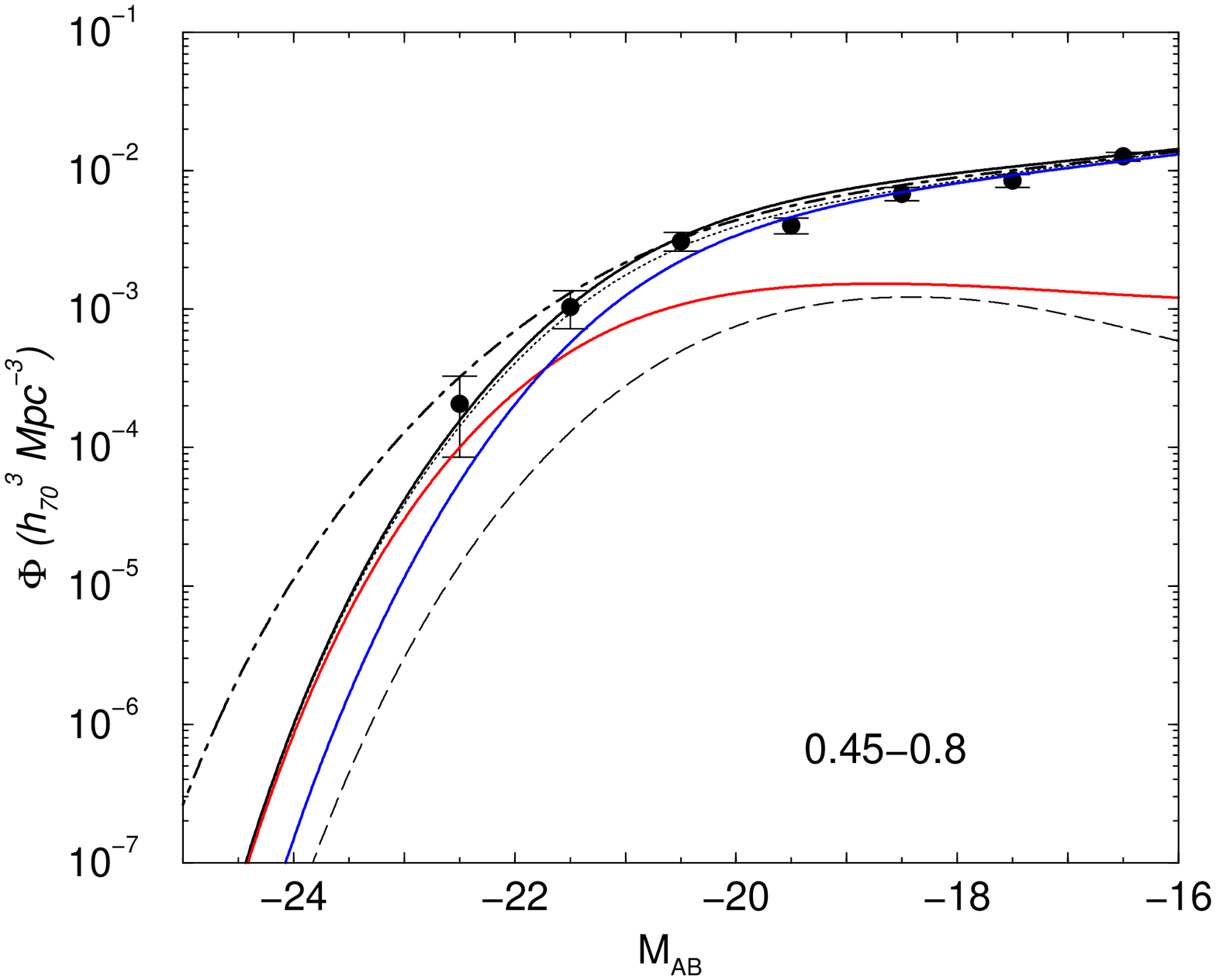,width=5cm,angle=-90}
\psfig{file=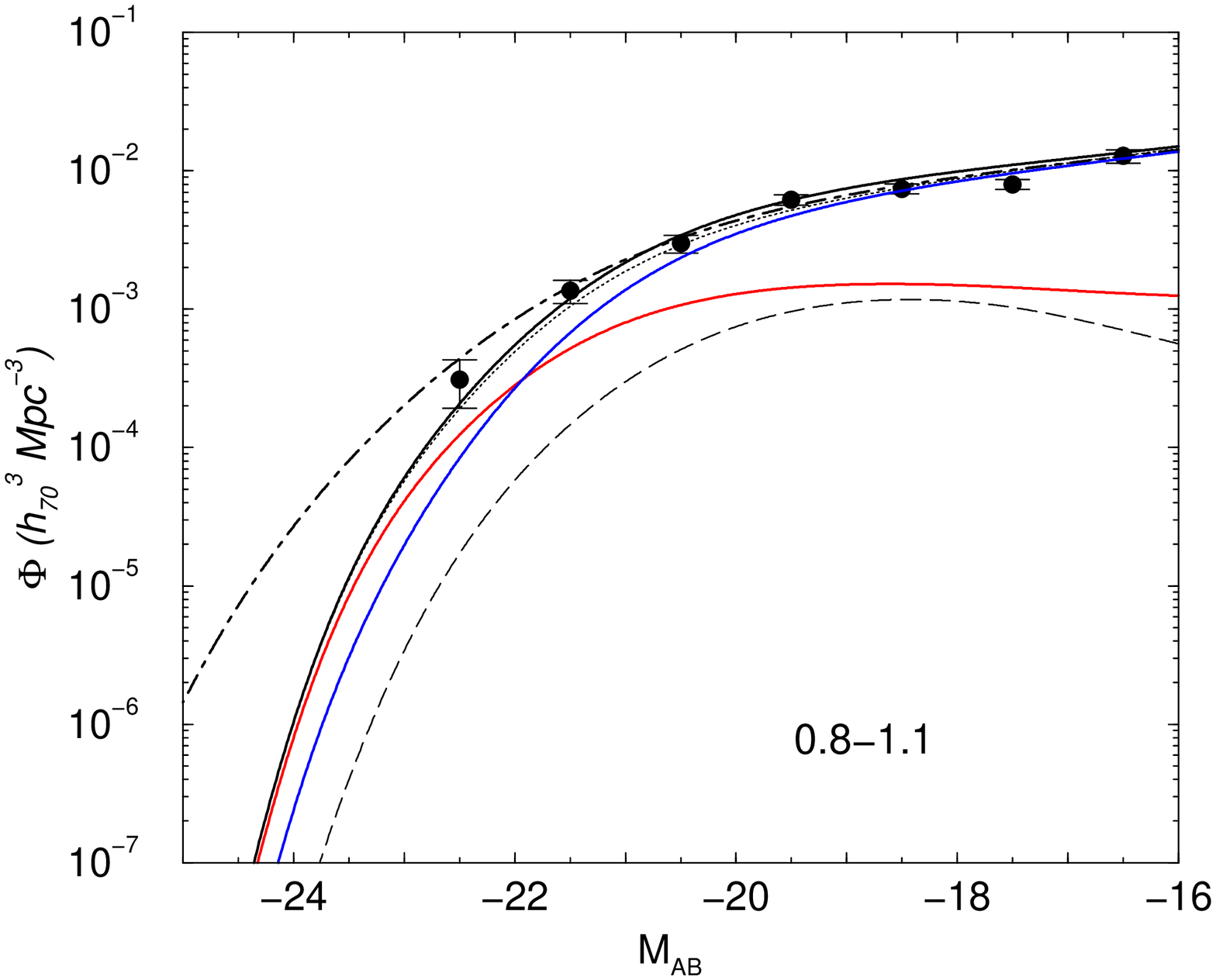,width=5cm,angle=-90}
}
\centerline{\psfig{file=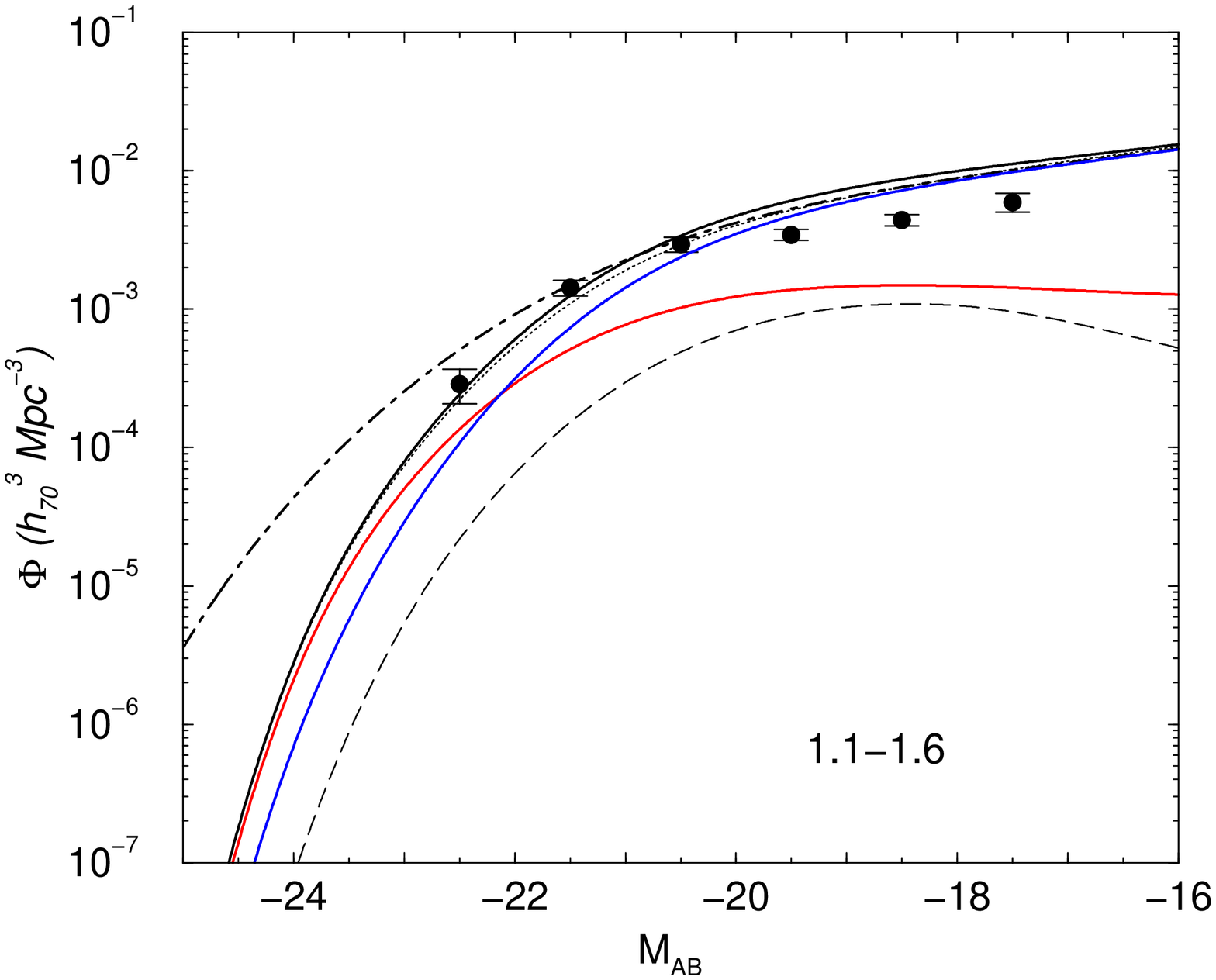,width=5cm,angle=-90}
\psfig{file=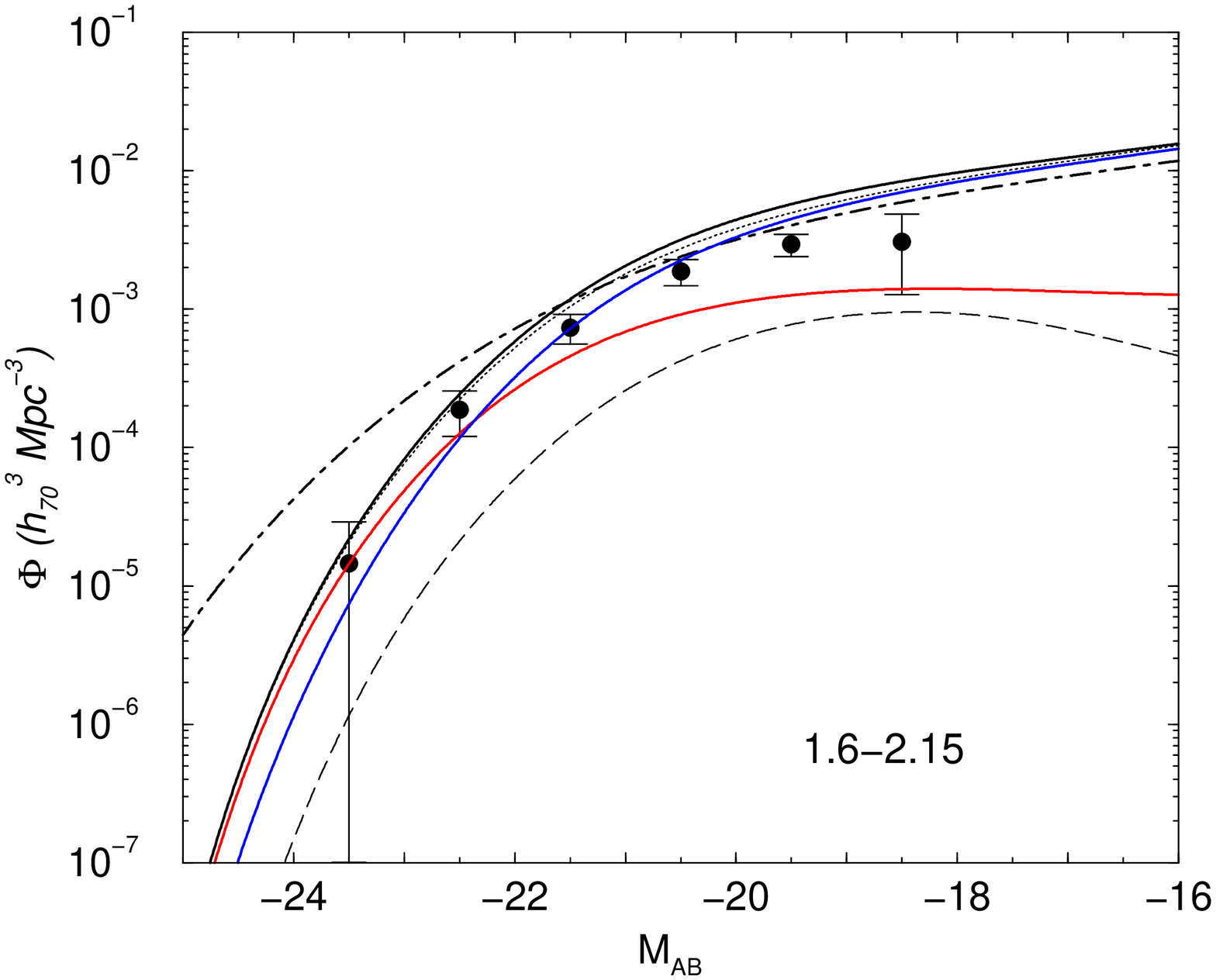,width=5cm,angle=-90}
\psfig{file=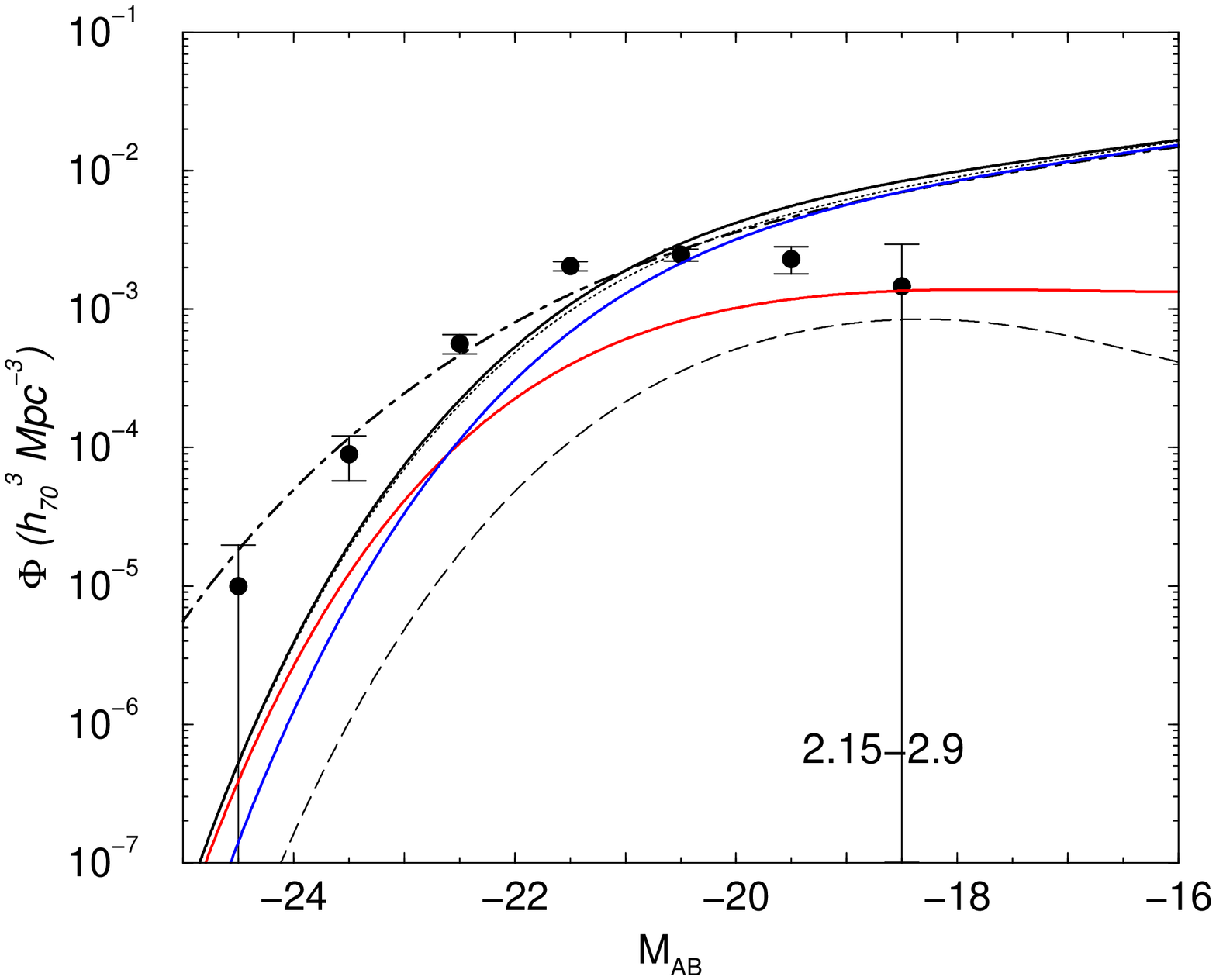,width=5cm,angle=-90}
}
\centerline{\psfig{file=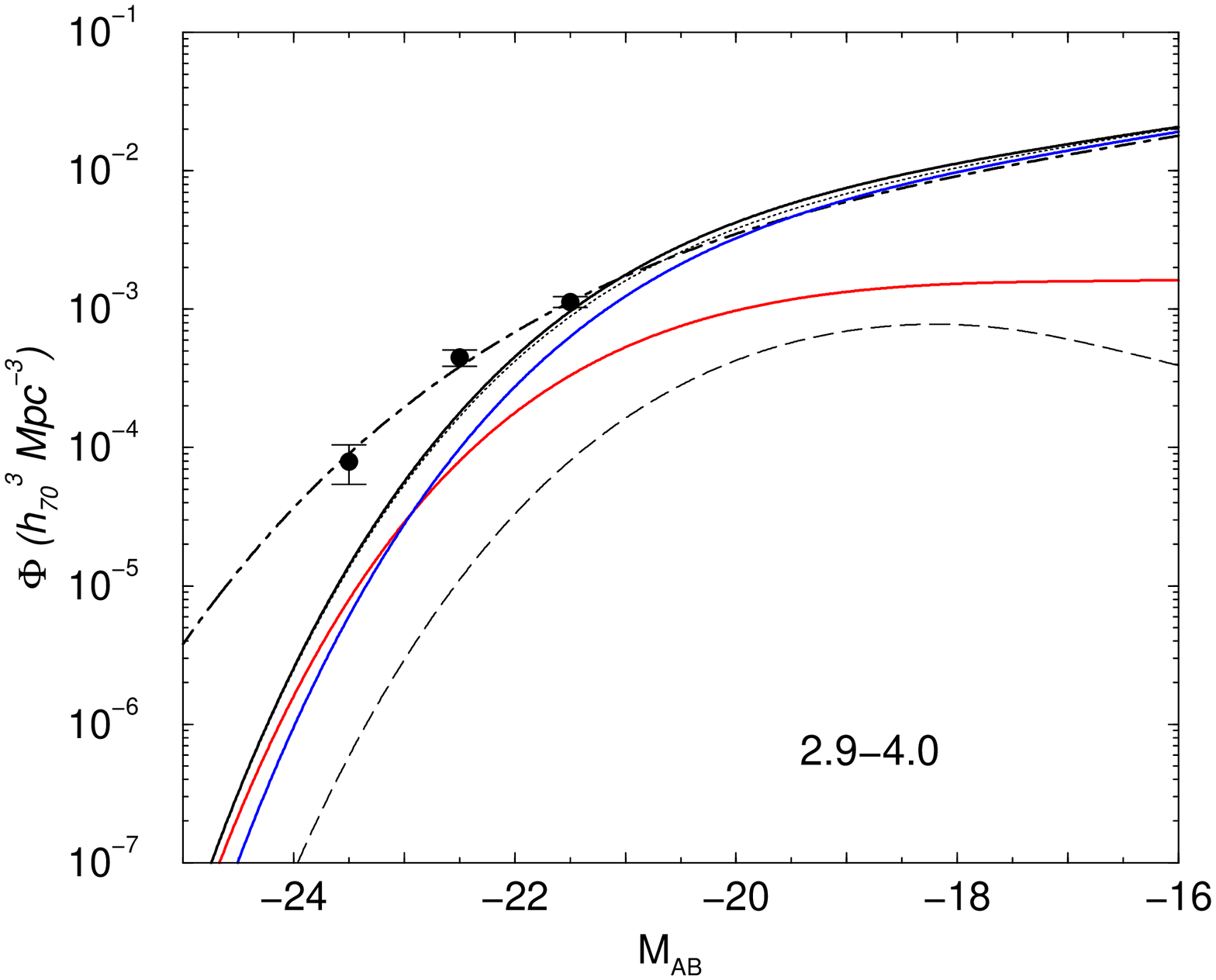,width=5cm,angle=-90}
\psfig{file=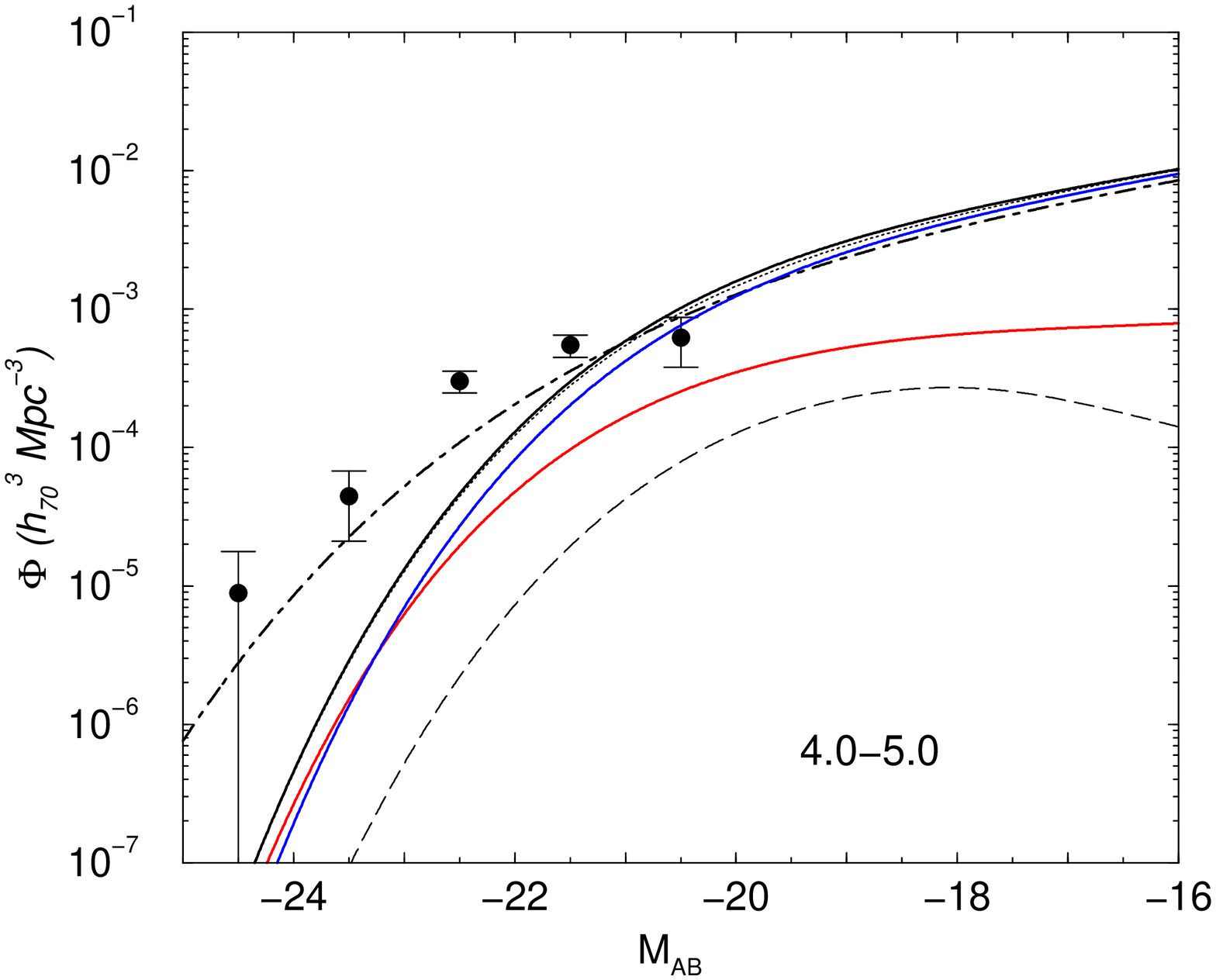,width=5cm,angle=-90}}
\caption{
Luminosity function in the B-band as a function of redshift (from Gabasch et al. 2004). 
The curves are same as Figure~7, except that we do not show model descriptions of early- and late-type
central and satellite LFs since the measured LF is related to the total galaxy sample only.
Note there difference between overall best fit model and the model preferred by this data alone,
especially at the highest redshift bin; this difference is also evident in Figure~5 where the
best fit region for this data set differs to some extent from the overall best fit region shown there.
}
\end{figure*}

\begin{figure*}
\centerline{\psfig{file=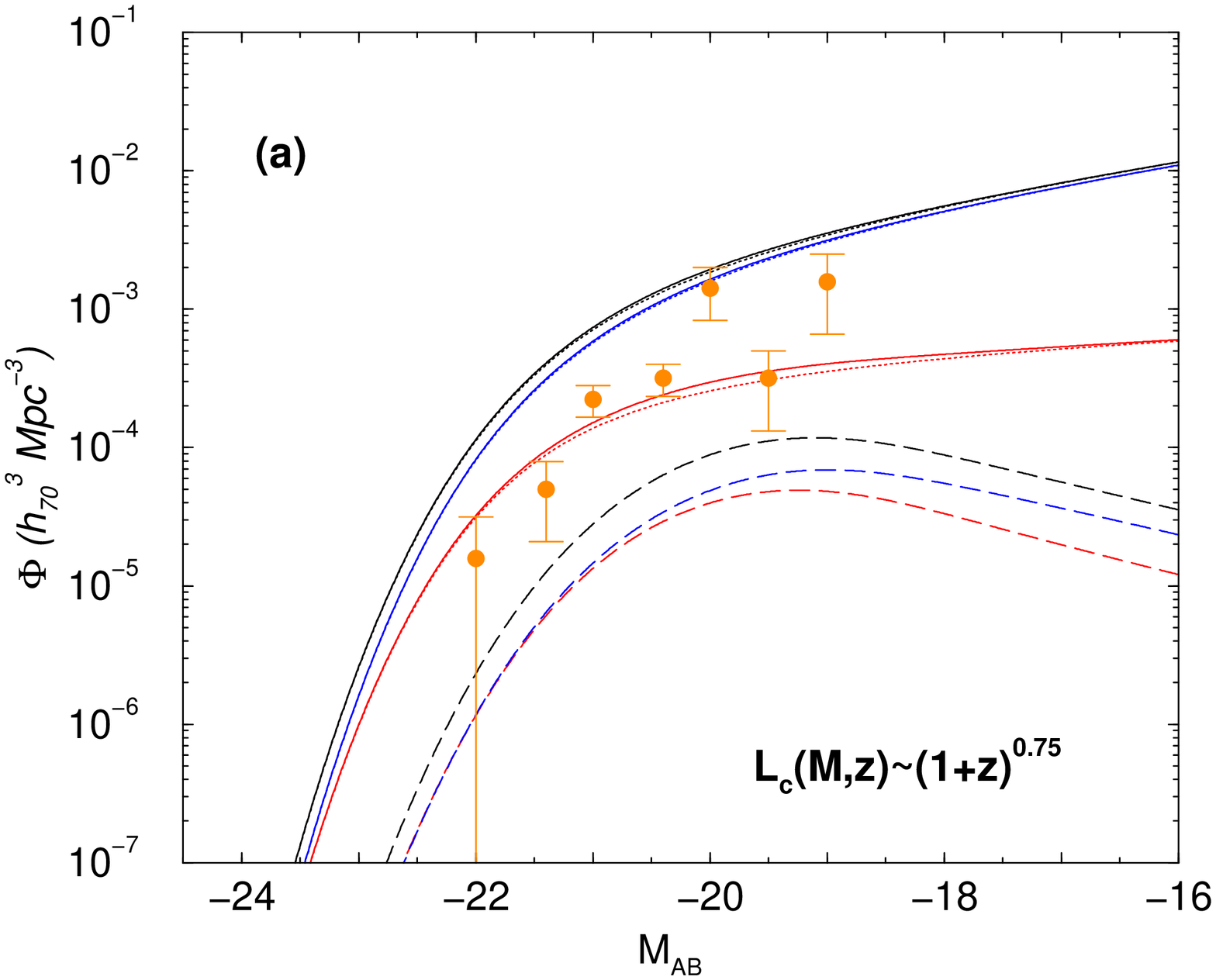,width=\hssize,angle=-90}
\psfig{file=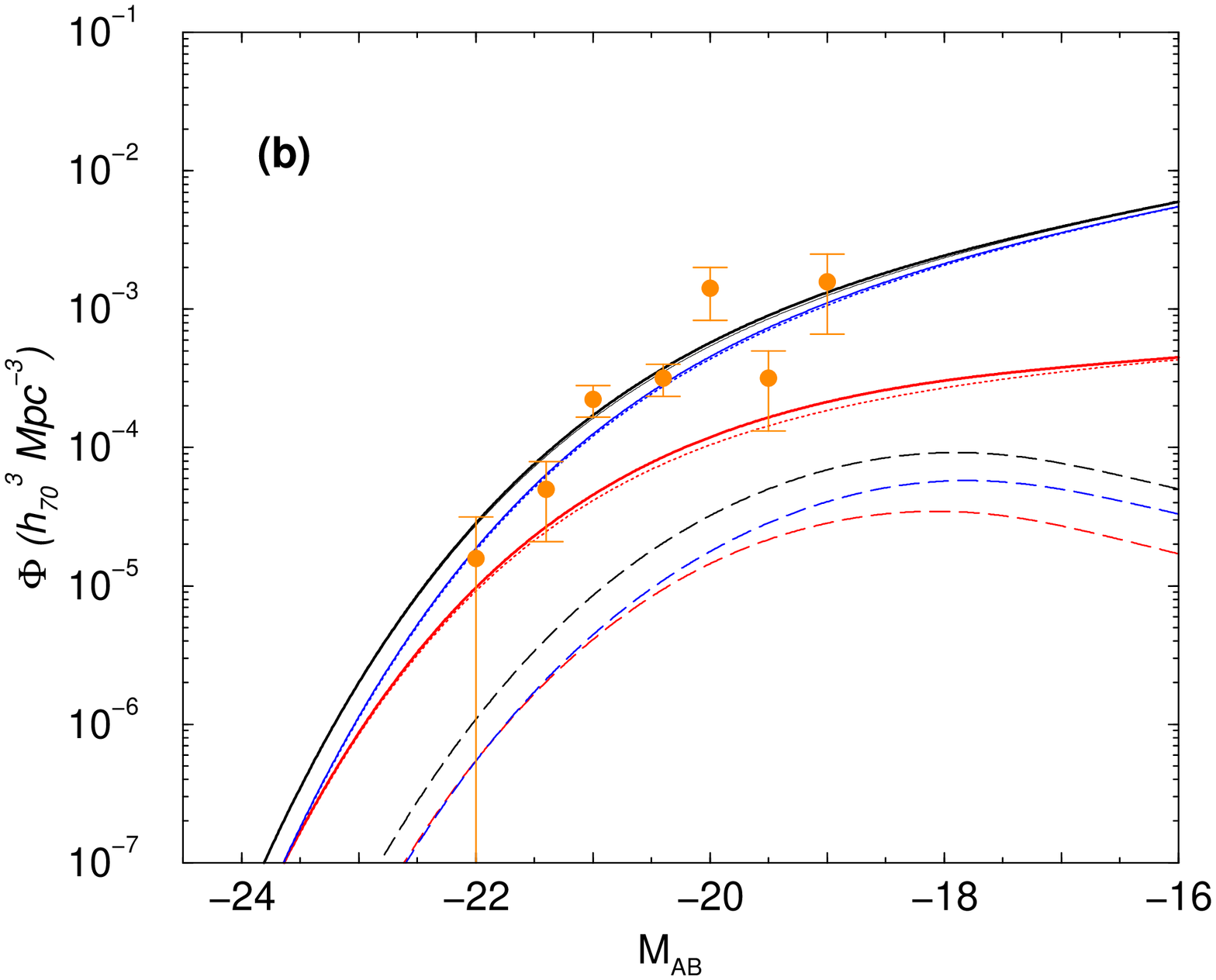,width=\hssize,angle=-90}}
\caption{The $z\sim 6$ UV LF. The measurements are from Bouwens et al. (2004a).
The curves show the model prediction based on pure-luminosity evolution, in panel (a),
and mass-dependent luminosity evolution, in panel (b), following Figure~1.
The lines follow the color and style scheme used in Figure~2. As shown in (a), pure luminosity evolution
matched to describe LFs out to $z \sim 3$ overestimates the observed density of galaxies,
while a mass-dependent luminosity evolution can provide an adequate description.
Note that there is a slight complication in this comparison;
the measurements are in the rest UV band while the predictions are based on a
$L_c(M,z=0)$ relation that is appropriate for rest B-band.}
\end{figure*}

\begin{figure*}
\centerline{\psfig{file=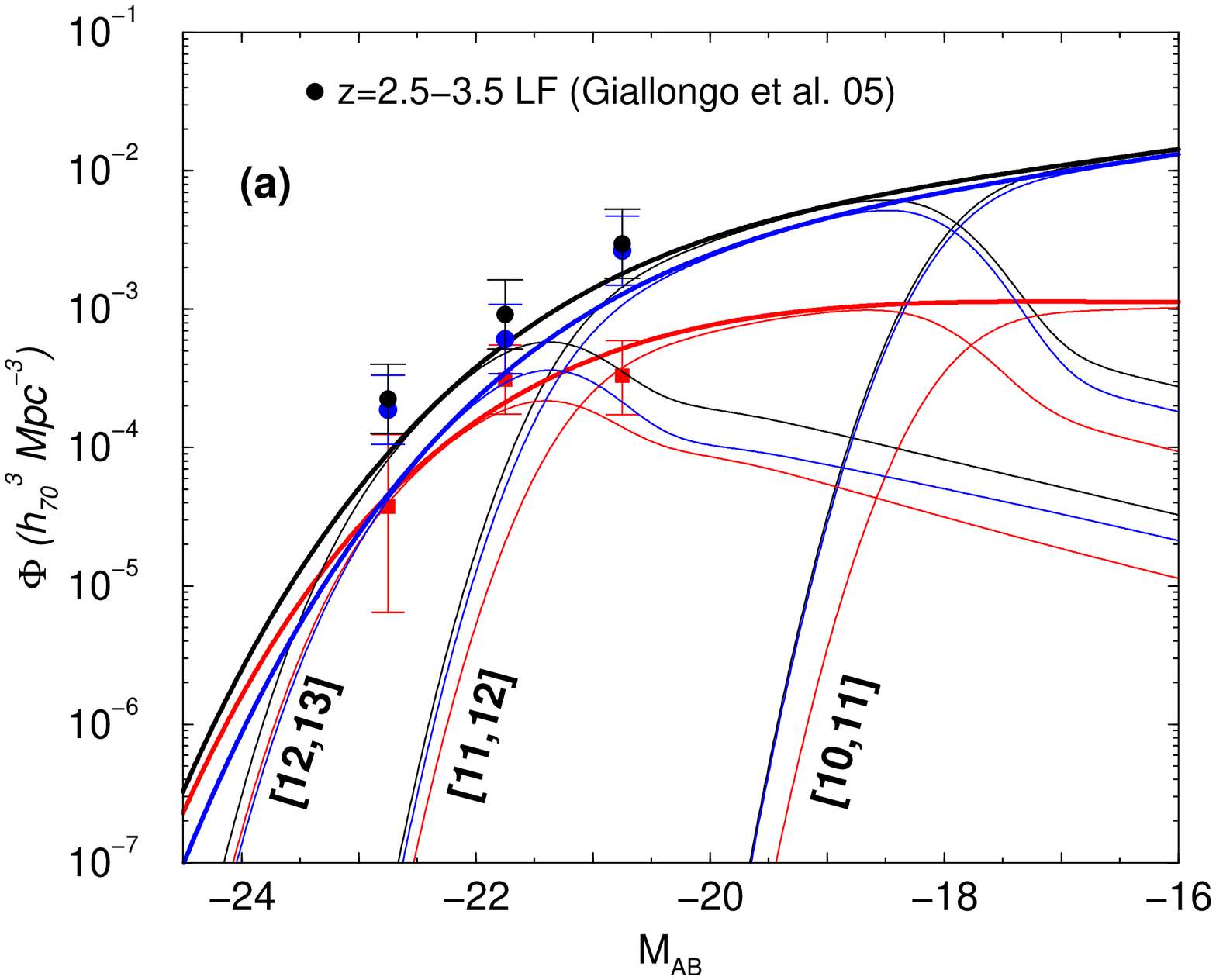,width=\hssize,angle=-90}
\psfig{file=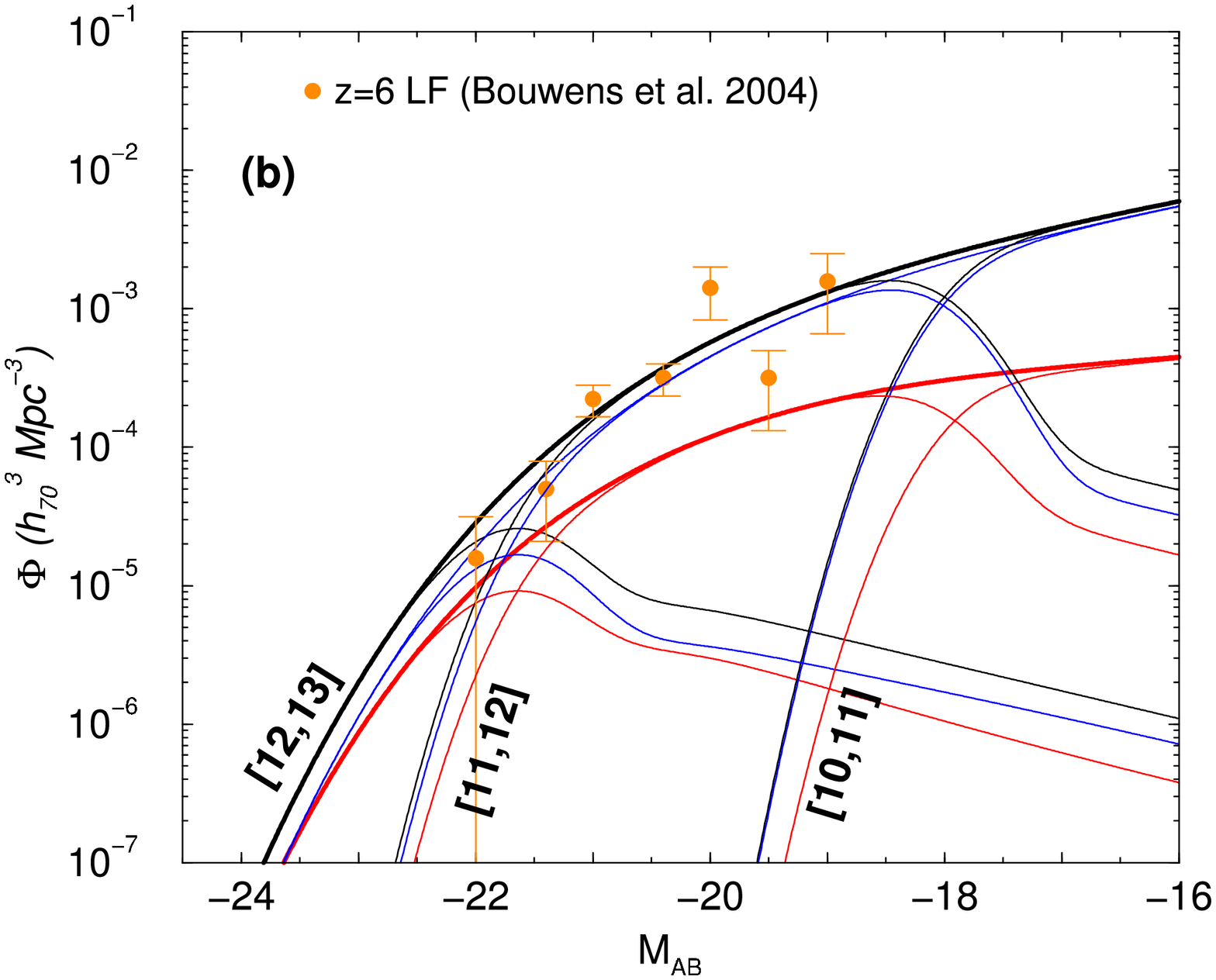,width=\hssize,angle=-90}}
\caption{ The mass dependence of the LF. In (a), we consider the $z\sim 3$ rest B-band LF from
Giallongo et al. (2005), while in (b), we consider the $z \sim 6$ rest UV LF from Bouwens et al.
(2004a). 
In both panels, the plotted curves are for the mass-dependent luminosity evolution considered
in Figure~4 (and show in Figure~1b). The mass ranges, in logarithmic values, are labeled on these plots.
The $z \sim 3$ LF is dominated by galaxies
in dark matter halos between $10^{11}$ to 10$^{13}$ M$_{\sun}$, while the $z\sim 6$ LF
is dominated by dark matter halos between $10^{11}$ M$_{\sun}$ and $10^{12}$ M$_{\sun}$.
}
\end{figure*}

\begin{figure*}
\centerline{\psfig{file=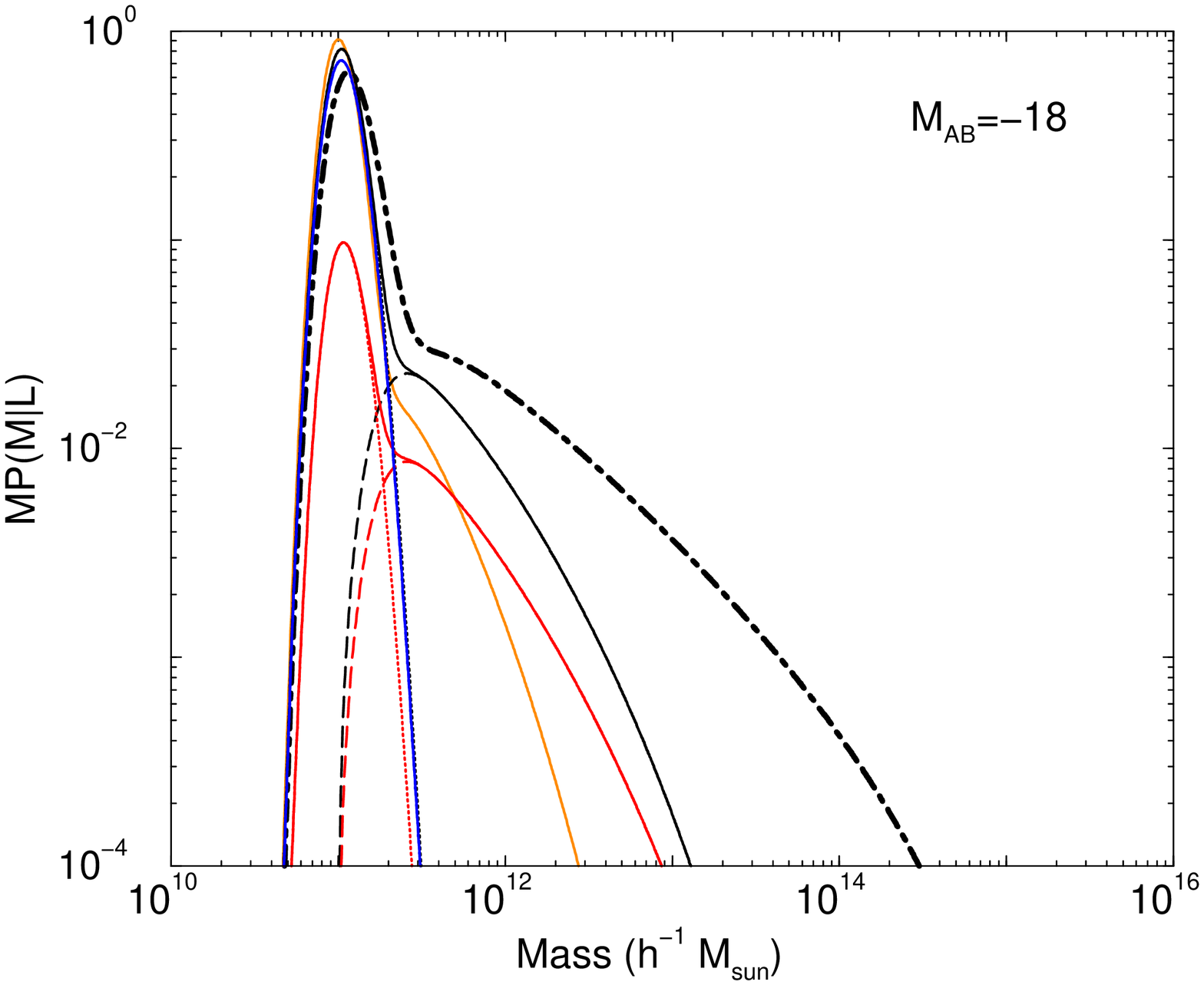,width=\hssize,angle=-90}
\psfig{file=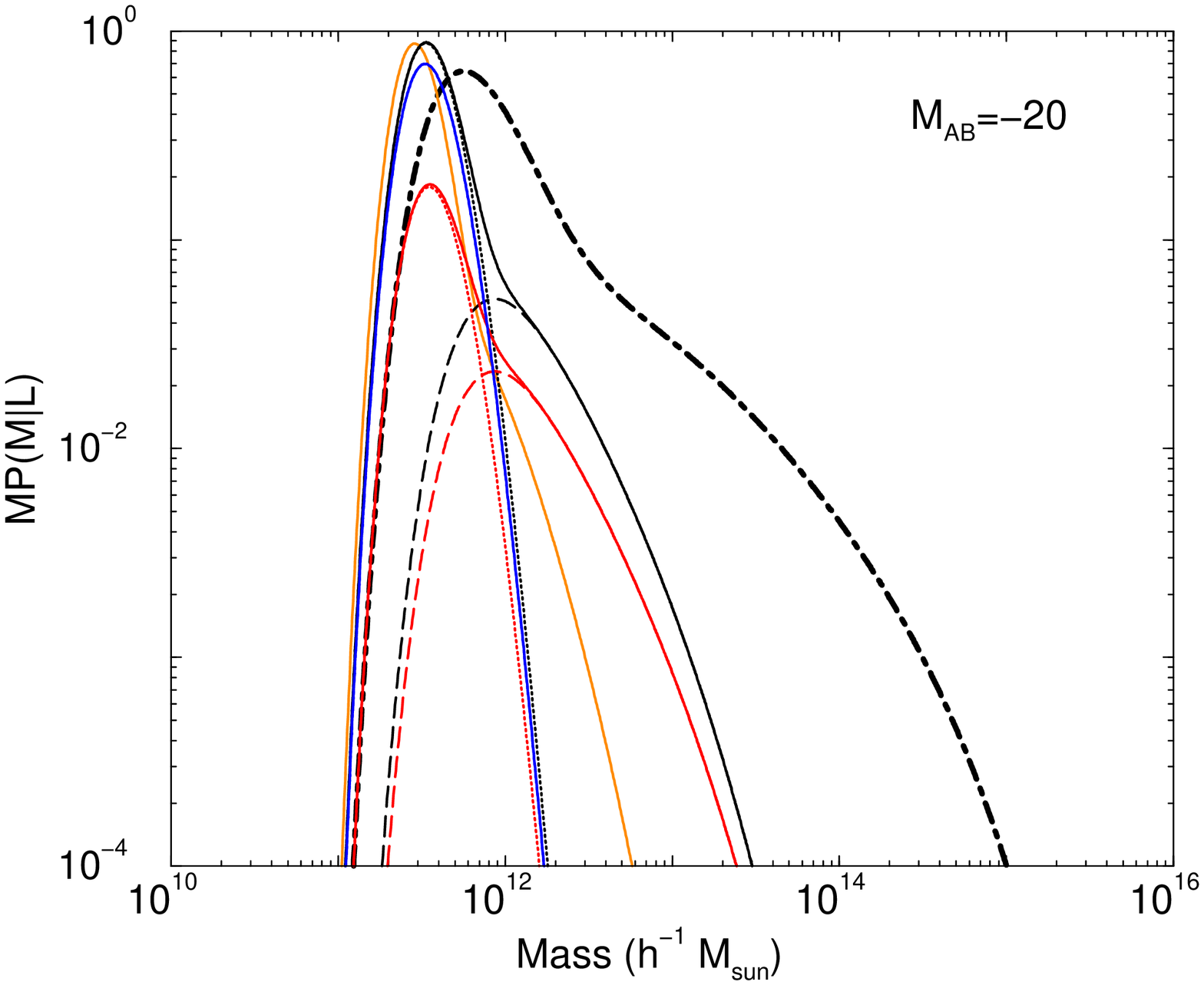,width=\hssize,angle=-90}}
\centerline{\psfig{file=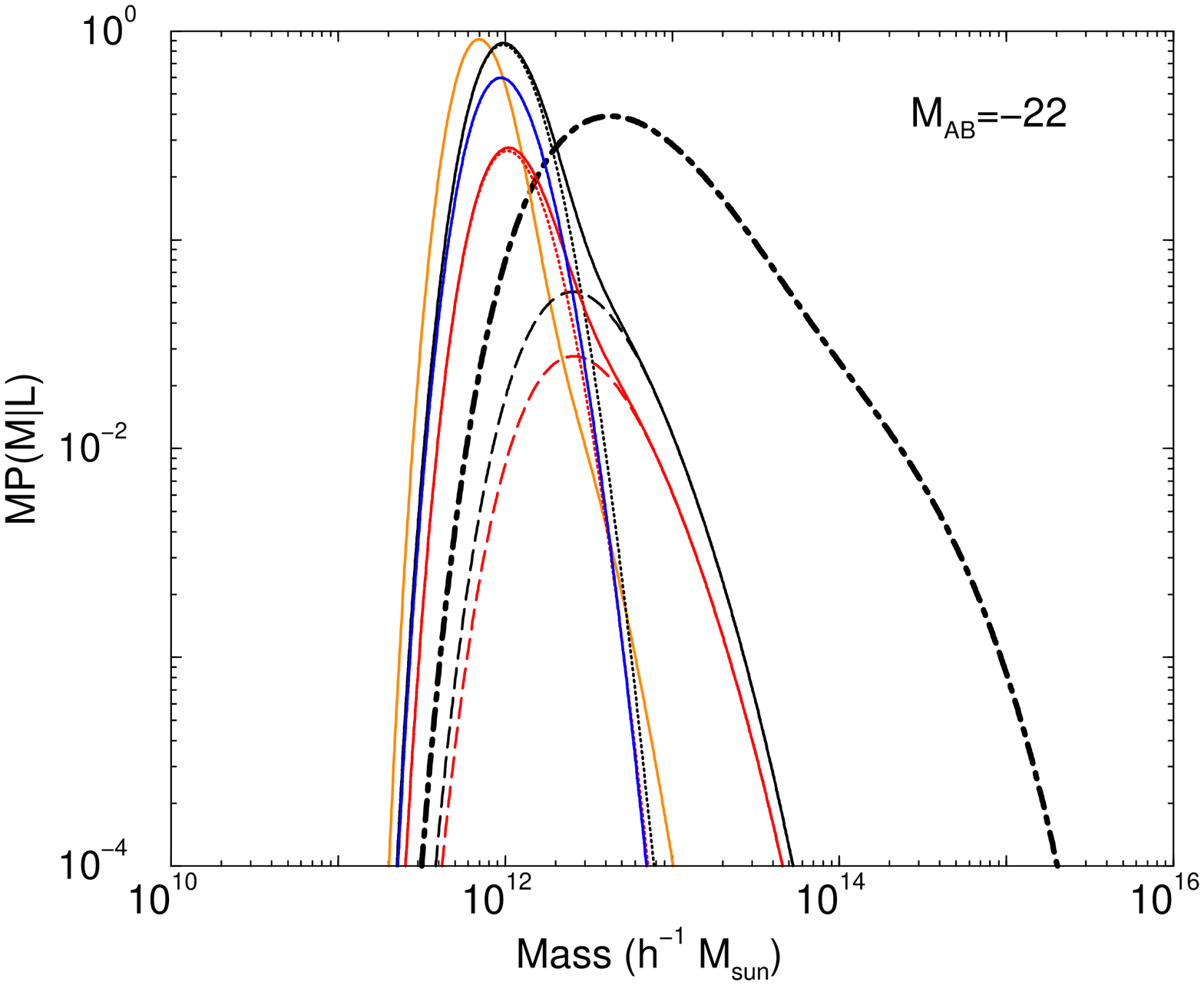,width=\hssize,angle=-90}
\psfig{file=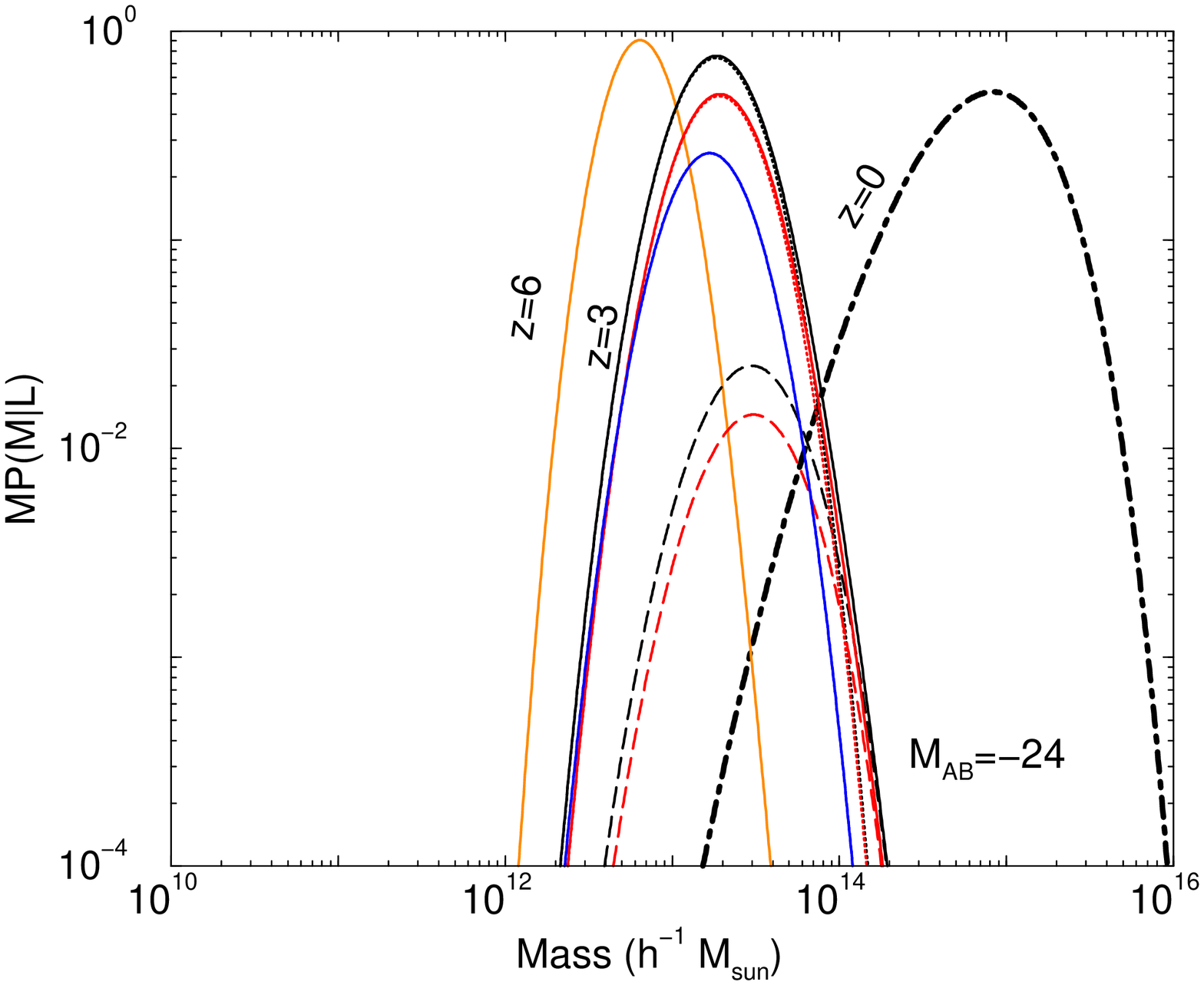,width=\hssize,angle=-90}}
\caption{The conditional probability distribution function of halo mass $P(M|L)$ to host a galaxy 
of the given luminosity at a redshift of 3, as a function of the halo mass. The black lines
are the total galaxy sample, while red and blue lines show the sample divided to early- and
late-type galaxies. The probabilities are calculated for the mass-dependent luminosity evolution model.
From top-left to bottom-right, we show these probabilities for $M_{\rm AB}=-18$, -20, -22, and
-24, respectively, in the B-band. The thick dot-dashed line is the same probability distribution for
the total galaxy sample at $z=0$, while the orange line shows the same at $z=6$.}
\end{figure*} 

Note that we are assuming here that the central galaxy luminosity of a given halo increases with redshift. This assumption
does not violate the fact that the halo occupation number is not changing with redshift (e.g., Yan et al. 2003; Coil et al. 2004), 
since the integral of $\Phi(L|M)$ over luminosities remain the same; all we have done is to shift the mean
of the log-normal distribution that describes central galaxy luminosity distribution to a higher 
luminosity when compared to the value today. The fact that the halo occupation number is the same at high redshifts,
when compared to today, should not be considered as a statement that galaxy properties do not change with redshift.

\section{High-Redshift Luminosity Functions}

Given our model for CLFs, we can now construct the LF by averaging CLFs over 
the halo mass distribution given by the mass function. Here, we use the
Sheth \& Tormen (1999; ST) mass function $dn/dM$ for dark matter 
halos. This mass function is in better agreement with numerical simulations (Jenkins et al. 2001), when compared to
the more familiar Press-Schechter (PS; Press \& Schechter 1974) mass function.
While there are differences in the ST mass function and the numerically simulated mass functions at the high mass end,
these differences do not affect these results as statistics of the LF at the present day are dominated by galaxies
in halos around $10^{13}$ M$_{\sun}$. As one moves to a higher redshift, statistics become dominated by
lower mass halos than today; at $z \sim 3$, statistics are dominated by halos with mass in the range between
few times $10^{11}$ M$_{\sun}$ and few times $10^{12}$ M$_{\sun}$.
Here, we make the assumption that the ST mass function is the correct
description for halo masses out to $z \sim$ 6 and above; any differences in the evolution of the mass function,
relative to ST description, could affect our conclusions regarding the luminosity evolution.
The same can also said of our assumption related to underlying cosmological model; a different
cosmology than the one considered here impacts the redshift evolution of the mass function differently than
the one we assume here. These differences, however, are at the few percent level, at most, and do not
strongly affect our results given the few tens percent uncertainties in the LF at high redshifts.

Given the mass function, the galaxy LF as a function  of $z$ is 
\begin{equation}
\Phi_i(L,z) = \int_0^\infty \Phi_i(L|M,z) \frac{dn}{dM}(z) dM \,,
\end{equation}
where $i$ is an index for early and late type galaxies.
The conditional luminosity function for each type involves the sum of central and satellites.
To compare with Giallongo et al. (2005), Willmer et al. (2005; DEEP2) and
Bell et al. (2004; COMBO-17) measurements, we will plot these two divisions, as well as the sum separately
for late-type (blue) and early-type (red) galaxies. In the case of Gabasch et al. (2004),
since the measurements did not consider the division to galaxy types, we will only consider the total
LF.

In Figure~2, we show the LF of galaxies at redshifts out to 3.5 from Giallongo et al. (2005). 
In Figure~2(a), we present a comparison to a model of the
$z=0$ LF of 2dFGRS data from Cooray (2005). In Figure~2(b), we assume no redshift evolution in the
$L_c(M,z)$ relation. The resulting LFs are then affected only by the redshift evolution of the dark matter halo
mass function, $dn/dM(z)$. At low redshifts, the dark matter halo mass function evolves such that
the number density of massive halos is rapidly 
decreasing while the density of halos at masses around and below $\sim 10^{12}$ M$_{\sun}$ is slightly increasing
relative to the mass function at $z=0$ (see, Reed et al. 2003). This evolution in the halo mass function is directly 
reflected on the high redshift galaxy 
luminosity function. Given that the $L_c(M,z=0)$ relation flattens at mass scales $\sim$ 10$^{13}$ M$_{\sun}$ (Cooray \& Milosavljevi\'c 2005a), 
the rapid decline in the number density of massive halos, corresponding to groups and clusters, does not lead to the same fractional
 decline in the bright-end of the galaxy LF relative to values today. Models based on the mass function evolution 
 alone, however, suggest a decline in the density of bright galaxies at high-redshifts when compared to densities measured today.

On the other hand, the observed galaxy LF at high redshifts indicates that the bright-end density is, in fact, increasing as one moves
to $z \sim 3$ from today. To compensate for the decline in the number density of dark matter halos that host galaxies, associated with
the redshift evolution of the mass function, the only possibility to increase the density of luminous galaxies
is to consider positive luminosity evolution; The galaxies {\it must} brighten
at high redshifts relative to luminosity values today. This brightening can be accomplished in several ways.
First, galaxies, regardless of the host halo mass, can brighten by a constant factor;
This description can be considered as a scenario involving pure luminosity evolution.
We consider this possibility by scaling the $L_c(Mz,z=0)$ relation by $(1+z)^\alpha$
with $\beta$ and $\eta$ both set to zero.
With $\alpha=0.75$, our model descriptions are plotted in Figure~3. For comparison,
 we also plot the measured LFs by Giallongo et al. (2005) where we show the total sample as well as the
 division to galaxy types.
The $L_c(M,z)$ relations related to this pure luminosity evolution scenario is shown in Figure~1(a).
The model can provide an adequate description, though one underestimates the density of
most luminous galaxies shown with the bright-end data points in Figures~3(b), (c) and (d), while overestimating
the faint-end density, for example, in Figure~3(c). To obtain a better fit to the bright-end density, one
can increase $\alpha$. This, however, comes at the expense of overestimating 
the number density of galaxies at the faint-end further.

A better description of the Giallongo et al. (2005) data may be that the luminosity evolution is
mass dependent. In Figure~1(b), we plot $L_c(M,z)$ relations under this alternative description, where we
allow the luminosity of halos above the flattening mass scale to grow rapidly with redshift. This is accomplished with
non-zero values for two parameters $\beta$ and $\eta$, with $\alpha=0$, though, 
alternative model descriptions that increase luminosities of central galaxies
in massive dark matter halos, while keeping the luminosity at the low-end  of the mass distribution essentially the same as 
today, can also be considered. As shown in Figure~4, where we also plot
measurements of Giallongo et al. (2005), with the mass-dependent
luminosity evolution description, the density of bright galaxies is increased
while keeping the faint-end density similar to values measured out to $z \sim 3$. It is likely that this model provides a 
more accurate
description of the data, though given various uncertainties in LFs out to a redshift of 3.5, 
we cannot distinguish reliably between the mass-dependent luminosity evolution and the pure luminosity evolution 
description from this data set alone.

\subsection{$L_c(M,z)$ evolution through model fits to data}

While Giallongo et al. (2005) data alone may not allow us to differentiate between 
the mass-dependent luminosity evolution and the pure luminosity evolution the DEEP2 (Willmer et al. 2005), COMBO-17
(Bell et al. 2004) and Gabasch et al. (2004) LFs may provide adequate statistics to consider detailed
model fits in combination. For this purpose, we describe the $L_c(M,z)$ relation as
\begin{equation}
\label{eqn:lcmz}
L_c(M,z) = L_0(1+z)^\alpha \frac{(M/M_1)^{a}}{[b+(M/M_1)^{cd(1+z)^\beta}]^{1/d}}\, ,
\end{equation}
where we have combined the two model descriptions to one by allowing variations in $\alpha$, pure luminosity
evolution, and $\beta$, mass-dependent evolution, but setting $\eta=1$ as $L_c(M,z)$ relation is not
strongly sensitive to $\eta$ given the appearance of $d$ and $1/d$ in the numerator of equation~(\ref{eqn:lcmz}).
We vary $\alpha$ and $\beta$ and consider likelihood model fits to LFs from DEEP2, COMBO-17, and
Gabash et al. (2004); given that DEEP2 and COMBO-17 probe the same redshift range,  the models
are distinguished mostly based on a comparison of those datasets with Gabasch et al. (2004) B-band
LFs out to $z \sim 5$. 

In Figure~5, we summarize our results in the $(\alpha,\beta)$ plane where we show the $1\sigma$, $2\sigma$, and 
$3\sigma$ allowed values of these parameters based on a comparison to DEEP2/COMBO-17 LFs and Gabasch et al. (2004)
LFs separately as well the combined constraints. When model fitting to the data, in the case of
DEEP2 and COMBO-17 LFs, we only consider the total LF and do not use the LFs of galaxy types;
In an upcoming paper, we will consider an analysis of galaxy statistics related to types including
an analysis of galaxy type-dependent clustering at $z \sim 0$ and the redshift dependence based on
galaxy type LFs from DEEP2 and COMBO-17.
While DEEP2/COMBO-17 allow $\beta >0$, in combination
with high redshift LFs, we can establish that at the 3 $\sigma$ level, $\beta <0$ with a preferred value around
-0.1; As shown in Figure~1(b), $\beta <0$ leads to an increase in the  luminosities of central galaxies
at large halo masses while the luminosities at low masses remain the same. While $\alpha =0$ is consistent with
the data, the constraint in the $(\alpha,\beta)$ plane lies in a degeneracy line that allow for
large negative values for $\alpha$, leading to a decrease in the luminosity of galaxies
under the $L_c(M,z)$ relation, compensated by the increase in luminosity in the
$L_c(M,z)$ relation associated with negative values for $\beta$. In either case, our models suggest one
strong conclusion: while pure-luminosity evolution is consistent, the high redshifts LFs require halo mass-dependent
evolution in the $L_c(M,z)$ relation.

In Figs.~6 and 7 we show the DEEP2 and COMBO-17 LFs and a comparison to model fits based on these
constraints, respectively. Here, for comparison, we also show the LF of galaxy types. In addition to the best fit model
parameters for $\alpha$ and $\beta$ based on the total sample, we also show the total LF related to
model fits using these data alone. The difference here is minor since out to $z \sim 1$, the difference
between overall best-fit parameters $\alpha$ and $\beta$ and the same parameters that best describe
these datasets best individually is minor. In Figure~8, we consider the LF out to $z \sim 5$ from Gabasch et al. (2004).
Here, again, we show the overall best-fit model as well the model that describe  Gabasch et al. (2004) LFs
best. In this case, the difference is significant since small variations in $\alpha$ and $\beta$ can lead to
large differences at high redshifts. For example, the overall best fit model, does not describe adequately the
Gabasch et al. (2004) LF at the highest redshift (results in an underestimate of the
number density of galaxies at the bright-end), though, this difference is accounted by the best fit model
preferred by these data alone.

While our models generally suggest that the central galaxy luminosity increases with redshift,
we do not find strong evidence for a pure-luminosity increase in the B-band, but rather a mass dependent increase.
It will be useful to see if the same increase repeats in other bands, such as the in K-band. Observations suggest that
in such redder bands, galaxy luminosities increase first with increasing redshift (passive evolution),
but then begin to decrease beyond a particular redshift that correspond to the formation  epoch
of galaxies where most stellar mass was assembled (e.g., Drory et al. 2003). In our models, the formation of
galaxies are strictly linked to the assembly of dark matter halos.  In the mass-dependent
luminosity evolution description, galaxies in dark matter halos below  a certain
characteristic halo mass scale, M$_c \approx 10^{11}$ M$_{\sun}$ in Figure~1(b),
do not increase in luminosity with an increase in the redshift. Thus, a
a decrease in galaxy luminosity beyond a certain redshift could be associated with the transition where
average halo mass at that epoch, given the halo mass function corresponding to that redshift,
becomes less than $M_c$. While we have not seen clear indications for such a transition in the B-band
high-z LFs, we plan to investigate this possibility in detail in near future using
high-z LFs in redder bands (e.g., Drory et al. 2003), where the $L_c(M,z)$ relation 
may become more sensitive to the mass-dependent luminosity evolution needed to explain the observations.

In addition to the redshift evolution of the total galaxy LF, the model description built using Cooray (2005)
model, but with an accounting of the redshift evolution in the
$L_c(M,z)$ relation is in good agreement with high redshift LFs of galaxy types. Note that we have not allowed for
a redshift variation in the fraction of red and blue galaxies, as a function of the halo mass.
As shown in Figures~3 and 4 in the case of Giallongo et al. (2005) data and Figures~6 and 7 with DEEP2 and COMBO-17 data, 
these models generally overestimate the LF of red galaxies
at the faint-end.
At $z > 1$, the fraction of blue galaxies increases; in Cooray (2005), the late-type (blue) fraction of central galaxies
increases at low-mass halos. Thus, at high redshifts, with the rapid disappearance of halos with masses above few times
$10^{13}$ M$_{\sun}$, the blue fraction begins to dominate. This, however, does not mean that all galaxies at redshifts out to 3 or so is
late-types. We certainly expect $\sim$ 50\% of bright galaxies at redshifts $\sim$ 3 to be early type, red galaxies.
The exact observed 
fraction of late- vs. early-type galaxies, as a function of redshift, can eventually be used to update this model and, especially, to understand whether there is a redshift evolution in the galaxy type fraction relative to values seen today.

\subsection{$z \sim 6$ LF}

While there is no measured LF of galaxies at $z\sim 6$ in the rest B-band, ignoring complications resulting from differences in the
rest wavelength, we can also compare our model predictions with the UV LF at $z\sim 6$ from Bouwens et al. (2004a).
In Figure~9, we summarize our results. Note that the pure luminosity evolution scenario
over predicts the LF at all luminosities of interest, while the mass-dependent luminosity evolution
provides a reasonable description of the data; note that any agreement or disagreement between these
models and measurements must be considered with the difference in color in mind.
This is due to the fact that, while the $L_c(M,z=0)$ relation is constructed as appropriate for the B-band,
the measurements at $z \sim 6$ is in the rest UV band. 

Given differences in model predictions compared to measurements in Figure~9, 
however, we believe that the two luminosity evolution descriptions, pure luminosity evolution at all mass scales
or mass-dependent luminosity evolution only at the high mass end, may be distinguished with $z \sim 6$ LFs, though these
two models give equally acceptable description of LFs out to a redshift of 3.5. Thus, based on the UV LF at $z \sim 6$, we suggest
that a favorable description of the high-redshift galaxies may be the evolution of luminosities based on the host halo mass.
As shown in Figure~9, the $z \sim 6$ LF is clearly dominated
by late-type blue galaxies; However this does not mean that there are no early-type galaxies at these
high redshifts.  The early-type red galaxies will only appear,
in the statistical sense, at the bright-end with a density close to that of blue galaxies.
At the faint-end, for each red-type galaxy, one should statistically expect a factor of 5 to 10 more
blue galaxies, depending on the luminosity.
The existence of bright and red galaxies at redshifts $\sim$ 6, as found with Spitzer  (Yan et al. 2005;
see, review in Stark \& Ellis 2005) does not contradict these models, as we do expect early-type
galaxies to be present even at these redshifts. While currently the observed sample is a few galaxies,
a precise determination of their number density, or the luminosity function, could strongly constrain
the redshift evolutionary properties of the early-to-late type fraction, which we have assumed to be a constant
in our models, so far.

\subsection{$z \sim 3$ to 6 halo masses from LFs}

Given that our model for the LF is constructed from CLFs --- the number of galaxies as a function of the halo mass ---
we can directly address an important question as to what  mass dark matter halos host galaxies seen at
$z \sim 3$ to 6. We show the mass dependence of the LF in Figure~10. At $z \sim 3$,
galaxies that are brighter than $M_{\rm AB} \sim$ -22  are hosted in dark matter halos with masses in
the range of $10^{12}$ to $10^{13}$ $h_{70}^{-1}$ M$_{\sun}$. To statistically detect dark matter halos
at $\sim$ $10^{11}$ M$_{\sun}$, one must study the LF down to an absolute magnitude of -18.
 In the case of $z \sim 6$ galaxy LF, all galaxies in the luminosity range corresponding to
 absolute magnitudes between -22 and -19 are hosted in dark matter halos between
$10^{11}$ to $10^{12}$ $h_{70}^{-1}$ M$_{\sun}$; the bright-end of the $z \sim 6$ LF corresponds to the
upper-end of this mass range.  

While these are approximate mass ranges,
using CLFs, we can quantify the mass distribution of $z\sim 3$ to $6$ galaxies exactly. Here,
we calculate the conditional probability distribution $P(M|L,z)$ that a galaxy of
a given luminosity $L$ at redshift $z$ is in a halo of mass $M$ (Yang et al. 2003b, Cooray 2005):
\begin{equation}
P(M|L,z)dM=\frac{\Phi(L|M,z)}{\Phi(L,z)} \frac{dn(z)}{dM} \; dM \, .
\end{equation}
In Figure~11, we summarize our results, where we plot probabilities at luminosities
that correspond to absolute magnitudes of -18 to -24. At the bright end of $M_{\rm AB}=-24$,
 at $z \sim 3$, galaxies are primarily in dark matter halos of mass $\sim$ 10$^{13}$ M$_{\sun}$.
In comparison, such galaxies are central galaxies in groups and clusters today with masses above
10$^{14}$ M$_{\sun}$. At $z\sim 6$, $M_{\rm AB}=-24$ galaxies are primarily in dark matter halos with
mass $\sim 5 \times 10^{12}$ M$_{\sun}$. Similarly, $M_{\rm AB}=-18$ galaxies at $z \sim 3$ are
found in dark matter halos with mass 10$^{11}$ M$_{\sun}$, though one finds a few percent probability
that some of these galaxies are satellites of dark matter halos with masses between  10$^{12}$ M$_{\sun}$
and  10$^{13}$ M$_{\sun}$.  The four panels, when combined, show the mass-dependent redshift evolution  of
the galaxy luminosity. Luminous galaxies at high redshifts
are found at lower mass halos than dark matter halo masses that corresponds to the same galaxy luminosity
today. At the faint-end, $M_{\rm AB} > -20$, regardless of the redshift, faint galaxies are
essentially found in dark matter halos with a similar range in mass, though at low redshifts, a 30\% or more fraction of
low-luminous galaxies could be satellites in more massive halos.

While we have simply used the LF to establish the mass scale of $z \sim3$ to 6 galaxies,
previous attempts have also been made to establish the dark matter halo masses associated with these galaxies.
These estimates on halo masses were primarily based on observed galaxy clustering at these
redshifts (e.g., Steidel et al. 1998; Bullock et al. 2002;
Moustakas \& Somerville 2002). To compare with these halo mass estimates, which essentially lead to an estimate of
the average halo mass, we calculate the  probability distribution  of halo mass associated with high-redshift galaxies:
\begin{equation}
P(M|z)dM=\frac{\int_{L_{\rm min}} dL\; \Phi(L|M,z)}{\int_{L_{\rm min}} \Phi(L,z)} \frac{dn(z)}{dM} \; dM \, ,
\end{equation}
with the low-end of luminosity integral set at $L_{\rm min}$. 
For example, to compare with Moustakas \& Somerville (2002) estimate on the average $z \sim 3$ Lyman Break Galaxy (LBG)
halo mass, we set $L_{\rm min}$ to be that corresponding to $M_{\rm AB} \sim -20.8$;
at this magnitude level, the number density of $z \sim 3$ galaxies is $\sim 5 \times 10^{-3}$ h$^3$ Mpc$^{-3}$, comparable to the
number density of LBG galaxies used in Moustakas \& Somerville (2002) together with clustering statistics (correlation length and bias) of galaxies down to this density.

Figure~12 shows the probability distribution  of mass related to this number density of galaxies at $z \sim 3$.
The probability peaks around $\sim$ 7 $\times 10^{11}$ h$^{-1}$ M$_{\sun}$ with the 1$\sigma$ range of 
$(4$ -- $21) \times 10^{11}$ h$^{-1}$ M$_{\sun}$, which is
consistent with  the average halo mass of $5.5 \times 10^{11}$ $h^{-1}$ M$_{\sun}$
 suggested in Moustakas \& Somerville (2002) based on detailed halo model descriptions of the bias factor;
A straight forward description of the bias, assuming a single LBG per each dark matter halo, leads to
a halo mass of  $8 \times 10^{11}$ h$^{-1}$ M$_{\sun}$ (Adelberger et al. 1998), with mass values similar to
this by others (Baugh et al. 1998; Giavalisco \& Dickinson 2001). These values, based on the two-point correlation
function, are in good agreement with the mass estimate here based on model fits to the $z \sim 3$ LF.

This agreement between mass estimates from two-point clustering statistic and the one-point LF of the LBG
distribution suggests that the LBG distribution follows the standard hierarchical clustering model.
Motivated by a slight discrepancy in the mass estimates from the clustering argument and an estimate based on the
velocity dispersion of spectroscopic lines (Pettini et all. 2001), Scannapieco \& Thacker (2003) suggested a
modified model for the clustering of LBGs that involve a time-dependent correction associated with the merging
history (also, Furlanetto \& Kamionkowski 2005). 
The agreement between mass estimates suggested here argues against an additional correction to the bias factor.
Recent estimates of the LBG halo masses based on H$\alpha$ spectroscopic observations suggest that LBG halo masses
at $z \sim 3$ is $M > 3 \times 10^{11}$ M$_{\sun}$ (Erb et al. 2003), 
in good agreement with previous clustering based estimates as well as the estimates based here from the LF.
While the approach based on galaxy clustering allows the average halo mass scale 
to be established,  while velocity dispersion measurements only lead reliably to a lower limit on the mass scale,
making use of CLFs, here, we have quantified the dark matter halo mass of $z \sim 3$ and 6 galaxies
as a probability distribution function.  This probability distribution function captures not only the average
mass function, but also the dispersion related to the average mass. In addition to the average halo mass,
 a proper measurement of this dispersion must be considered before discrepancies are highlighted.

\subsection{$z >3$ UV LFs}

In Figure~13, we compare our predictions with several measurements in the literature on the UV LF at high redshifts.
At $z \sim 3$, Lyman-break galaxy (LBG) LF is measured by Steidel et al. (1999). We find reasonable agreement, though
we emphasize that our models are constructed for the rest B-band instead of rest UV-band related to these observations.
For reference, we also show the LF of galaxies at redshifts 8 and 10; while no measurements currently exist at these
high redshifts, the agreement, at least out to $z \sim 6$ suggests that our predictions may be directly testable in the near future
using deep IR images in near-IR wavelengths. Relative to $z \sim 6$ LF, the $z\sim 10$ LF predicts a factor of $\sim$ 10 
lower number density of galaxies at $M_{\rm AB} \sim -20$. Based on $i$-band dropouts, the surface density of
$z \sim 6$ galaxies, is roughly $\sim 0.5 \pm 0.2$ galaxies per square arcmin.  To detect a $z \sim$ 10 galaxy, it may be that
one must search over an area of $\sim$ 15 to 30 square arcmins, on average. The strong clustering of high-z galaxies,
discussed below, may affect search for these  galaxies.

In Figure~14, for comparison with existing measurements, we plot the cosmic luminosity density as a function of redshifts.
These luminosity densities are calculated via $\rho_L(z) \propto \int_{L_{\rm min}} L \Phi(L,z) dL$,
where we consider two values for the minimum luminosity. In Figure~14(a), we
consider a fainter cut off, at $M_{\rm AB} \sim -16$,  and compare with measurements of the
luminosity density at rest B-band from the literature. Our predictions generally agree at low redshifts, though, over predicts the
density measured by Dahlen et al. (2005) using LFs constructed from GOODS data.
In Figure~14(b), we set the low luminosity end of the integral to be roughly $0.3 L_\star$ ($M_{\rm AB} \sim -20$) 
at $z \sim 3$ to be consistent with most measurements by Bouwens et al. (2004a, 2004b, 2005). 
In this panel, we compare with measurements of the luminosity density at high redshifts in the rest UV-band.
Note that our underlying model here is designed for rest B-band and we do not attempt to include any corrections due to color differences
when comparing with observations at rest UV wavelengths. 
Our models suggest that the luminosity density at $z \sim$ 10 should be an order of magnitude
below what is suggested in Bouwens et al. (2005), under the assumption that they detect 
3 $z \sim 10$ dropouts in deep HST NICMOS fields
in a search area over $z \sim 15$ square arcmins. Based on our LFs, we expect at most a single $z \sim 10$ galaxy in  such
a small survey area. If a significant density of $z \sim 10$ galaxies were to exist, one would require  a sharp
increase in the evolution of galaxy luminosities at halo mass scales around $\sim 10^{11}$ M$_{\sun}$ than the luminosity evolution
we have suggested so far to explain $z \sim 3$ to 6 galaxy LFs.

\begin{figure}
\centerline{\psfig{file=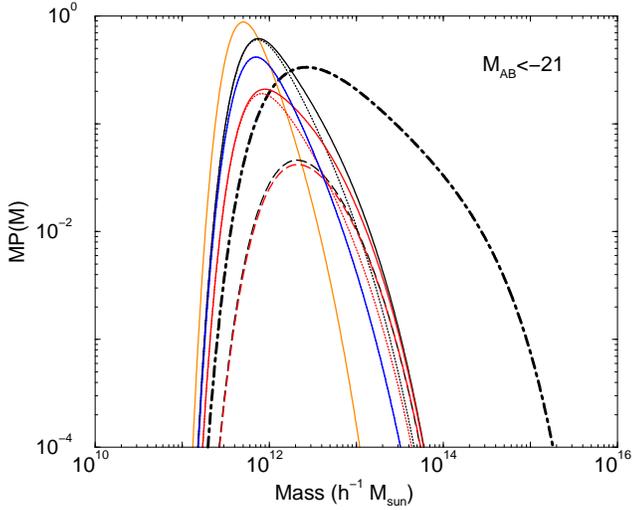,width=\hssize,angle=-90}}
\caption{The probability distribution function of halo mass $P(M,z=3)$ hosting 
galaxies down to number density of galaxies of $\sim  5 \times 10^{-3}$ $h^{3}$ Mpc$^{-3}$
at $z=3$. For comparison, we also show the same distribution, down to the same luminosity,
at today (dot-dashed line), and at a redshift of 6 (solid-orange line).
The probability distribution peaks around a mass of $7 \times 10^{11}$ h$^{-1}$ M$_{\sun}$
and is consistent with the average LBG halo mass determined by Moustakas \& Somerville (2002)
based on the clustering properties of galaxies down to the same number density of galaxies.}
\end{figure}

\begin{figure}
\centerline{\psfig{file=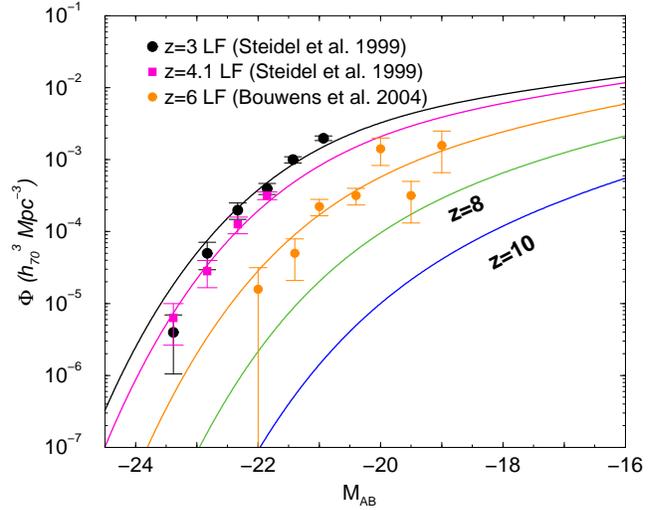,width=\hssize,angle=-90}}
\caption{The $z=3$ to 10 galaxy LFs; the plotted data show the measured LFs at $z \sim 3$ and 4.1 by Steidel et al. (1999)
and at $z \sim 6$ by Bouwens et al. (2004a). The solid lines show the expected LF (strictly speaking, in the rest B-band),
as a function of redshift with redshift values chosen, from top to bottom, of 3, 4, 6, 8 and 10.
The decline in the density of galaxies at redshifts greater than 3 is a reflection of the rapid decline in the
number density of dark matter halos; this decrease should not be explained as an effect associated with
negative luminosity evolution, since in our model, luminosities at the bright-end does evolve, though the number of halos hosting such \
bright galaxies is decreasing.}
\end{figure}

\begin{figure*}
\centerline{\psfig{file=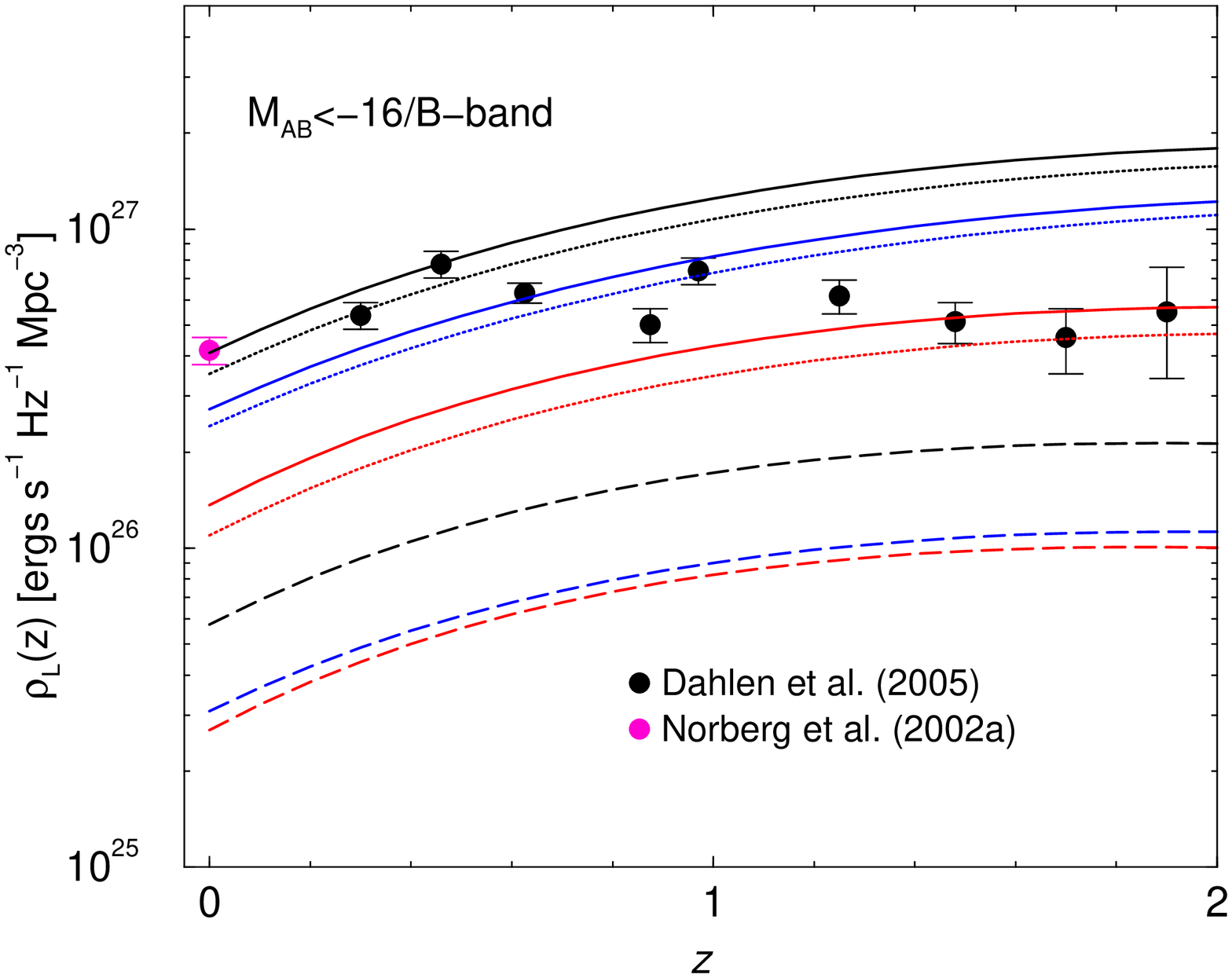,width=\hssize,angle=-90}
\psfig{file=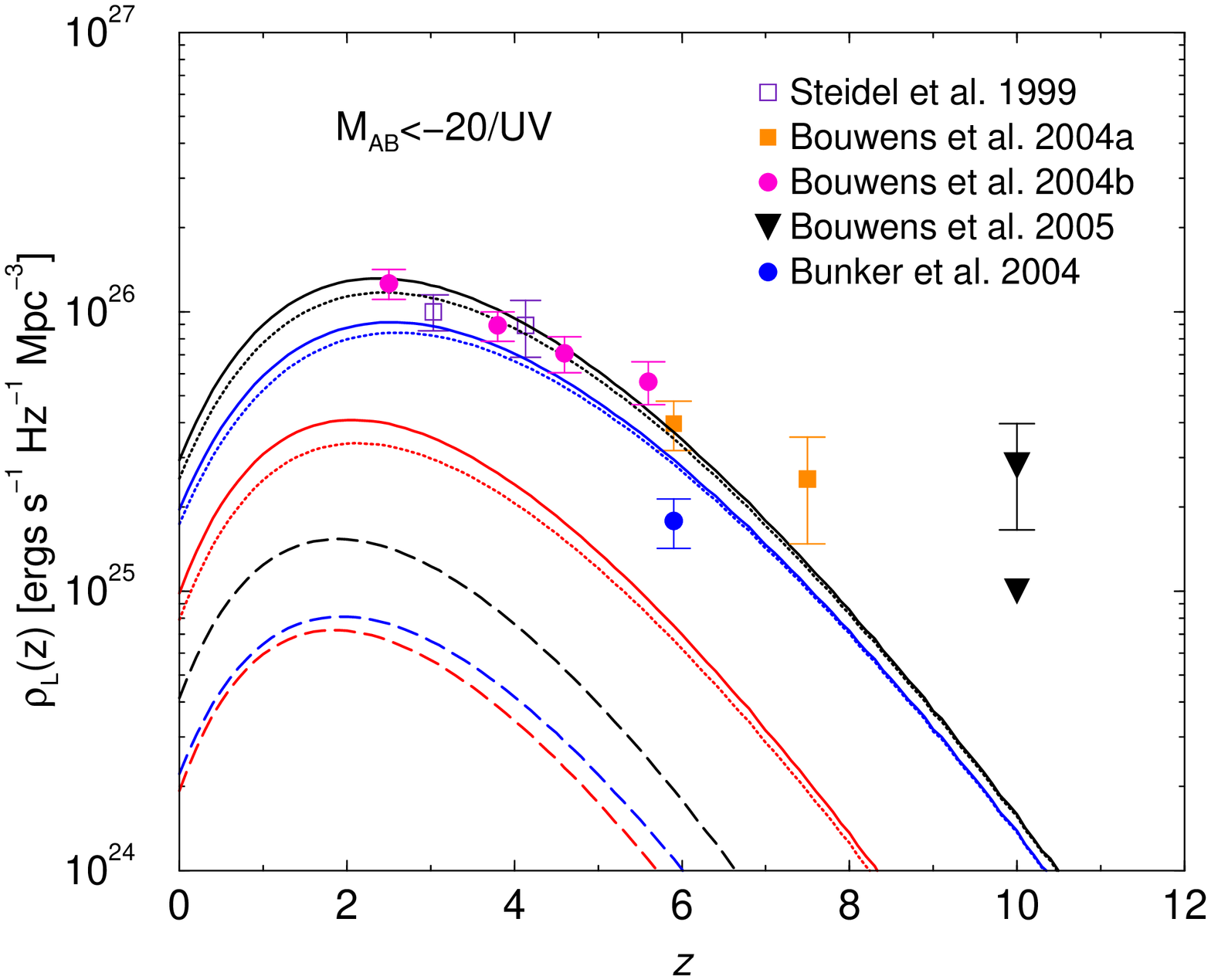,width=\hssize,angle=-90}}
\caption{The cosmic luminosity density  as a function of the redshift. The plotted curves are the expectations based
on the mass-dependent luminosity evolution model. In (a), we integrate to a the faint-end magnitude and compare with
predictions from the literature at rest B-bands. By design, we recover the luminosity density at $z \sim 0$
measured from the 2dFGRS survey (Norberg et al. 2002a), though we over predict the luminosity density at
redshifts $\sim$ 1 to 2 when compared to measurements by Dahlen et al. (2005). In (b), we consider a brighter cut off
in luminosity chosen to be roughly consistent with the cut-off in the measured luminosity densities by Bouwens et al. (2004a).
Now, we compare with measurements at high redshifts in the rest-UV wavelengths (see, figure panel for references). 
In converting to UV luminosity densities, 
we have ignored any color corrections resulting from the $L_c(M,z)$ relation designed to model
rest B-band LFs.}
\end{figure*}

\subsection{Galaxy Bias}

While our models can generally describe the LF of galaxies at redshifts out to 6,
another useful quantity to compare with observed data is the galaxy bias, as a function of the luminosity. Using the conditional
LFs, we calculate the luminosity and redshift dependent galaxy bias as
\begin{equation}
b(L,z) = \int b_{\rm halo}(M,z) \frac{\Phi_i(L|M,z)}{\Phi_i(L,z)} \frac{dn}{dM}(z) dM \, ,
\end{equation}
where $b_{\rm halo}(M,z)$ is the halo bias with respect to the linear density field (Sheth, Mo \& Tormen 2001; also, 
Efstathiou et al. 1988; Cole \& Kaiser 1989;
Mo et al. 1997) and $i$ denotes the galaxy type.

In Figure~15, we show the galaxy bias as a function of the luminosity. We also divide the sample to galaxy
types and redshifts at $z=0$, 3 and 6. At redshift of 3, we plot several estimates of the LBG bias;
we convert the bias--number density relations, from e.g., Bullock et al. 2002, to plot bias as a function of luminosity
 based on the expected number density of $z \sim 3$ galaxies, down to the given luminosity,
 given our model description for the LF.
The $b(L)$ relation provides a more direct approach to compare how galaxy bias evolves with redshift, than using
the bias factor as a function of the number density, though the latter is what is measured from the data;
With adequate statistics, in fact, it should be possible to measure
$b(L)$ relation at high redshifts directly from the data as has been at $z=0$ with SDSS (Zehavi et al. 2004) and
with 2dFGRS (Norberg et al. 2002b). As shown in Figure~11, the $z \sim 6$ galaxies
are biased 
by factors of $\sim$ 3 or higher relative to the linear density field. While there are no published
measurements of galaxy clustering at $z \sim 6$, the existence of clustered large-scale structures
has been noted through searches for Lyman-$\alpha$ emitters at these redshifts (e.g., Ouchi et al. 2005; Wang et al. 2005).
While our models were developed to discuss the LF and clustering of galaxies, one can easily 
modify the current prescriptions to describe statistics of sources such as Lyman-$\alpha$ emitters; we plan to
model Ly-$\alpha$ galaxy statistics at redshifts 4 to 7 in an upcoming paper.

\begin{figure}
\centerline{\psfig{file=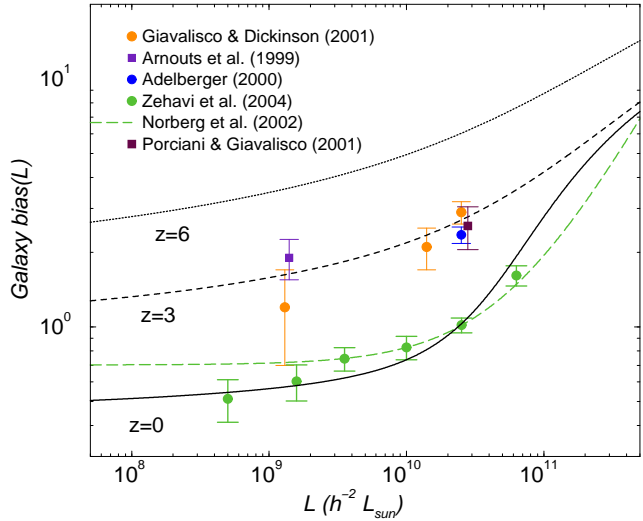,width=\hssize,angle=-90}}
\caption{The galaxy bias as a function of luminosity. We show bias measurements at $z=0$ in SDSS (Zehavi et al. 2004)
and with 2dFGRS (Norberg et al. 2002b). The $z \sim 3$ bias values come from the literature; we have converted the
LBG number density and bias relations (e.g., Bullock et al. 2002)
to luminosity--bias relation based on the expected number density of galaxies
down to a certain luminosity using the predicted LF of galaxies at $z \sim 3$. We also show a precision related to the
expected clustering bias of galaxies at $z \sim 6$; at this redshift, galaxies are significantly biased with typical
bias values $\sim$ 3.
}
\end{figure}

\section{Summary and Conclusions}

To summarize our discussion involving high-redshift galaxy LFs,  our main results are:

(1) Galaxy luminosities must evolve with redshift; while to describe rest B-band LFs out to $z \sim 3$, one can consider
either a mass-independent evolution, with luminosities increasing as $(1+z)^{\alpha}$, 
or a mass-dependent evolution scenario (see, Figure~1b); galaxy LFs from DEEP2 (Willmer et al.2005), 
COMBO-17 (Bell et al. 2004) and Gabasch et al. (2004) are more compatible with
a mass-dependent evolution model with the model parameter $\beta <0$ at the 3$\sigma$ confidence level. 
In this scenario, galaxies that are present in halos above $\sim 10^{12}$ M$_{\sun}$
brighten by factor of 4 to 6 between now and $z \sim 6$. This suggests that the star formation rate,
per given dark matter halo mass,  was increasing to high redshifts and that the star formation was more efficient in the past
relative to low redshifts. A conclusion similar to what we generally suggest here was also
reached by Dahlen et al. (2005) based on the rest B-band LF constructed from GOODS data.

(2) While the bright-end density of galaxies increases with redshift, the faint-end density remains 
essentially the same out to $z \sim 3$. The suggested evolution in then $L_c(M)$ relation with redshift
compensates for the decrease in dark matter halo density.  Another important reason why the number density of
faint galaxies does not decrease rapidly out to a $z \sim 2$ is that $L_c(M)$ relation flattens at a halo mass scale around
$10^{13}$ M$_{\sun}$; at low redshifts, the redshift evolution of the dark matter halo mass function mostly results in a decrease
in the number density of massive halos, while the number density around $10^{13}$ M$_{\sun}$ is not affected. 
At $z > 2$, the exponential cut-off associated with the halo mass function moves to mass scales around and below $10^{13}$ 
M$_{\sun}$.
This leads to a decrease in the number density of halos hosting central galaxies over  the range in luminosity of interest.
Thus, the turn over in the LF at the bright-end between redshifts 2 to 3 is a reflection on the lack of an
adequate density of dark matter halos with masses around $10^{13}$ M$_{\sun}$ to host bright galaxies.
One does not need to invoke different or multiple scenarios to explain why the density of bright galaxies
first increases and then decreases at redshifts greater than 2 to 3. Even in the presence of a decrease in the number density of
bright galaxies as a function of redshift, the suggested luminosity evolution continues to be present.

(3) The mass scale of galaxies at redshifts $\sim$ 3 is distributed between $\sim 4 \times$ 10$^{11}$ M$_{\sun}$ to
$2 \times$ 10$^{12}$ M$_{\sun}$ (Figure~12). The brightest galaxies at $z \sim 3$ (with $M \sim -24$) are found
in dark matter halos with masses slightly below 
10$^{13}$ M$_{\sun}$. Today, these galaxies are found in dark matter halos with masses
around 10$^{15}$ M$_{\sun}$, or as central galaxies in massive clusters. At a redshift of $\sim 6$, these galaxies
were in dark matter halos of mass few times 10$^{12}$ M$_{\sun}$. The $z \sim 3$ halo masses based on the one-point LF,
as determined with model fits here, agree with previous estimates in the literature based on the two-point
correlation function and the bias factor of these galaxies (e.g., Adelberger et al. 1998; Moustakas \& Somerville 2002)
as well as recent estimates of the halo masses for $z \sim 3$ LBGs based on spectroscopic observations (Erb et al. 2003)
and that a disagreement alluded in the literature (e.g., Scannapieco \& Thacker 2003; Furlanetto \& Kamionkowski 2005)
is not present. These agreements, especially between the LF and galaxy bias, also
suggest that no major correction to the bias factor
 of LBGs is necessary from additional effects such as merging (though such corrections may be necessary to
interpret other galaxy samples). 

(4) The number density of $z \sim 10$ galaxies are roughly a factor of 10 lower than the density of galaxies at $z \sim 6$;
if a higher density is found, the $L_c(M,z)$ relation must evolve rapidly at redshifts above 6 than suggested here to
explain $z \sim 3$ to 6 LFs. If our predictions our correct,
one must search roughly an area of $\sim$ 30 sqr. arcmins to find a galaxy at a redshift of 10.
The high redshift galaxies are extremely biased with respect to the linear dark matter density field; the bias
factor of galaxies at $z \sim 6$ varies from less than 3 to 10 for brightest galaxies. Our predictions for clustering
bias factors, as a function of luminosity, are in general agreement with estimated values based on the correlation length
(Figure~11), at least out to $z \sim 3$.

To conclude, we have presented a description of the
LF of galaxies as a function of redshift, in the rest B-band, using a simple empirical model. The approach
has the main advantage that one can extract underlying reasons associated with the observed evolution of the LF.
Here, we have characterized this evolution in terms of the $L_c(M,z)$ relation --- the luminosity of
central galaxies of dark matter halos as a function of halo mass and redshift. Given the rapid
decline in the number density of dark matter halos, in order to explain the presence of bright galaxies, we have suggested that
the $L_c(M,z)$ relation must evolve such that galaxies have a higher luminosity, when compared to today, at the
high mass end of the halo distribution.

The $z=0$, $L_c(M)$ relation was explained in Cooray \& Milosavljevi\'c (2005a) in terms of
dissipationless merging of galaxies in dark matter halos centers. The flattening of the $L_c(M)$
relation at halo mass scales $\sim 10^{13}$ M$_{\sun}$ was suggested as due to a decrease in the efficiency
at which dynamical friction  decays satellite orbits to merge with the central galaxy;
the dynamical friction time scale for the merging of satellites in halos with masses above few times
$10^{13}$ M$_{\sun}$ is more than Hubble time. As one moves to a high redshift, the flattening mass scale is expected to move
to a lower mass scale given that the age of the Universe, or the age of the halo since it formed, is lower than
the age today. Such a behavior is captured in our empirical model for the $L(M,z)$ relation based on
mass-dependent luminosity evolution.
While our approach has allowed us to capture the evolution associated with the $L_c(M,z)$ relation,
what physical processes govern this behavior is yet to be understood. 
Any reasonable explanation for the brightening of galaxies at high redshifts must consider the merging history
in addition to the underlying astrophysical reason why galaxies were brighter in the past when compared to today.
The $L_c(M,z)$ relation is certainly a key aspect in
galaxy formation and evolution, and we expect approaches involving numerical simulations (e.g., Kay et al. 2002)
and semi-analytic models (e.g., Benson et al. 2003) will be used to understand if what we have suggested on 
the evolution of this relation is consistent or not with astrophysical processes associated with galaxy formation and evolution.

{\it Acknowledgments:} 
The author thanks Emanuele Giallongo, Chris Willmer, and Armin Gabasch for electronic tables of the high-redshift LFs described 
in this paper and members of Cosmology and Theoretical Astrophysics groups at Caltech and UC Irvine for useful discussions. 
This paper was completed while the author was at the Aspen Center for Physics in Summer of 2005.


\begin{thebibliography}{}

\bibitem[Adelberger et al.<1998>]{Adelberger}
Adelberger, K. L., Steidel, C. C., Giavalisco, M., Dickinson, M., Pettini, M. \& Kellogg, M. 1998, \apj, 505, 18

\bibitem[Baldry et al.<2004>]{Baldry:04}
  Baldry, I. K., Glazebrook, K., Brinkmann, J., Ivezic, Z., Lupton, R. H., Nichol, R. C. \& Sza;ary, A. S. 2004, \apj, 600, 681


\bibitem[Balogh et al.<1998>]{Bal:98}
Baugh, C. M., Cole, S., Frenk, C. S. \& Lacey, C. G. 1998, \apj, 498, 504

\bibitem[Balogh et al.<2004>]{Bal:04}
 Balogh, M.~L., Baldry, I. K., Nichol, R., Miller, C., Bower, R., Glazebrook, K. 2004, \apj, 615, L101


\bibitem[Bell et al.<2004>]{Bell:04}
Bell, E. F., et al. 2004, ApJ, 609, 752

\bibitem[Benson et al.<2001>]{Benson:01} Benson, A.~J., Pearce, F.~R., Frenk, C.~S., Baugh, C.~M., \& Jenkins, A.
 2001, \mnras, 321, L7




\bibitem[Berlind et al.<2002>]{Berlind:02}
Berlind, A.~A.,  Weinberg, D. H., Benson, A. J. et al. 2003, \apj, 593, 1



\bibitem[Blanton et al.<2004>]{Blanton:04}
Blanton, M.~R., Eisenstein, D.~J., Hogg, D.~W., \& Zehavi, I.\ 2004, astro-ph/0411037

\bibitem[Bouwens et al.<2004>]{Bouwens:04}
  Bouwens, R., Illingworth, G. D., Thompson, R. I. et al. 2004a, \apj, 606, L25

\bibitem[Bouwens et al.<2004>]{Bouwens:04}
  Bouwens, R., Thompson, R.I., Illingworth, G. D. et al. 2004b, \apj, 616, L79

\bibitem[Bouwens et al.<2005>]{Bouwens:04}
  Bouwens, R., Illingworth, G. D., Thompson, R. I. \& Franx, M. J. 2005, \apj, 624, L5

\bibitem[Bullock et al.<2002>]{Bul:00}
  Bullock, J.~S., Wechsler, R.S., \& Somerville, R. S. 2002, \mnras, 329, 246

\bibitem[Bunker et al.<2004>]{Bun:04}
  Bunker, A. J., Stanway, E. R., Ellis, R. S. \& McMahon, R. G. 2004, \mnras, 355, 374

\bibitem[Coil et al.<2004>]{Coi04}
  Coil, A. L. et al. 2004, \apj, 609, 525


\bibitem[Cole \& Kaiser<1989>]{Cole:89}
Cole, S. \& Kaiser, N. 1989, \mnras, 237, 1127

\bibitem[Colless et al. <2001>]{Col:01}
  Colless, M. et al. 2001, MNRAS, 328, 1039


\bibitem[Cooray et al. <2000>]{Cooetal00}
        Cooray A., Hu W., \& Miralda-Escud\'e, J.\ 2000, \apj, 535, L9

\bibitem[Cooray <2002>]{Cooetal02}
        Cooray A. 2002, \apj, 576, L105

\bibitem[Cooray \& Sheth<2002>]{Cooray:02}
Cooray, A., \& Sheth, R.\ 2002, Phys.\ Rep., 372, 1 (astro-ph/0206508)

\bibitem[Cooray \& Milosavljevi\'c<2005a>]{Cooray:05}
Cooray, A., \& Milosavljevi\'c, M.\ 2005a, ApJ in press (astro-ph/0503596)

\bibitem[Cooray \& Milosavljevi\'c<2005b>]{Cooray:05}
Cooray, A., \& Milosavljevi\'c, M.\ 2005b, ApJ in press (astro-ph/0504580)

\bibitem[Cooray <2005>]{Cooray:05}
Cooray, A. 2005, \mnras submitted (astro-ph/0505421)

\bibitem[Croton et al.<2004>]{Cro:04}
  Croton, D.~J. et al. 2004\, \mnras in press (astro-ph/0407537)

\bibitem[Dahlen et al. <2005>]{Dah05}
Dahlen, T., et al. 2005, \apj in press (astro-ph/0505297)

\bibitem[Davis et al.<2003>]{Davis:03}
Davis, M. et al. 2003, SPIE, 4834, 161



\bibitem[De Lucia et al.<2004>]{DeLucia:04}
De Lucia, G., Kauffmann, G., Springel, V., White, S.~D.~M., Lanzoni, B., Stoehr, F., Tormen, G., \& Yoshida, N.\ 2004, \mnras, 348, 333



\bibitem[Drory et al.<2003>]{Drory:03} Drory, N., Bender, R., Feulner, G., Hopp, U., Marston, C., Snigula, J.,
 \& Hill, G.~J.\ 2003, \apj, 595, 698

\bibitem[Efstathiou et al.<1988>]{Efs88}
  Efstathiou, G., Frenk, C. S., White, S. D. M. \& Davis, M. 1988, \mnras, 235, 715

\bibitem[Erb et al.<2003>]{Erb:03}
Erb, D. K., Shapley, A. E., Steidel, C. C. et al. 2003, \apj, 591, 101

\bibitem[Faber et al.<2005>]{Faber:05}
Faber, S. M. et al. 2005, ApJ submitted, astro-ph/0506044

\bibitem[Furlanetto \& Kamionkowski<2005>]{Furl:05}
Furlanetto, S. R. \& Kamionkowski, M. 2005, MNRAS submitted, astro-ph/0507650

\bibitem[Gabasch et al.<2004>]{Gabasch:04}
  Gabasch, A. et al. 2004, A\&A, 421, 41

\bibitem[Giallongo et al.<2005>]{Giallongo:05}
  Giallongo, E., Salimbeni, S., Menci, N. et al. 2005, \apj, 622, 116

\bibitem[Giavalisco \& Dickinson<2001>]{Gia:01}
  Giavalisco, M. \& Dickinson, M. 2001, \apj, 550, 177




\bibitem[Huang et al.<2003>]{Huang:03}
Huang, J.-S., Glazebrook, K., Cowie, L.~L, \& Tinney, C.\ 2003, \apj, 584, 203

\bibitem[Jenkins et al.<2001>]{Jenkins:01}
Jenkins, A., Frenk, C.~S., White, S.~D.~M., Colberg, J.~M., Cole, S., Evrard, A.~E., Couchman, H.~M.~P., \& Yoshida, N.\ 2001, \mnras, 321, 372

\bibitem[Kay et al.<2002>]{Kay:02}
Kay, S.~T., Pearce, F.~R., Frenck, C.~S., \& Jenkins, A.\ 2002, \mnras, 330, 113


\bibitem[Kravtsov et al.<2004>]{Kravtsov:04}
Kravtsov, A.~V., Berlind, A.~A., Wechsler, R.~H., Klypin, A.~A., Gottl{\" o}ber, S., Allgood, B., \& Primack, J.~R.\ 2004, \apj, 609, 35

\bibitem[Lilly et al.<1996>]{Lilly}
  Lilly, S. J., Le Fevre, O., Hammer, F. \& Crampton, D. 1996, \apj, 460, L1


\bibitem[Lin, Mohr, \& Stanford<2004>]{Lin:04}
Lin, Y., Mohr, J.~J., \& Stanford, A.\ 2004, \apj, 610, 745



\bibitem[Moustakas \& Somerville<2002>]{Mou02}
  Moustakas, L. A. \& Somerville, R. S. 2002, \apj, 577, 1

\bibitem[Norberg et al.<2002>]{Norberg:02}
Norberg, P., et al.\ 2002a, \mnras, 336, 907

\bibitem[Norberg et al.<2002>]{Norberg:02b}
Norberg, P., et al.\ 2002b, \mnras, 332, 827

\bibitem[Oguri \& Lee<2004>]{Oguri:04}
  Oguri, M. \& Lee, J.\ 2004, \mnras, astro-ph/0401628

\bibitem[Ouchi et al.<2005>]{Ouch:05}
  Ouchi, M., Shimasaku, K., Akiyama, M. et al. 2005, \apj, 620, L1



\bibitem[Pettini et al.<2001>]{Pettini:01}
Pettini, M. et al. 2001, \apj, 554, 981

\bibitem[Press \& Schechter<1974>]{Press:74}
  Press, W. H., \& Schechter, P.\ 1974, \apj, 187, 425

\bibitem[Reed et al.<2003>]{Reed:03}
  Reed, D., Gardner, J., Quinn, T., et al. 2003, \mnras, 346, 565



\bibitem[Scannapieco \& Pettini<2003>]{Scan:03}
Scannapieco, E., \& Thacker, H. J. 2003, \apj, 590, L69


\bibitem[Schechter<1976>]{Schechter:76}
  Schechter, P.\ 1976, \apj, 203, 297

\bibitem[Scoccimarro et al. <2001>]{Scoetal01}
      Scoccimarro R., Sheth R., Hui L., Jain B., 2001, ApJ, 546, 20

\bibitem[Seljak<2000>]{Seljak:00}
Seljak, U.\ 2000, \mnras, 318, 203

\bibitem[Sheth \& Tormen<1999>]{Sheth:99}
 Sheth, R.~K., \& Tormen, G.\ 1999, \mnras, 308, 119


\bibitem[Sheth, Mo, \& Tormen<2001>]{Sheth:01}
Sheth, R.~K., Mo, H.~J., \& Tormen, G.\ 2001, \mnras, 323, 1


\bibitem[Spergel et al.<2003>]{Spergel:03} Spergel, D.~N., et al.\ 2003, \apjs, 148, 175

\bibitem[Stark \& Ellis<2005>]{Stark:05} Stark, D. P. \& Ellis, R. S. 2005, in
 in ``First Light \& Reionization'', eds. E. Barton \& A. Cooray, New Astronomy Reviews, in press
(astro-ph/0508123)

\bibitem[Steidel et al.<1998>]{Steidel:98} Steidel, C., Adelberger, M., Dickinson, Giavalisco,  M., Pettini, M. \& Kellogg, M. 1998,
 \apj, 492, 428

\bibitem[Steidel et al.<1999>]{Steidel:99} Steidel, C., Adelberger, K. L., Giavalisco, M., Dickinson, M. \& Pettini, M. 1999,
 \apj, 519, 1



\bibitem[Trentham \& Tully<2002>]{Trentham:02} Trentham, N., \& Tully, R.~B.\ 2002, \mnras, 335, 712


\bibitem[Vale \& Ostriker<2004>]{Vale:04}
Vale, A., \& Ostriker, J.~P.\ 2004, \mnras, 353, 189




\bibitem[Wang et al.<2005>]{Wan05}
Wang, J. X., Malhotra, S. \& Rhoads, J. E. 2005, \apj, 622, L77



\bibitem[Willmer et al.<2005>]{Willmer:05}
  Willmer, C. N. A., et al. ApJ submitted, astro-ph/0506041

\bibitem[Wolf et al.<2003>]{Wolf:03}
Wolf, C., Meisenheimer, K, Rix, H.-W., Borch, A., Dye, S. \& Kleinheinrich, M. 2003, A\&A, 401, 73

\bibitem[Wolf et al.<2001>]{Wolf:01}
Wolf, C., Meisenheimer, K \& R\"oser, H.-J. 2001, A\&A, 365, 660

\bibitem[Yan et al. <2003>]{Yan03}
  Yan, R., Madgwick, D. S. \& White, M. 2003, \apj, 598, 848


\bibitem[Yan et al.<2005>]{Yan:05}
Yan, H. et al. 2005, ApJ in press, astro-ph/0507673

\bibitem[Yang et al.<2003a>]{Yang:03}
Yang, X., Mo, H.~J., Kauffmann, G., \& Chu, Y.~Q.\ 2003a, \mnras, 339, 387

\bibitem[Yang, Mo, \& van den Bosch<2003>]{Yang:03b}
Yang, X., Mo, H.~J., \& van den Bosch, F.~C.\ 2003b, \mnras, 339, 1057

\bibitem[Yang et al.<2005>]{Yang:05}
Yang, X., Mo, H.~J., Jing, Y.~P., \& van den Bosch, F.~C.\ 2005, \mnras, 358, 217

\bibitem[York et al.<2000>]{York:00}
  York, D.~G., et al.\ 2000, \aj, 120, 1579


\bibitem[Zheng et al.<2004>]{Zheng:04}
  Zheng, Z., et al.\ 2004, preprint (astro-ph/0408564)





\end{thebibliography}
\end{document}